\documentclass{vldb}
\usepackage{graphicx}
\usepackage{balance}  % for  \balance command ON LAST PAGE  (only there!)
\usepackage{colortbl}
\usepackage{color}
\usepackage{subcaption}
\usepackage{fixltx2e}
\usepackage{booktabs}
\usepackage{times}
\usepackage{arydshln}
\usepackage{tabularx}
\usepackage{mdframed}
\usepackage{amssymb}
\usepackage{mathrsfs}
\usepackage[linesnumbered,ruled,vlined]{algorithm2e}
\usepackage{environ}
\usepackage{amsmath}

\usepackage{amssymb}

\usepackage{fancyvrb}

\usepackage{graphicx}

\usepackage{float}

\usepackage{caption}

\usepackage{framed}

\usepackage{dsfont}
\usepackage{multirow}

\usepackage{balance}

\usepackage[table]{xcolor}

\usepackage{soul}

\usepackage[normalem]{ulem}

\usepackage{hyperref}

\newenvironment{compact_enum}
{\setlength{\leftmargini}{1em}
\begin{enumerate}
  \setlength{\labelsep}{.3em}
  \setlength{\itemsep}{.4em}
  \setlength{\parskip}{0pt}
  \setlength{\parsep}{0pt}}
{\end{enumerate}}

\newenvironment{compact_item}
{\setlength{\leftmargini}{1em}
\begin{itemize}
  \setlength{\labelsep}{.3em}
  \setlength{\itemsep}{.4em}
  \setlength{\parskip}{0pt}
  \setlength{\parsep}{0pt}}
{\end{itemize}}

\def\semijoin{\ltimes}

\usepackage{tikz}

\usepackage{commentary-private} %make comments private

\newif\ifDisplayBasicVersion
\DisplayBasicVersionfalse           %hide basic version

\newif\ifDisplayFullVersion
\DisplayFullVersiontrue             %show full version
\NewEnviron{gotobasic}{
\ifDisplayBasicVersion
\textcolor{black}{\BODY}
\fi}

\NewEnviron{gotofull}{
\ifDisplayFullVersion
\textcolor{black}{\BODY}
\fi}

\usepackage{array}
\newcolumntype{L}[1]{>{\raggedright\let\newline\\\arraybackslash\hspace{0pt}}m{#1}}

\newcommand{\yannis}[1]{\reminder{green}{Yannis}{#1}}
\newcommand{\matthias}[1]{\reminder{blue}{Matthias}{#1}}
\newcommand{\chunbin}[1]{\reminder{red}{Chunbin}{#1}}

\newtheorem{example}{Example}[section]
\newtheorem{base}{Base}[section]
\newtheorem{definition}[base]{Definition}

\definecolor{Gray}{gray}{0.85}
\newcommand{\gt}[1]{\small \texttt{\textbf{#1}} \normalsize}
\newcommand{\rijoin}[0]{\stackrel{\rightarrow}{\Join}}
\newcommand{\riinter}[0]{\stackrel{\rightarrow}{\cap}}
\newcommand{\risemijoin}[0]{\stackrel{\rightarrow}{\rtimes}}
\newcommand{\nonl}{\renewcommand{\nl}{\let\nl\oldnl}}% Remove line number for one line
\newcommand{\smallfootnote}[1]{\footnote{\small{#1}}}

\newcommand{\fastr}{GQ-Fast }
\newcommand{\fastrs}{GQ-Fast's }
\newcommand{\fastrUA}{GQ-Fast-UA }
\newcommand{\fastrUAB}{GQ-Fast-UA(Binary) }
\newcommand{\fastrUAM}{GQ-Fast-UA(Map) }
\newcommand{\fastrUAS}{GQ-Fast-UA(Non-Share) }
\begin{document}

% ****************** TITLE ****************************************

\title{Fast In-Memory SQL Analytics on Graphs}
\date{}
%Fast In-Memory SQL Analytics on Relationships between Entities

\numberofauthors{1}

 \author{
 \alignauthor
 Chunbin Lin, Benjamin Mandel, Yannis Papakonstantinou, Matthias Springer\\
       \affaddr{CSE, UC San Diego}\\
       \email{\{chunbinlin, bmandel, yannis, mspringer\}@cs.ucsd.edu}\\
 }

\maketitle

\begin{abstract}
%What kind of queries
We study a class of graph analytics SQL queries, which we call \emph{relationship queries}. Relationship queries are a wide superset of fixed-length graph reachability queries and of tree pattern queries. A relationship query performs selections, joins and semijoins (WHERE clause subqueries) over tables that correspond to entities (i.e., vertices) and binary relationships (i.e., edges). Intuitively, it discovers target entities that are reachable from source entities specified by the query. It usually also finds aggregated scores, which correspond to the target entities and are calculated by applying aggregation functions on measure attributes, which are found on (1) the target entities, (2) the source entities and (3) the paths from the sources to the targets. We present real-world OLAP scenarios, where efficient relationship queries are needed. However, row stores, column stores and graph databases are unacceptably slow in such OLAP scenarios. We briefly comment on the straightforward extension of relationship queries that allows accessing arbitrary schemas.

The \emph{\fastr} in-memory analytics engine utilizes a bottom-up fully pipelined query execution model running on a novel data organization that combines salient features of column-based organization, indexing and compression. Furthermore, \fastr compiles its query plans into executable C++ source code. Besides achieving runtime efficiency, \fastr also reduces main memory requirements because, unlike column databases, \fastr selectively allows more dense forms of compression including heavy-weighted compressions, which do not support random access.

%Chunbin:The performance of FastR
We used \fastr to accelerate queries for two OLAP dashboards in the biomedical field.
\begin{gotofull}
The first dashboard runs queries on the PubMed dataset and the second one on the SemMedDB dataset.
\end{gotofull}
It outperforms Postgres by 2-4 orders of magnitude and outperforms MonetDB and Neo4j by 1-3 orders of magnitude when all of them are running on RAM. Our experiments dissect the \fastrs advantage between (i) the use of compiled code, (ii) the bottom-up pipelining execution strategy and (iii) various data structure choices. Other analysis and experiments show the space savings of \fastr due to the appropriate use of compression methods, while we also show that the runtime penalty incurred by the dense compression methods decreases as the number of CPU cores raises.
\yannis{To add: Finally, we comment on future work that will allow relationship queries to be subqueries of larger queries, hence allowing any query to be supported, while relationship (sub)queries will have the benefit of higher performance.}
\end{abstract}

\chunbin{I defined new command ``$\setminus$fastr'' in order to provide a convenient way to change the name of our algorithm}
\chunbin{Texts with red color are in basic version, while that with blue color are in full version}

\yannis{To Chunbin: We should also highlight the graph database aspect on the title and
the abstract. Last time we did not do it because we were afraid of
missing references to graph database works but this time there should
not be such fear. Comparing to graphs, take a look at the Neo4j graph
database. We should add a column for its run-time performance in the
graph that compares Postgres, MonetDB etc. I think it is worse than
MonetDB and probably not much better than Postgres. Ben Good was aware
of it and they did not use it. Chunbin:yes}

\yannis{High level issues: (1) Complete the definition of RQNA (2) Investigate compiled code related works, compare, add citations and related work discussion. For example, Cristoph Koch has done a lot of work on compiled code. (3) Complete the query processing section. It is total mess. (4) Align notations. All sections must use the same notation. Have a pattern on why you use certain styles. E.g., I'm trying to figure out what is your convention for using mathcal. I cannot see one.(5) The benefit over OMC seems unreasonably high. A larger problem presently is that the codes of PMC and OMC are at a very high level, where the reviewer can correctly argue that there may all sorts of inefficiencies hiding in the details. For example, when we write ``Store the obtained values to $\mathcal{R}_2$" what exactly happens in the code?  There should be appendix material (which can become extended version) that shows the complete PMC and OMC code. Why do we do $\mathcal{R} \mathcal{R}_2$? Why didn't we place the $\mathcal{A}$ values directly into $\mathcal{R}$ and we now have to copy? What kind of data structures are those? Are they efficient data structures, like the arrays that we are using in FastR? If so, how big are the allocations? If they arenot arrays what are they? (*) Read carefully, edit and make sure the whole paper is well defined and has reasonable flow. Chunbin: I fixed them.}

\section{Introduction}
\label{sec: introduction}

%Chunbin: What is a relationship query.
A common type of Online Analytical Processing (OLAP) queries on graphs
\begin{gotofull}\cite{chen2008graph,chen2009graph,zhao2011graph,wang2014pagrol}\end{gotofull}
\begin{gotobasic}\cite{chen2008graph,zhao2011graph,wang2014pagrol}
\end{gotobasic}
proceeds in three steps: First, the query selects a set of source entities that satisfy certain user-provided properties. Then, the query discovers target entities that are reachable from the source entities. If it is an SQL query (as is the case in this paper), it ``navigates" between entities via join operations. During the navigation the query collects measure attributes about the nodes and the edges. In the last step, these measures are aggregated and assigned to the target entities. The resulting aggregate measures typically indicate metrics of ``relevance" or ``importance" of the discovered entities in the context created by the source entities. We call such queries {\em relationship queries}.

\begin{figure}
\center
\includegraphics[width=0.4\textwidth]{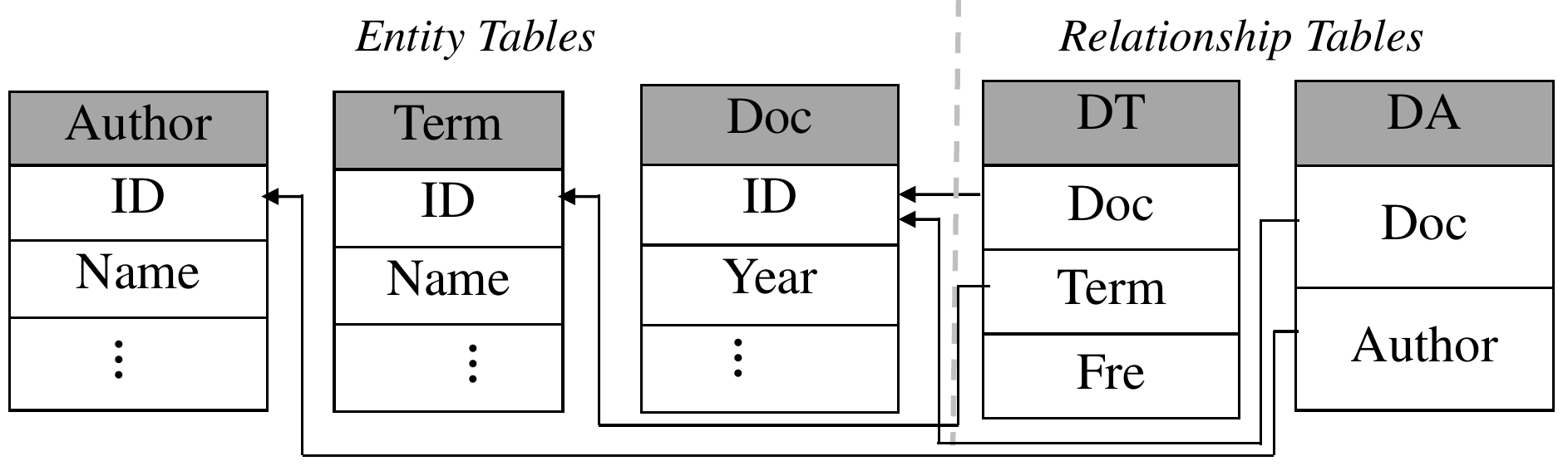}
\caption{\textbf{Entities and relationships in Pubmed}.
%Tables \gt{Term}, \gt{Doc} and \gt{Author} are entity tables while tables \gt{DT} and \gt{DA} are relationship tables.
}
\label{figure:pubmed-schema}
\end{figure}

%Chunbin: Introduce the entity and relationship tables
In the interest of further specifying the subset of SQL queries that corresponds to relationship queries, we classify the tables of an SQL schema into two categories, which correspond to the entities and the relationships of the E/R model
\begin{gotobasic}
\cite{garcia2008database}
\end{gotobasic}
\begin{gotofull}
\cite{sapia1999extending,golfarelli1998conceptual,garcia2008database}
\end{gotofull}: {\em relationship tables} and {\em entity tables}. Each tuple of an entity table corresponds to a real-life entity, which can be thought of as a vertex in a graph. Each entity table has an ID (primary key) attribute. Relationship tables, on the other hand, capture many-to-many relationships. This paper focuses on binary relationships. Therefore a (binary) relationship table has two \textit{foreign key attributes} referencing the IDs of respective entity tables. A relationship table may also have \textit{measure attributes}. Each tuple of a relationship table can be thought of as one edge between the two entities referenced by the foreign key attributes. The measure attributes can be thought of as attributes of the edge.  Figure~\ref{figure:pubmed-schema} is a relational representation of entities and relationships in Pubmed, the premier public biomedical database\smallfootnote{http://www.ncbi.nlm.nih.gov/pubmed}.
For example, the \gt{Term} entity table has one tuple for each term of the PubMed MeSH (Medical Subject Heading) \smallfootnote{https://www.nlm.nih.gov/bsd/disted/meshtutorial/themeshdatabase/} ontology. The relationship table \gt{DT} has two foreign key attributes  \gt{Doc} and \gt{Term} referring to the \gt{ID} column of entity tables \gt{Doc} and \gt{Term}. The measure attribute \gt{Fre} stands for ``frequency". A tuple $(d,t,f)$ indicates that the term $t$ appears in the document $d$, $f>0$ times.

%Chunbin: definition of relationship queries and an example query
An SQL query over relationship tables and entity tables, is a relationship query if (1) in its algebraic form, it involves $\sigma$, $\pi$, $\Join$, $\ltimes$ operators and an optional aggregation operator $\gamma$ at the end; (2) each join and semijoin condition involves an equality between (primary or foreign) key attributes and (3) aggregations group-by on a single primary key or foreign key.
%i.e., they group-by on an entity set.
Notice, the set of relationship queries includes bounded-length graph reachability (path finding) queries and tree pattern queries, where the graph is SQL-encoded, i.e., the edges are defined by foreign keys.
The restrictions (1-3) are commonly met in practice. In particular, most joins are based on primary keys and foreign keys) and do not prevent relationship queries from having wide applications. Furthermore, we show that restriction~(3) can be removed at a minor cost.
Section~\ref{sec:queries} formally defines the wide class of relationship SQL queries and argues for its prevalence. It provides a variety of real-life applications and queries relating to PubMed~\cite{kilicoglu2008semantic} and the SemMedDB~\cite{kilicoglu2012semmeddb} biomedical knowledge databases.

Next we illustrate a relationship query. Given the very high volume of publications (more than $23$M) and authors (more than $6.3$M) in PubMed, one may want to issue the following \emph{Author Similarity (AS)} relationship query that finds authors that publish on the same topics with the author with ID \gt{7}. \begin{gotofull}More precisely, the query finds the authors (identified by \gt{da2.Author}) whose publications (identified by \gt{d.ID}) relate to the MeSH terms (identified by \gt{dt1.Term}) in the publications \gt{da1.Doc} of the author \gt{7}.\end{gotofull} Furthermore, each discovered author is given a weight/similarity score, by first computing the similarity of her publications' terms to the terms of the publications of author \gt{7}.   The similarity function is a typical cosine-based: Conceptually there is a vector $v^i=[t_1^i,\ldots,t_m^i]$ for each author $a_i$; $t_j^i$ is the number of times that the term $t_j$ appeared in the publications of $a_i$. Then the similarity between authors $a_1$ and $a_2$ is the cosine of the vectors $v^1$ and $v^2$. Additional weight is given to recent publications.%
\smallfootnote{{For simplicity, we ignore that in practice the weight function adjusts frequency to whether the actual content of the paper is available or just the abstract. In practice, we also use a log-based adjustment of frequency.}}

\begin{tabbing}
AS query:\\
\=\gt{SELECT da2.Author,SUM(dt1.Fre$\times$dt2.Fre)/(2017-d.Year)}\\
\>\gt{FROM }\=\gt{(}\=\gt{(}\=\gt{(DA da1 JOIN DT dt1 ON da1.Doc=dt1.Doc)}\\
\>\>\>\>\gt{JOIN DT dt2 ON dt1.Term = dt2.Term)}\\
\>\>\>\gt{JOIN Doc d ON dt2.Doc=d.ID)}\\
\>\>\gt{JOIN DA da2 ON dt2.Doc=da2.Doc}\\
\>\gt{WHERE da1.Author = 7} \\
\>\gt{GROUP BY da2.Author}\\
\end{tabbing}

\yannis{to Chunbin: This sentence should be moved (and modified) to Section~\ref{sec:queries} (or wherever we talk about the application) since it is not clear yet what live feedback and search results it talks about: ``For example, it can provide live feedback and update search results (authors and terms such as diseases, substances, etc.) instantaneously while the user is typing and modifying search criteria." chunbin: agree}

Given the analytical nature of relationship queries, column-oriented database systems are much more efficient than row-stores and graph databases, as our experiments verified (see Section~\ref{sec:experiments})
\begin{gotofull}
~\cite{stonebraker2005c,abadi2008column,abadi2008query,abadi2009column,nes2012monetdb,nguyen2015join}.
\end{gotofull}
\begin{gotobasic}
~\cite{stonebraker2005c,abadi2008column,abadi2009column,nes2012monetdb}.
\end{gotobasic}
 Nevertheless, the obtained performance is often not sufficient for online queries and interactive applications. For example, the AS query takes $1.24$ hours on the column database MonetDB~\cite{nes2012monetdb}, $8.20$ hours on the row database Postgres\smallfootnote{http://www.postgresql.org/}
\begin{gotofull}
~\cite{momjian2001postgresql},
\end{gotofull}
, and 2.05 hours on the graph database Neo4j\smallfootnote{http://neo4j.com/},
\begin{gotofull}
~\cite{Neo4j},
\end{gotofull}
even though we fully cached the data in main memory in all of them. In contrast, \fastr executes the AS query in just 5.66 seconds, which presents an improvement of many orders of magnitude. Furthermore, we show that the space requirements of \fastr are generally lower. Section~\ref{sec:queries} provides multiple additional examples of OLAP relationship queries.

\begin{figure*}
\includegraphics[width=1\textwidth]{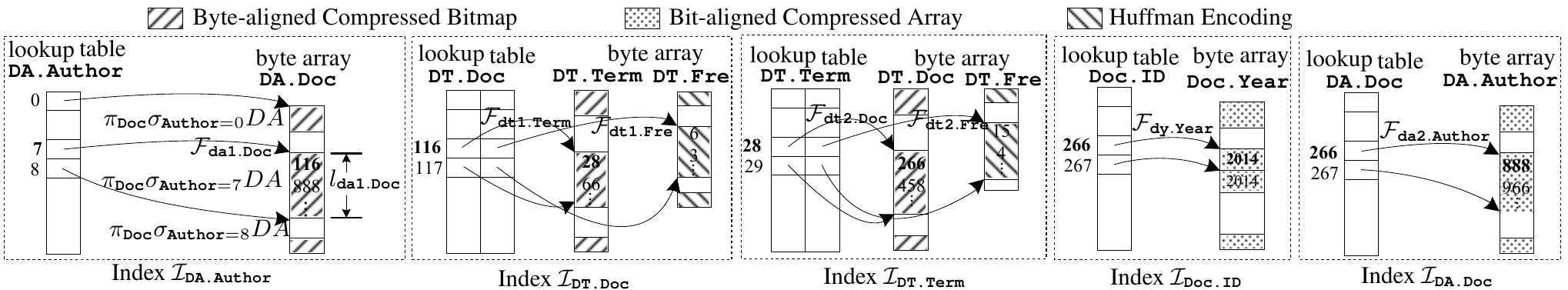}
\caption{\textbf{Example of fragments and query processing}.Fragment $\pi_{\gt{Doc}}\sigma_{\gt{Author}=7} \gt{DA}$ is encoded with byte-aligned bitmap while fragments $\pi_{\gt{Term}}\sigma_{\gt{Doc}=116} \gt{DT}$ and $\pi_{\gt{Fre}}\sigma_{\gt{Doc}=116} \gt{DT}$ are encoded by Bit-aligned compressed array and Huffman encoding respectively.}
\label{figure:example_fragments}
\end{figure*}

%Chunbin: fragment-structure
% Check how repetitive is this against the data structure section. Write it to only capture an example.
\fastr achieves such superior performance by combining a novel fragment-based data structure with a novel type of compiled execution plans.
Generally, for each binary relationship table $R(F_1,$ $F_2,M_1,\ldots,M_m)$ where $F_1,F_2$ are foreign keys and each $M_i$ is a measure, \fastr conceptually (but not physically) makes two copies $R_1$ sorted by $F_1$ and $R_2$ sorted by $F_2$.  Physically, \fastr makes an index $\mathcal{I}_{R.F_1}$ corresponding to $R_1$ and an $\mathcal{I}_{R.F_2}$ corresponding to $R_2$. Assuming $F_1$ has $n$ unique values $t_1,\ldots, t_n$, \fastr produces $n$ fragments $\pi_{F_2}\sigma_{F_1=t_i}(R_1)$ and $n\times m$ fragments   $\pi_{M_j}\sigma_{F_1=t_i}(R_1)$. A fragment is a list of values potentially with duplicates.
Figure~\ref{figure:example_fragments} shows example indices and fragments for the schema in Figure~\ref{figure:pubmed-schema}. The index $\mathcal{I}_{\gt{DA.Author}}$ associates the \gt{Author} entity \gt{7} with the \gt{DA.Doc} fragment $\pi_{\gt{Doc}}\sigma_{\gt{Author}=7}(\gt{DA})$. Note that, this new organization allows \fastr to directly obtain values (eg, document IDs) that are associated to a given value (eg, the given author ID \gt{7}), without accessing row ids.

For space saving reasons explained later, the fragments of an attribute are stored consecutively but encoded separately in large byte arrays. For example, the \gt{DA.Doc} byte array (of the $\mathcal{I}_{\gt{DA.Author}}$ index) contains the values of the \gt{DA.Doc} attribute. Therefore, at a first glance, it appears as if the \gt{DA.Doc} is identical to a column in a column database. However, there is an important difference, which is motivated by this key observation: \fastr query plans do not need random access within any individual fragment of the \gt{DA.Doc} byte array. Rather, when a fragment (say, $\pi_{\gt{Doc}}\sigma_{\gt{Author}=7} \gt{DA}$) is accessed, all its data will be used. Based on this observation, \fastr allows dense encodings of each individual fragment, such as Huffman encoding~\cite{huffman1952method,storer1988data,ChungW99}. Such encodings do not allow random access within the fragment. Note that a typical fragment has relatively small size and can typically fit in the L1 cache or, at least, in the L2 cache. Hence, its decoding is not penalized with random accesses to the RAM.

Still, plans for relationship queries need efficient access to whole fragments, i.e., need to efficiently find their start and end. This is achieved by the offset-based lookup tables of the \fastr indices. Section~\ref{sec:data-structure} further discusses aspects of the \fastr indices, which lead to runtime performance and space savings.

%Chunbin: FastR algorithm preview
The \fastr query plans run on the indices (exclusively). They essentially employ a bottom-up pipelined execution model to avoid large intermediate results. In addition, \fastr employs a code generator to compile the query plan into executable C++ code. The compiled code utilizes simple for-loops to access the elements in a fragment. Such tight for-loops without function calls create high instruction locality which eliminates the instruction cache-miss problem. In addition, such simple loops are amenable to compiler optimization, and CPU out-of-order speculation
\begin{gotofull}
~\cite{abadi2008query,nes2012monetdb}.
\end{gotofull}
\begin{gotobasic}
~\cite{nes2012monetdb}.
\end{gotobasic}
\yannis{to Chunbin: If a little space, bring back the Abadi paper. It's more well-known than papers we cite in other places in the paper.}

\begin{figure}[h!]
\centering
\includegraphics[width=1\columnwidth]{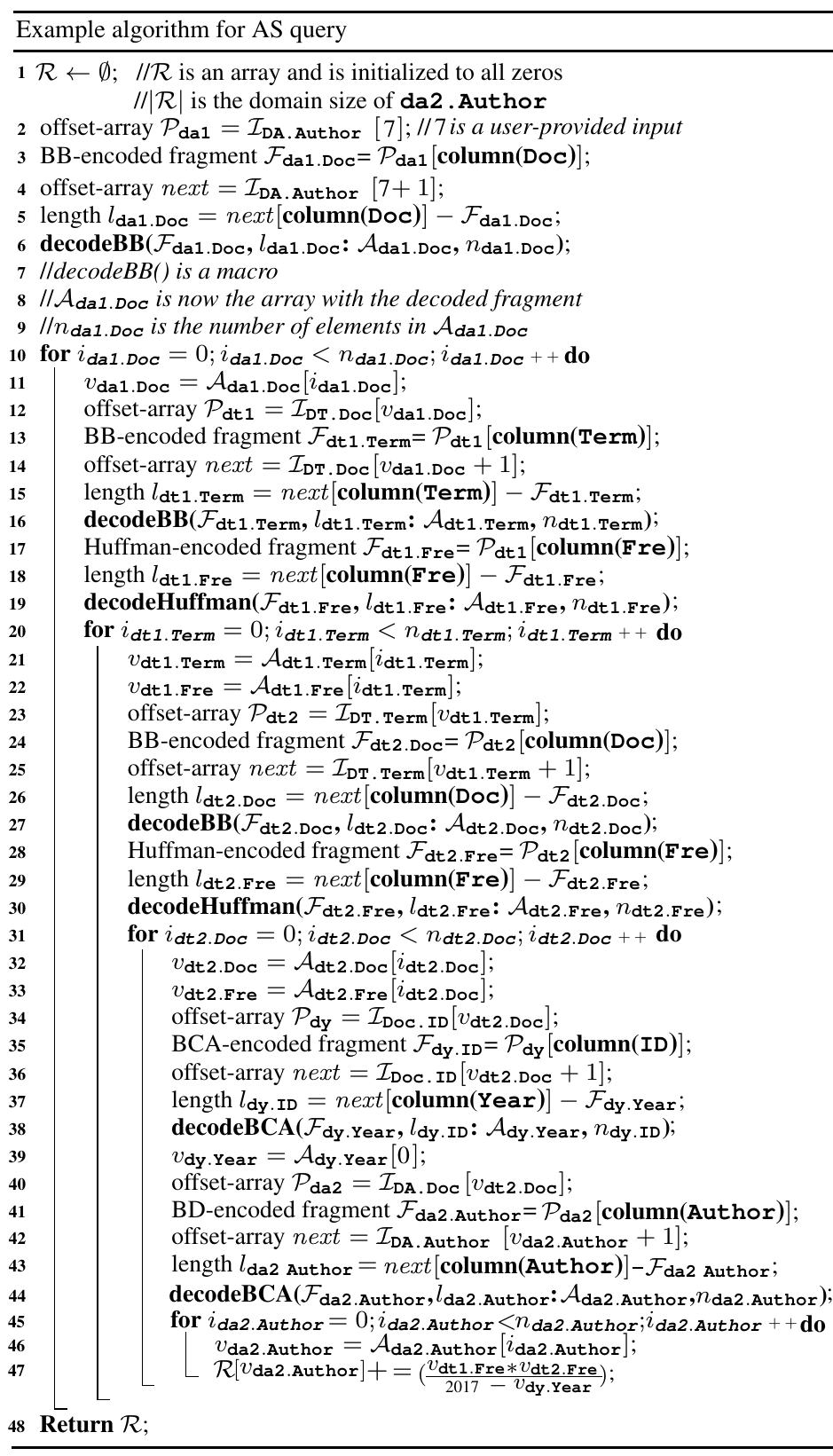}
%\vspace{-6mm}
\caption{\textbf{Example \fastr code for the AS query}}
\label{FastR_querySA}
\end{figure}

% Align with the large code example
% obtains => decodes
%To give a preview of the FastR, we provide a running example to answer the AS query.
As an example, consider the code of Figure~\ref{FastR_querySA} that is the result of code generation for the AS query. The Figure uses variable names that indicate their relationship to the tuple variables (\gt{da1}, \gt{dt1}, \gt{dt2}, \gt{d} and \gt{da2}) of the query. Notice that the AS query has only joins and aggregation. Other cases, involving semijoins and intersection are discussed later.

First, in (Lines~2 to~9) \fastr uses the $\mathcal{I}_{\gt{DA.Author}}$ index to find that the document-fragment $\pi_{\gt{Doc}}\sigma_{\gt{Author}=7}(\gt{DA}\!\mapsto\!\gt{da1})$ starts at the position offset $\mathcal{F}_{\gt{da1}.\gt{Doc}}$ (see Figure~\ref{figure:example_fragments}) which is the offset in the 7-th position of the \gt{DA.Author} lookup table. The encoded fragment length $l_{\gt{da1}.\gt{Doc}}$ is computed by subtracting the offset of the 7-th position from the offset of the 8-th position. \fastr proceeds to decode the fragment into the (preallocated) array $\mathcal{A}_{\gt{da1}.\gt{Doc}}$.

Then, in the inner loop of Lines~10 to~19, for each document ID $v_{\gt{da1.Doc}}\in \mathcal{A}_{\gt{da1}.\gt{Doc}}$, \fastr uses the index $\mathcal{I}_{\gt{DT.Doc}}$ to find the term-fragment $\pi_{\gt{dt1.Term}}\sigma_{\gt{dt1.Doc}=v_{\gt{da1.Doc}}}$ $(\gt{DT}\!\mapsto\!\gt{dt1})$, which starts from the position offset $\mathcal{F}_{\gt{dt1}.\gt{Term}}$ of the \gt{DT.Term} byte array (see example of Figure~\ref{figure:example_fragments}); similarly, it also finds the frequency-fragment $\pi_{\gt{dt1.Fre}}\sigma_{\gt{dt1.Doc}=v_{\gt{da1.Doc}}}(\gt{DT}\!\mapsto\!\gt{dt1})$. For example, since the document id \gt{116} appears in the decoded $\pi_{\gt{Doc}}\sigma_{\gt{Author}=7}\gt{DA}$, \fastr finds the corresponding offsets $\mathcal{F}_{\gt{dt1}.\gt{Term}}$ and $\mathcal{F}_{\gt{dt1}.\gt{Fre}}$ in the 116-th position of the \gt{DT.Doc} lookup table. Then \fastr decodes the identified fragments into $\mathcal{A}_{\gt{dt1}.\gt{Term}}$ and $\mathcal{A}_{\gt{dt1}.\gt{Fre}}$.

Intuitively, at this point, the two outer loops have identified the documents, terms and frequencies associated with the tuple variables \gt{da1} and \gt{dt1} of the query. Similarly, the inner loops identify the attributes associated with the tuple variables \gt{dt2}, \gt{d} and \gt{da2}. The innermost loop utilizes the current values of the attributes that appear in the \gt{SELECT} clause in order to update the sum aggregates in the array $mathcal{R}$. Notice that since the group-by list is a single key, it is sufficient to use an array for the final aggregation. The size of such array is the domain size of \gt{Author.ID}. Assuming consecutive author IDs, this simply the number of authors.

\begin{gotobasic}See \cite{lin2016fastr} for a detailed run through the code of the innermost loops.\end{gotobasic}

\begin{gotofull}In particular, for each term $v_{\gt{dt1.Term}}\in \mathcal{A}_{\gt{dt1}.\gt{Term}}$, \fastr decodes a fragment $\pi_{\gt{dt2.Doc}}\sigma_{\gt{dt2.Term}=v_{\gt{dt1.Term}}}(DT\mapsto dt2)$ in the \gt{DT.Doc} byte array  starting from position offset $\mathcal{F}_{\gt{dt2}.\gt{Doc}}$ and a fragment $\pi_{\gt{dt2.Fre}}\sigma_{\gt{dt2.Term}=v_{\gt{dt1.Term}}}(DT\mapsto dt2)$ in the \gt{DT.Fre} byte array  starting from position offset $\mathcal{F}_{\gt{dt2}.\gt{Fre}}$ on index $\mathcal{I}_{\gt{DT.Term}}$. For example, for term id $28$, \fastr gets the offsets $\mathcal{F}_{\gt{dt2}.\gt{Doc}}$ and $\mathcal{F}_{\gt{dt2}.\gt{Fre}}$ in the 28-th position in the \gt{DT.Term} lookup table.

For each document $v_{\gt{dt2.Doc}}\in \mathcal{A}_{\gt{dt2}.\gt{Doc}}$, \fastr decodes a fragment $\pi_{\gt{dy.Year}}\sigma_{\gt{dy.ID}=v_{\gt{dt2.Doc}}}(Doc\mapsto dy)$ in the \gt{Doc.Year} byte array, which starts from position offset $\mathcal{F}_{\gt{dy}.\gt{Year}}$ on index $\mathcal{I}_{\gt{Doc.ID}}$, and also decodes a fragment $\pi_{\gt{da2.Author}}\sigma_{\gt{da2.Doc}=v_{\gt{dt2.Doc}}}$ $(DA\mapsto da2)$ in the \gt{DA.Author} byte array, which starts from position offset $\mathcal{F}_{\gt{DA}.\gt{Doc}}$ on index $\mathcal{I}_{\gt{DA.Doc}}$. For example, for document id $266$, \fastr gets the offsets $\mathcal{F}_{\gt{dy}.\gt{Year}}$ and $\mathcal{F}_{\gt{da2}.\gt{Author}}$ in the 266-th position in the \gt{Doc.ID} and \gt{DA.Doc} lookup tables respectively.

Finally, for each author $v_{\gt{da2}.\gt{Author}}$, \fastr calculates the aggregated score $\sum (v_{\gt{da1}.\gt{Fre}}\times v_{\gt{da2}.\gt{Fre}})/(2017-v_{\gt{dy}.\gt{Year}})$ and inserts it into the result set $\mathcal{R}$.
\end{gotofull}

\paragraph*{Contributions}
\yannis{Some bullet must refer to the implementation.}
\yannis{The bottom-up pipelined execution has to come in. The data structure requirement for bottom-up pipelined is to have lookup structures that return the required data efficiently. (Hence, the OMC data structure could also be used for the bottom-up pipelined.) Then it is an independent issue/contribution whether the fragments are stored compressed.}
\yannis{to self: Need to state somewhere that (a) it is a memory-based database, (b) utilizes compiled code, (c) adjusts data structures. Possibly split the advantage between navigation and aggregation. Cast the next sentence in the positive.}

\fastr introduces multiple coordinated novelties that lead to runtime performance, space savings, or both, in the case of graph analytics.

\begin{compact_item}
\item Section~\ref{sec:queries} formally \emph{defines relationship queries}, a class of OLAP graph
queries that are prevalent in practice, as illustrated in two real-world use cases.

\item Section~\ref{sec:data-structure} introduces a new \emph{fragment-based data organization}, which combines features of the column data organization and indexing. This data organization gets rid of row ids. Based on the observation that plans need to efficiently find a fragment but do not need random access within the fragment, \fastr allows for aggressive compressions of fragments, such as compressed bitmaps and Huffman encoding. Therefore, more data can fit in the main memory. We study the space cost of the used compression algorithms and show that it is relatively easy to choose between them.

\item Section~\ref{sec:qp} shows how \fastr query plans are \emph{compiled into code} (C++).%
Furthermore, this code uses \emph{bottom-up pipelining} to evaluate the query, effectively avoiding the materialization of intermediate results. More precisely, whereas a column database (see discussion in Section~\ref{sec:related}) would
create intermediate results for row IDs and/or columns of intermediate
results, in \fastrs bottom-up pipelining, each intermediate
result is a scalar variable of the compiled C++ code, hence being
efficiently stored and accessed via a CPU register. The combination of compiled code and bottom-up pipelining further amplifies benefits of compiled code, such as high instruction locality and out-of-order speculation.

\item The data organization (Section~\ref{sec:data-structure}), as well as the plans, are further \emph{optimized by tuning to the dense IDs}, i.e., tuning to the fact that the entities' IDs are consecutive integers.

\item We present a comprehensive set of experiments on three real-life datasets, i.e., PubMed-M, PubMed-MS and SemmedDB. The results show that \fastr is $10^2\sim 10^4$ more efficient than MonetDB, Neo4j and Postgres when they all run on main memory.

\item As is obvious from the above contributions list, the runtime performance advantage of \fastr is affected by multiple factors: bottom-up pipelining, compiled code, dense IDs, aggressive compression.%
\smallfootnote{This is alike column databases, where multiple contributions converged
to their performance benefits.} Therefore, a next series of experiments isolates the effect of individual factors producing multiple important observations:
First, the vast majority of the runtime performance advantage is due to the combination of compilation
with bottom-up pipelining, while the dense IDs optimization has a secondary role.
Second, the use of aggressive compressions
will become more useful as the number of cores in
CPUs increases rapidly (hence reducing the decoding penalty), while the RAM-to-CPU bandwidth improves
significantly slower.
%The experiments show that the use of multiple cores reduces the penalty of decompression since dense compressions benefit from the extra CPU cycles that multicores provide, while they are less demanding on RAM transfer.

\item We deployed \fastr in two real world use cases, around the PubMed and SemMedDB data. An online and interactive demo system is also provided {\small\url{http://137.110.160.52:8080/demo}} for the reader to experience the benefit of \fastr in online graph analytics.

\end{compact_item}

\begin{gotofull}
\noindent \textbf{Roadmap} Section~\ref{sec:related} describes related work. It emphasizes the necessary background on compression. It also illustrates the limitations of column databases by providing a detailed step-by-step description of how an exemplary column database would answer the AS query. Section~\ref{sec:queries} introduces the schema and relationship queries, respectively. Section~\ref{sec:architecture} illustrates the architecture of \fastr. We describe the structure of the \fastr index and theoretically analyzes different compressions in Section~\ref{sec:data-structure}. Section~\ref{sec:qp} describes the \fastr code generator. Finally, Section~\ref{sec:experiments} conducts comprehensive experiments to measure the performance in three real-life datasets.
\end{gotofull}

\yannis{Ideas on how to shorten column databases: Intermediate results are columns, with late tuple materialization. Eg, the middle step after there is an $mathcal{R}_1$ column and we obtain the next list of columns. The example query obviously benefits from indices. An alternate idea is to have two copies (OMC).}

\section{Related Work}
\label{sec:related}
\begin{gotofull}
We illustrate the operation of an exemplary column database during execution of the AS query, highlighting relevant optimizations~\cite{journals/ftdb/AbadiBHIM13,abadi2009column} and limitations. (Readers familiar with column databases can skip this part). We employ the modern column database storage strategy--using virtual IDs instead of explicit row ids (Explicit row ids blot the size of data) for each separately storing column. To access the i-th value in column $C$  simply requires a random access at the location $startOf(C)+i\times width(C)$. Note that, columns here have fixed width, since columns are dictionary-encoded first if they are not integers.

In a convention followed across the full paper, the leaves of algebraic expressions have the form ``\emph{table}~$\mapsto$~\emph{variable}". Consequently, the other operators use the ``\emph{variable}.\emph{attribute}" notation. For readability, the variables have the same names with the tuple variables of the \gt{FROM} clause of the respective queries.

First, we describe the case where there is no sorting and no indices for the \gt{DA} and \gt{DT} tables.
\begin{compact_enum}
%\item Obtain $\mathcal{R}_1=\pi_{\gt{Doc}} \sigma_{\mathit{Author}=7}(\gt{DA}\mapsto \gt{da1})$: First scan the column \gt{DA.Author} to find the set of row IDs $RID_1$ associated with the author \gt{7}. Then retrieve the document ID's for every row in $RID_1$ using random access on \gt{DA.Doc}. $RID_1$ and $\mathcal{R}_1$ are materialized.
\item Obtain $\mathcal{R}_1=\pi_{\gt{Doc}} \sigma_{\mathit{Author}=7}(\gt{DA}\mapsto \gt{da1})$: First scan the column \gt{DA.Author} to find the set of row IDs $RID_1$ associated with the author \gt{7}. Then performing random accesses on column \gt{DA.Doc} to get the document ID's for every row in $RID_1$. Note that $\mathcal{R}_1$ is materialized.

\yannis{to Chunbin: The reader can reasonably ask here why does the column database materialize the $RID_1$/ Instead, it could find the associated \gt{DA.Doc} to a given row id as soon as it finds this row id. Then it can throw away the row id, without materializing an array of ID's. Take two measures to avoid this problem: The first measure, break the steps. Step 1 should break into a Step 1 that executes the selection. The selection delivers RID's, since in the late tuple materialization policy of column databases one never makes whole tuples. Then the new Step 2 does the projection. It is OK to abuse the notation and describe Step 2 $\pi_{Doc}RID_1$. Do the same for the other steps. If the description becomes too large and repetitive we can cut it later. The second measure is to become precise about the cited work: Is there a particular publication or publications that indicate that each operator is executed by itself? I.e., that the meaning of a selection is to produce some data structure with the qualifying ID's. We should quote this in particular. If it is not stated exactly like that, email me how exactly it is stated (perhaps point me to a pdf/page) and I will see how to write it. If we cannot produce such ``proof by reference" we can ask Stratos. If this also fails, we can demote the ``RID materialization" from the list of limitations, to avoid being criticized. Chunbin: I removed the row ids in step 1.}

\item Obtain $\mathcal{R}_2=\pi_{\gt{dt1.Term}, \gt{dt1.Fre}}(\mathcal{R}_1\Join(\mathit{DT}\mapsto \gt{dt1}))$. For each document ID $d\in \mathcal{R}_1$, scan the column \gt{DT.Doc} to find the set of row IDs $RID_2$ referring to $d$. Retrieve the term ID and term frequency for every row in RID using random access on \gt{DT.Term} and \gt{DT.Fre}, respectively. $RID_2$ and $\mathcal{R}_2$ are materialized.
%\yannis{to Chunbin: When you break this into many steps have separate projections that have one column at a time.}

\item Obtain $\mathcal{R}_3=\pi_{\gt{dt2.Doc},\mathcal{R}_2.\gt{Fre}\rightarrow\gt{Fre}_1,\gt{dt2.Fre}\rightarrow\gt{Fre}_2}(\mathcal{R}_2\Join(\mathit{DT}\mapsto \gt{dt2}))$. For each term ID $t\in \mathcal{R}_2.Term$, scan column \gt{DT.Term} to find the set of row IDs $RID_3$ referring to $t$. Retrieve the term ID and term frequency for every row in $RID_3$ using random accesses on \gt{DT.Doc} and \gt{DT.Fre}, respectively. $RID_3$ and $\mathcal{R}_3$ are materialized.

\yannis{to Chunbin: There is something more going on here: The $Fre_1$ values are replicated so that they can align with Doc and $Fre_2$. Correct? this should be said.}

\item Obtain $\mathcal{R}_4=\pi_{\mathcal{R}_3.\gt{Doc},\mathcal{R}_3.\gt{Fre}_1,\mathcal{R}_3.\gt{Fre}_2, \gt{d.Year}}(\mathcal{R}_3\Join(\mathit{Document}\mapsto \gt{d}))$. For each document ID $d\in \mathcal{R}_3.Doc$, scan column \gt{Document.ID} to find the set of row IDs $RID_4$ referring to $d$. Retrieve the publishing year for every row in RID using random access on \gt{Document.ID}. $RID_4$ and $\mathcal{R}_4$ are materialized.

\item Obtain $\mathcal{R}_5=\pi_{\gt{da2.Author},\mathcal{R}_4.\gt{Fre}_1, \mathcal{R}_4.\gt{Fre}_2, \mathcal{R}_4.\gt{Year}}(\mathcal{R}_4\Join\mathit{DA}\mapsto \gt{da2}))$. For each document ID $d\in \mathcal{R}_4.Doc$, scan column \gt{DA.doc} to find the set of row IDs $RID_5$ referring to $d$. Retrieve the author ID for every row in RID using random access on \gt{DT.Author}. $RID_5$ and $\mathcal{R}_5$ are materialized.

\item Calculate $\mathcal{R}_6=\gamma_{\gt{Author}, \gt{SUM}(\gt{Fre}_1\cdot\gt{Fre}_2)/(2017-\gt{Year})}(\mathcal{R}_5)$. Scan $\mathcal{R}_5$ to get the set of distinct authors and associated scores $\sum (\mathit{Fre}_1 \cdot \mathit{Fre}_2)/(2017-\mathit{Year})$.
\end{compact_enum}
The above example exhibits a number of inefficiencies.

\begin{compact_enum}
\item Expensive scanning operations are executed. Step 1$\sim$5 obtain explicit or implicit row ids via whole column scanning.

\item Intermediate results are maintained. Steps 1$\sim$5 materialize all the temporary output results $\mathcal{R}_1,\ldots\mathcal{R}_5$ and also row ids $RID_2$, $\ldots RID_5$, which is both time and space consuming.
%Row ids are utilized to get values in other columns in corresponding rows efficiently.
\end{compact_enum}
To speedup the performance of the above example, many existing optimizations can be applied, e.g., late tuple materialization, block iteration, invisible joins and column-specific compression schemes~\cite{abadi2006integrating,abadi2008column,abadi2008query,abadi2009column,nes2012monetdb,journals/ftdb/AbadiBHIM13}. In this paper, these optimizations are all utilized in our baselines. Note that, we do not build hash index in our baselines, since we mainly work on relationship tables, each individual column in a relationship table is not a primary key and has many duplicates.

For instance, to eliminate whole column scan, binary search can be utilized~\cite{bertino2012indexing, conf/vldb/LehmanC86} if the column is sorted.
Since we adopt the virtual IDs store strategy, all the columns should be organized in one order. That is binary search only works for one column. However, relationship queries might perform lookup operation in two reference columns. To avoid whole column scan in both columns, two copies of data with different sorting should be maintained (as we did for \fastr and OMC, more details can be found in \ref{sec:experiments}).
%column store does not use hash index.
\end{gotofull}
%\vspace{-1mm}

\begin{gotobasic}
\noindent \textbf{Column Database Optimizations}
A number of optimizations (late tuple materialization, block iteration, invisible joins and column-specific compression schemes) have been widely studied in order to improve the performance of column databases for OLAP query workloads~\cite{abadi2006integrating,abadi2008column,abadi2009column,nes2012monetdb,journals/ftdb/AbadiBHIM13}. These optimizations are used by our baseline compiled-code approach, named \textit{OMC}, which is used to separate the benefit of compiled code from the benefits of the data structure and plan execution novelties of \fastr. The extended version~\cite{lin2016fastr} provides a detailed example of how a compiled-code column database (i.e., the OMC), would execute the AS query.
\end{gotobasic}
\smallskip
\noindent \textbf{Data Compression Schemes}
Column databases do not consider heavy-weight data compression schemes, since they do not allow partial decompression - hence requiring that the whole column (or block) should be decompressed whenever data of the column are needed. The archetypical heavy-weight compression scheme is Huffman encoding~\cite{huffman1952method,storer1988data,ChungW99}. In the context of our comparisons, OMC does not use heavy-weight compression schemes, while \fastr does, when beneficial.
\begin{gotofull}We list below the compression schemes used in column databases and in \fastr.\end{gotofull}
\begin{gotobasic}We list below the compression schemes used in column databases and in \fastr, while \cite{lin2016fastr} provides a standalone description of each compression.\end{gotobasic}

\smallskip

\begin{gotobasic}
\noindent\textbf{Dictionary encoding.}
The widely utilized (light-weight) dictionary encoding maintains a global table to map each wide value to an integer~\cite{iyer1994data,abadi2006integrating}. \fastr also applies dictionary encoding to map string values to integers upon load time. So, \fastr only processes integer columns in the query processing phase.
\end{gotobasic}
\begin{gotofull}

\noindent\textbf{Dictionary encoding.}
The most widely utilized dictionary encoding  maintains a global mapping table to map each wide value to an integer. For example, a simple dictionary for a string-typed column of states
might map ``Alabama'' to 0, ``Alaska'' to 1, ``Arizona'' to 2, and so on~\cite{iyer1994data,roth1993database,zukowski2006super,abadi2006integrating}. \fastr applies a dictionary encoding to map string-type columns to integers upon load time, so \fastr only processes integer columns in query processing phase. \fastr also applies a variant of bit-aligned dictionary encoding on fragments, which will be discussed in Section ~\ref{sec:data-structure}.
\end{gotofull}

\smallskip

\begin{gotobasic}
\noindent\textbf{Run-length encoding.}
The traditional Run-Length Encoding (RLE) expresses the repeats of the same value as pairs (value, run-length)~\cite{storer1988data,journals/ftdb/AbadiBHIM13,abadi2006integrating}.
However, it is inappropriate for projection-selection queries $\pi_B\sigma_{A=a}R$, which are part of larger join queries, since it does not tell where the run starts. Hence databases also consider a (value, start-position, run-length) scheme. Both OMC and \fastr use a variant of the latter. OMC, in particular, compresses the repeats of sorted columns as (value, start-position). Assuming the encodings for values $a$ and $a+1$ are $(a,s)$ and $(a+1, s')$ respectively, the answer of $\pi_B\sigma_{A=a}R$ can be directly found in column $B$ from position $s$ to position $s'-1$.
\end{gotobasic}
\begin{gotofull}

\noindent\textbf{Run-length encoding.}
The traditional run-length encoding (RLE) expresses the repeats of the same element as (value, run-length)~\cite{storer1988data,journals/ftdb/AbadiBHIM13,abadi2006integrating}.
Assume a column $R.A$ of R(A,B) is compressed by this standard RLE, then the whole column should be decompressed to answer the projection-selection query $\pi_B\sigma_{A=a}R$, which is an important component in relationship queries. To avoid decompressing the whole column, we consider the variant of RLE -- compressing the repeats of the same element as (value, start-position). Assume the corresponding encoded pairs for values $a$ and $a+1$ are $(a,s)$ and $(a+1, s')$ respectively. Then the answers of $\pi_B\sigma_{A=a}R$ are the values in column $B$ from position $s$ to position $s'-s$. OMC adopts this variant of RLE on sorted columns, e.g., \gt{DT.Term}.
\end{gotofull}

\begin{gotobasic}
The extended version~\cite{lin2016fastr} discusses additional light-weight compression schemes, which are not beneficial for relationship queries.
\end{gotobasic}

\begin{gotofull}

\noindent\textbf{Bitmap encoding.}
A bitmap index consists of a set of bit-vectors. Each bitvector indicates the occurrences of one distinct value of the
indexed attribute. That is, for each distinct value $t$, one bitvector is maintained. The i-th position of the bitvector is set to `1' if $t$ appears at the i-th position in the original column and `0' otherwise. It is efficient to perform bitwise operations -- AND, OR, XOR and NOT on bitmap index. To further reduce the size of the bitmaps, literature~\cite{lemire2010sorting,wu2006optimizing,antoshenkov1995byte} applies run-length encoding(RLE) on them. Interestingly, The bitewise operations can still be issued directly on the compressed bitmaps without decompressing. Two outstanding techniques are byte-aligned bitmap compression (BBC)~\cite{antoshenkov1995byte} and word-aligned hybrid (WAH) compression~\cite{wu2006optimizing, wu2004performance}. However, the bitmap scheme is recommended to apply only on columns with low cardinality, say around 50~\cite{abadi2006integrating}. OMC does not use bitmap encoding, since the cardinalities of columns in our datasets are generally high and RLE achieves better performance compared with bitmap.
\end{gotofull}

\begin{figure*}[htbp]%
\centering
\includegraphics[width=0.9\textwidth]{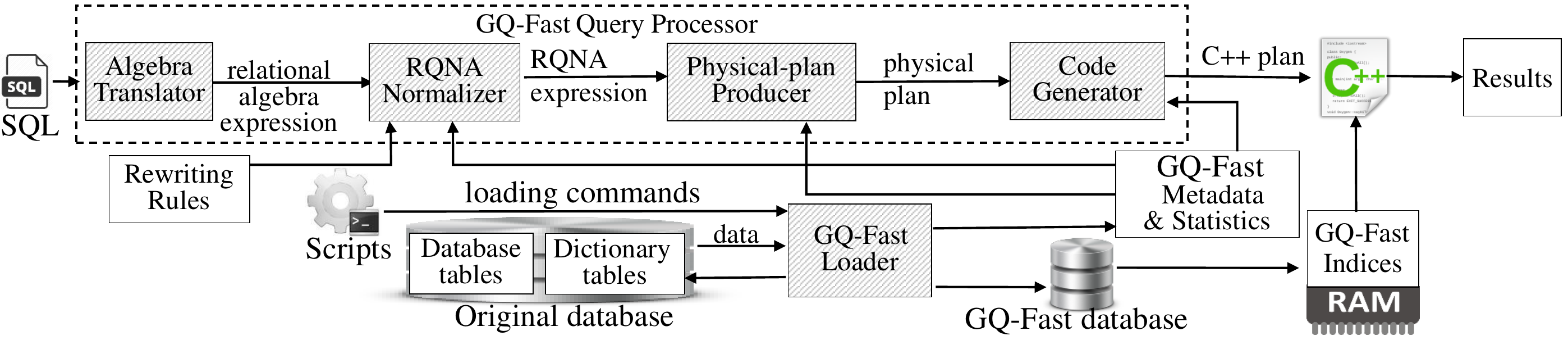}
%\vspace{-3mm}
\caption{\textbf{Architecture of relationship query processing of \fastr}. The \textit{\fastr Loader} produces \fastr database and metadata, the \textit{\fastr Query Processor} receives relationship queries and produces answers.}
\label{fig:architecture}
\end{figure*}
%\vspace{1mm}

\subsection{Compiling Code for SQL Queries}
\begin{gotobasic}
Generating and compiling code has been studied widely~\cite{ahmad2009dbtoaster,neumann2011efficiently,myers2015compiled}. For example, DBToaster\cite{ahmad2009dbtoaster} compiles SQL queries into C++ code in order to provide fast delta processing in incremental view maintenance cases.  \cite{myers2015compiled} compiles path-counting queries to produce in-memory pipelined hash-join query plans.
\end{gotobasic}
\begin{gotofull}
Generating and compiling code has been studied widely in order to improve the performance~\cite{ahmad2009dbtoaster,neumann2011efficiently,myers2015compiled}. For example, DBToaster\cite{ahmad2009dbtoaster} compiles SQL queries into C++ code for  view maintenance problem in order to provide fast delta processing. \cite{neumann2011efficiently} compiles TPC-H style analytics queries to LLVM bytecode and C. In addition, \cite{myers2015compiled} generates in-memory pipelined hash-join query plans in C++ directly from Datalog queries for path-counting queries. In this paper, we provide a code generator to generate C++ code for relationship queries.
\end{gotofull}
%\vspace{-1mm}

\begin{gotobasic}

\noindent\textbf{Graph Databases}
Multiple graph databases are currently in use: Apache Giraph\smallfootnote{http://giraph.apache.org}, GPS~\cite{Gps}, GraphLab\smallfootnote{http://graphlab.org}, and Neo4j. The first three do not offer declarative query interface and cannot express simple relational idioms~\cite{myers2015compiled}, while Neo4j provides its own graph Cypher query language. Indeed, Neo4j can express relationship queries. Therefore we employed Neo4j as one more baseline system in our experiments.
\end{gotobasic}
%\begin{gotobasic}
%Row-oriented databases employ iterator model~\cite{lorie1974xrm,graefe1993query} to provide a top-down pipelined method by passing one tuple at a time to the next operator in order to avoid intermediate results. However,  iterator model shows poor performance on modern CPUs due to lack of locality, frequent instruction mis-predictions and too many function calls~\cite{neumann2011efficiently}. To address the limitation, modern column databases choose one the following design choices: (1) passing blocks of tuples (batch oriented processing) between operators, which greatly reduces the number of function invocations~\cite{padmanabhan2001block,neumann2011efficiently}, (2) materializing all intermediate results to eliminate the need of calling an input operator repeatedly, which simplifies operator interaction~\cite{manegold2009database,nes2012monetdb}, or (3) choosing a middle ground of the above two by passing large vectors of data and evaluating queries in a vectorized manner on each chunk~\cite{zukowski2005monetdb}. However, none of them can reach the speed of hand-written code~\cite{neumann2011efficiently}. \fastr employs a bottom-up pipelined method and utilizes a code generator to produce query-aware physical plans, which can achieve the performance of hand-written code and also get rid of too many function calls and materializing intermediate results.
%\end{gotobasic}

\vspace{-2mm}
\begin{gotofull}
\subsection{Graph Databases}
Relationship queries are a superset of graph reachability queries, so one possible solution is to apply existing graph databases.
Graph databases have become ubiquitous with the advancement and popularity of social networking. Some prominent ones include Apache Giraph~\cite{Giraph},  GPS~\cite{Gps}, GraphLab~\cite{GraphLab}, and Neo4j~\cite{Neo4j}~\cite{guo2014benchmarking}. The first three adopt vertex-centrix programming models, which supports asynchronous computation and prioritized scheduling, but do not offer declarative query interface and can not express simple relational idioms~\cite{myers2015compiled}. Neo4j provides its own graph Cypher query language, which can express relationship queries. Therefore, we employ Neo4j as a baseline system in our experiments. \cite{lin2016fastr} provides the Cypher statements for the SQL relationship queries of this paper.
\subsection{Execution Model}
To answer an SQL query, database systems first translate the query into a physical query plan, then evaluate this physical query plan to produce result. The traditional way to evaluate this physical query plan is the iterator model \cite{lorie1974xrm,porto2007qef,graefe1993query}. In the iterator model,  every physical algebraic operator conceptually generates a tuple stream from its input, and iterating over this tuple stream by repeatedly calling the \textit{next} function of the operator. This iterator model is a simple interface, which allows for easy combination of arbitrary operators, and is a good choice when query processing is dominated by I/O and CPU consumption is less important. As main memory grows, query performance is more and more determined by the CPU costs. The iterator model shows poor performance on modern CPUs due to lack of locality and frequent instruction mis-predictions\cite{neumann2011efficiently}.

Since CPU costs are a critical issue for modern main-memory database systems, the iterator model is not the optimal solution any more. Some systems tried to reduce the high calling costs of the iterator model by passing blocks of tuples (batch oriented processing) between operators~\cite{padmanabhan2001block,neumann2011efficiently}. This greatly reduces the number of function invocations, but causes additional materialization costs. The MonetDB system \cite{manegold2009database,nes2012monetdb} materializes all intermediate results, which eliminates the need
to call an input operator repeatedly. Besides simplifying operator interaction, materialization has other advantages, but it also causes significant costs. The MonetDB/X100 system \cite{zukowski2005monetdb} chooses a middle ground by passing large vectors of data and evaluating queries in a vectorized manner on each chunk. This offers good performance, but still does not reach the speed of hand-written code\cite{neumann2011efficiently}. In this paper, \fastr produces  C++ hand-written like code directly from the physical query plan of relational queries. \fastr iterators each fragment by using simple loop, since the size of each fragment is calculated in the code generating phase by using meta data.
\end{gotofull}

\section{Architecture}
\label{sec:architecture}

Applications use \fastr as an OLAP-oriented database that accompanies their original, transaction-oriented database. Figure~\ref{fig:architecture} is an overview of \fastrs architecture.

The \emph{\fastr Loader} receives loading commands and, in response, it retrieves data from a relational database (or databases) and creates a \fastr database\smallfootnote{Currently, \fastr does not allow incremental update of the loaded data. In the two biomedical knowledge analysis applications, in which \fastr is currently used, this is not an important limitation since datasets change relatively rarely. Hence, it is sufficient to reload an entire \fastr database periodically. Future work will lift this restriction.} and relevant metadata, which contains information about fragments and their encodings. The schema of the \fastr database has to follow certain conventions; see Section~\ref{sec:queries}. The data of a \fastr database is stored into in-memory data structures (see Section~\ref{sec:data-structure}).

Then the \emph{\fastr Query Processor} receives an SQL query and outputs its result. It consists of several subcomponents. The \textit{Algebra Translator} translates an SQL query into a relational algebra expression, which is then transformed into a \emph{Relationship Query Normalized Algebra (RQNA) expression} (RQNA will be formally defined in Section~\ref{sec:queries}) by the \textit{RQNA Normalizer} by applying rewriting rules.
\begin{gotofull}
Given an SQL query $q$ in its algebraic format, the RQNA Normalizer applies the following rewriting rules to transform it into RQNA: (1) push every possible selection and projection down to the corresponding tables; projections are upon selections, (2) rewrite into a \textit{left-deep}.
\end{gotofull}
The \textit{RQNA Normalizer} is also a verifier who verifies whether an input SQL query is a relationship query by checking the restrictions according to metadata.  Afterwards, the valid RQNA expression is transformed into a physical-level plan by the \textit{Physical-plan Producer}. The \textit{Code Generator} consumes the physical plan and metadata and produces a C++ program, which is then compiled and ran on the \fastr index to get final results. \fastr also provides (as JDBC does) the ability to prepare a query statement once and execute it multiple times, changing each time the parameters.

\section{Relationship Queries and Algebra Expressions}
\label{sec:queries}
\begin{figure}
\center
\includegraphics[width=0.4\textwidth]{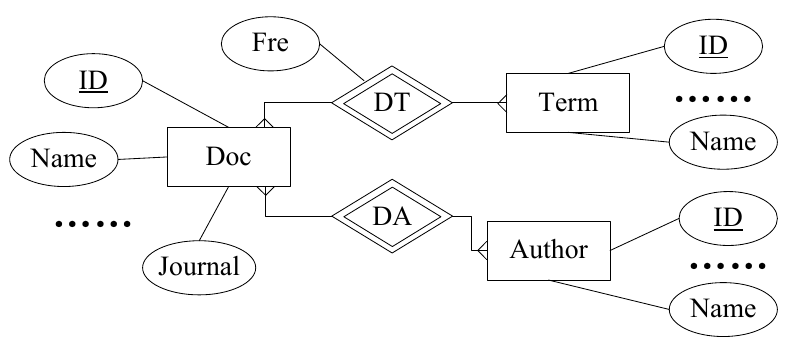}
%\vspace{-5mm}
\caption{\textbf{PubMed ER schema}. \gt{Document} and \gt{Term} (Resp. \gt{Author}) have n-to-n relationship.}
\label{fig:ER}
\end{figure}

\yannis{to self: make clear disctinction of what is harmless convention vs actual limitation. }
\begin{gotofull}
%chunbin: define entity tables and relationship tables.
FastR supports SQL schemas. It classifies tables into \emph{entity} tables $E_1, \ldots,$ $E_m$ and \emph{relationship} tables $R_1, \ldots, R_n$. The naming relates to the well-known E/R schema design technique~\cite{garcia2008database}. Intuitively, entity tables correspond to entities of the E/R design, whereas the relationship tables correspond to many-to-many relationships. For example, Figure~\ref{fig:ER} illustrates the E/R design of the PubMed database.  The tables \gt{Doc}, \gt{Term}, \gt{Author} and \gt{Journal} are entity tables. The relationship tables \gt{DT} and \gt{DA} capture the many-to-many relationship between \gt{Doc} and \gt{Term}, and \gt{Doc} and \gt{Author}, respectively.
\end{gotofull}

%chunbin: entity tables have ID attribtue
\begin{gotobasic}
\noindent \textbf{Schema} The introduction presented the essentials of entity and relationship tables. Note that the \fastr Loader uses consecutive IDs for the ID (primary key) attributes of entity tables by applying dictionary encoding~\cite{iyer1994data}\smallfootnote{Alternately, if the original database features integer primary keys with reasonably compact ID range, \fastr may be instructed to simply use the IDs of the original database and avoid having dictionary tables.
}
This paper focuses on the frequent case where the relationships are binary and the combination of the two foreign keys is unique.The extended version~\cite{lin2016fastr} generalizes to multiway relationships.
\end{gotobasic}

\begin{gotofull}Entity tables have to follow a single convention: Each FastR entity table must have an integer primary key (aka ID) attribute. In practice, this convention does not limit generality, since database schema designers often follow it. This convention is also recommended when translating an E/R design into a schema \cite{garcia2008database}.
In our examples, the ID attributes are always named \gt{ID}.
\end{gotofull}

\begin{gotofull}
%chunbin: relationship tables have foreign keys and measures
In the general case, each relationship table $R(F_1, \ldots, F_d, M_1,$ $ \ldots, M_k)$ has $d$ \emph{foreign key attributes} $F_1,\ldots,F_{d}$, where each foreign key $F_i$ refers to the ID of an entity $E_i$. The intuition, according to E/R design, is that the foreign key $F_i$ corresponds to the connection between the many-to-many relationship (that corresponds to the relationship table $R$) and the entity $E_i$. To make this connection clear in our examples, the foreign key attribute $F_i$ has the same name as the entity $E_i$. \yannis{to Chunbin: The foreign keys are named \gt{Doc} but the entity table is named \gt{Document}. Turn \gt{Document} to \gt{Doc} everywhere. Chunbin:done. Yannis: I still found a few. I may have missed some.} E.g., in the PubMed example, the \gt{Term} foreign key of the \gt{DT} table points to the \gt{ID} of the \gt{Term} table. The relationship table may also have $k$ attributes $M_1, \ldots, M_{k}$ that are not foreign keys. They correspond to the attributes of the many-to-many relationship in the E/R design. We call $M_1, \ldots, M_{k}$ \emph{measure attributes}. For example, the \textit{Fre} attribute of the \gt{DT} table is a  measure attribute.
%chunbin: one-to-many relationship
As is typical in E/R-based schema design, a one-to-many relationship between entities does not have a corresponding relationship table. Instead, it is captured by including a foreign key attribute to the entity table corresponding to the entity on the ``many'' side of the relationship. For example, the entity table \gt{Doc} has a \textit{Journal} foreign key attribute, which captures the many-to-one relationship between documents and journals.
\end{gotofull}

\begin{figure}
\small
\[
\begin{array}{lllr}
\mathit{RQNA}    & \Rightarrow & \gamma_{k; f_1(.) \mapsto N_1,\ldots,f_n(.) \mapsto N_{n_{\gamma}}} \mathit{Join} & (1) \\
 		   &   & \multicolumn{1}{r}{\rm{attributes\ named}\ \mathit{k}\ \rm{are\ primary\ or\ foreign\ keys}}\\
                & | & \mathit{Join} &(2)\\
\mathit{Join}    & \Rightarrow &  \mathit{Join} \Join_{j.k_1=v.k_2} (\pi_{\bar{A}} ( T \mapsto v ))  &(3)\\
		    &  & \multicolumn{1}{r}{j\ \rm{is\ a\ variable\ defined\ by}\ \mathit{Join}}\\
                 & | & \pi_{\bar{A}} (\sigma_c ( T \mapsto v )) &(4)\\
                 & | & \pi_{\bar{A}} (( T \mapsto v ) \ltimes_{v.k_1= x.k_2} \mathit{Context}) &(5)\\
		    &   & \multicolumn{1}{r}{x\ \rm{is\ a\ variable\ defined\ by}\ \mathit{Context}}\\
\mathit{Context} & \Rightarrow & \pi_{v.k} \mathit{Join} &(6)\\
                 & | & \pi_{v.k}\sigma_{c_1} (T_1 \mapsto v) \cap\ldots\cap \pi_{v.k}\sigma_{c_n} ( T_n \mapsto v) &(7)\\
\end{array}
\]
%\vspace{-5mm}
\caption{\textbf{Grammar describing RQNA expressions}.}
\label{fig:RQNA-BNF}
\end{figure}

\yannis{good for intro also.Chunbin: I added it in the Introduction}
\noindent \textbf{Queries} A relationship query (in its algebraic form) involves $\sigma$, $\pi$, $\Join$, $\ltimes$ operators and an optional aggregation at the end. It meets the following restrictions (a) join and semijoin conditions are equalities between (primary or foreign) key attributes and (b) aggregations group-by on a primary key or foreign key. The set of relationship queries includes graph reachability (path finding) queries, where the edges are defined by foreign keys. More generally, it includes tree pattern queries, followed by aggregation.

\begin{figure*}[h!]
\center
\includegraphics[width=1\textwidth]{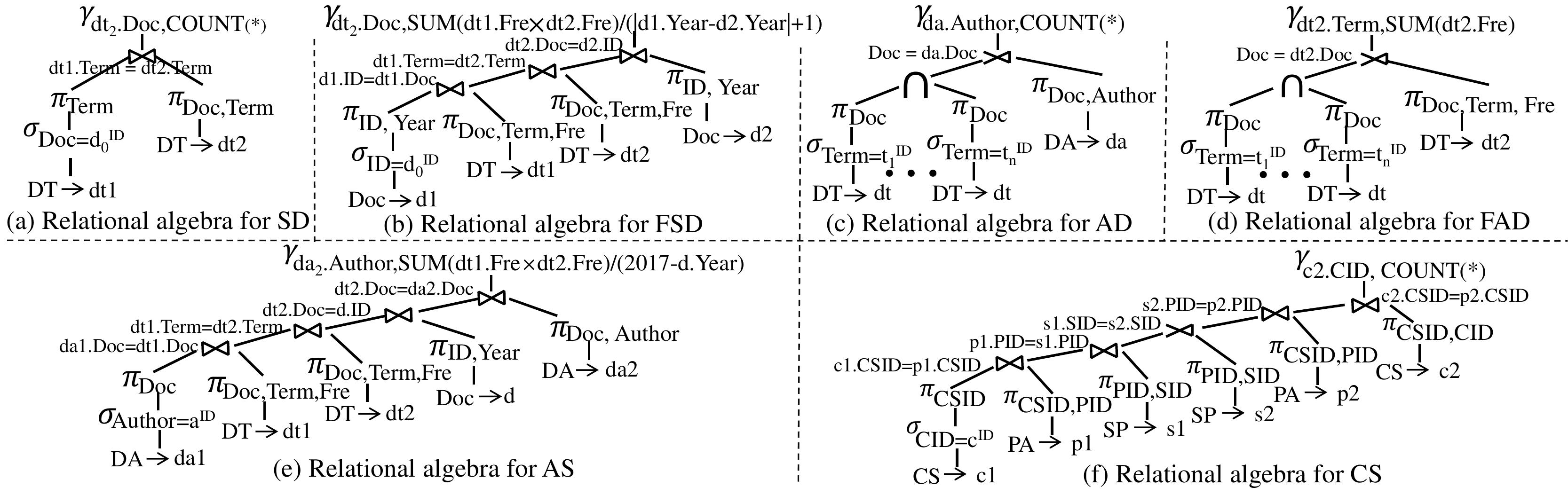}
%\vspace{-5mm}
\caption{\textbf{Relational algebra expressions}.}
\label{fig:xmpl:RQNA}
\end{figure*}

To efficiently answer relationship queries, \fastr first translates them into RQNA (Relationship Query Normalized Algebra) expressions (see syntax in Figure~\ref{fig:RQNA-BNF}). In the simplest case, an RQNA expression is a left-deep series of joins with a selection and aggregation: In Line~\textbf{4} the RQNA expression starts with a selection  $\sigma_c (T \mapsto v)$, that qualifies some entities - we call them the \emph{context} entities.%
\smallfootnote{In a trivial case the condition may be set to true, effectively being absent. We do not discuss this case.}
Consequently the RQNA expression performs a series of left-deep joins (Line~\textbf{3}) that navigate to entities that relate to the qualifying entities. An RQNA expression may be just the left-deep join (Line~\textbf{2}) or a group-by on the key attribute $k$ (Line~\textbf{1}), followed by many aggregations.
In more complex cases, the SQL query (as shown in examples next) contains ``\gt{IN} (nested query), whereas the \gt{IN} translates into a semijoin (Line~\textbf{5}). The nested query is itself a relationship query (Lines~\textbf{6}), which is recursively a relationship query or a result of an intersection (Lines~\textbf{7}).

Next we illustrate relationship queries using the datasets of \fastrs applications: PubMed and SemmedDB. These queries are also used in the experiments and the queries on PubMed dataset are demonstrated in our OLAP dashboard {\small\url{http://137.110.160.52:8080/demo}}.
\yannis{to Chunbin: when you make the extended version remember to remove phrases like the next one. Chunbin:done}

\yannis{to Chunbin: reintroduce the following sentence in the extended. Chunbin: done}

\eat{\begin{gotofull}
Figure~\ref{figure:pubmed-schema} shows a relational representation of a part of PubMed. It contains three entity tables \gt{Doc}, \gt{Term} and \gt{Author}, and two relationship tables  \gt{DT} and \gt{DA}.
\end{gotofull}}

\yannis{yannis to self: move it}

\begin{gotofull}
\begin{figure*}[h!]
\center
\includegraphics[width=1\textwidth]{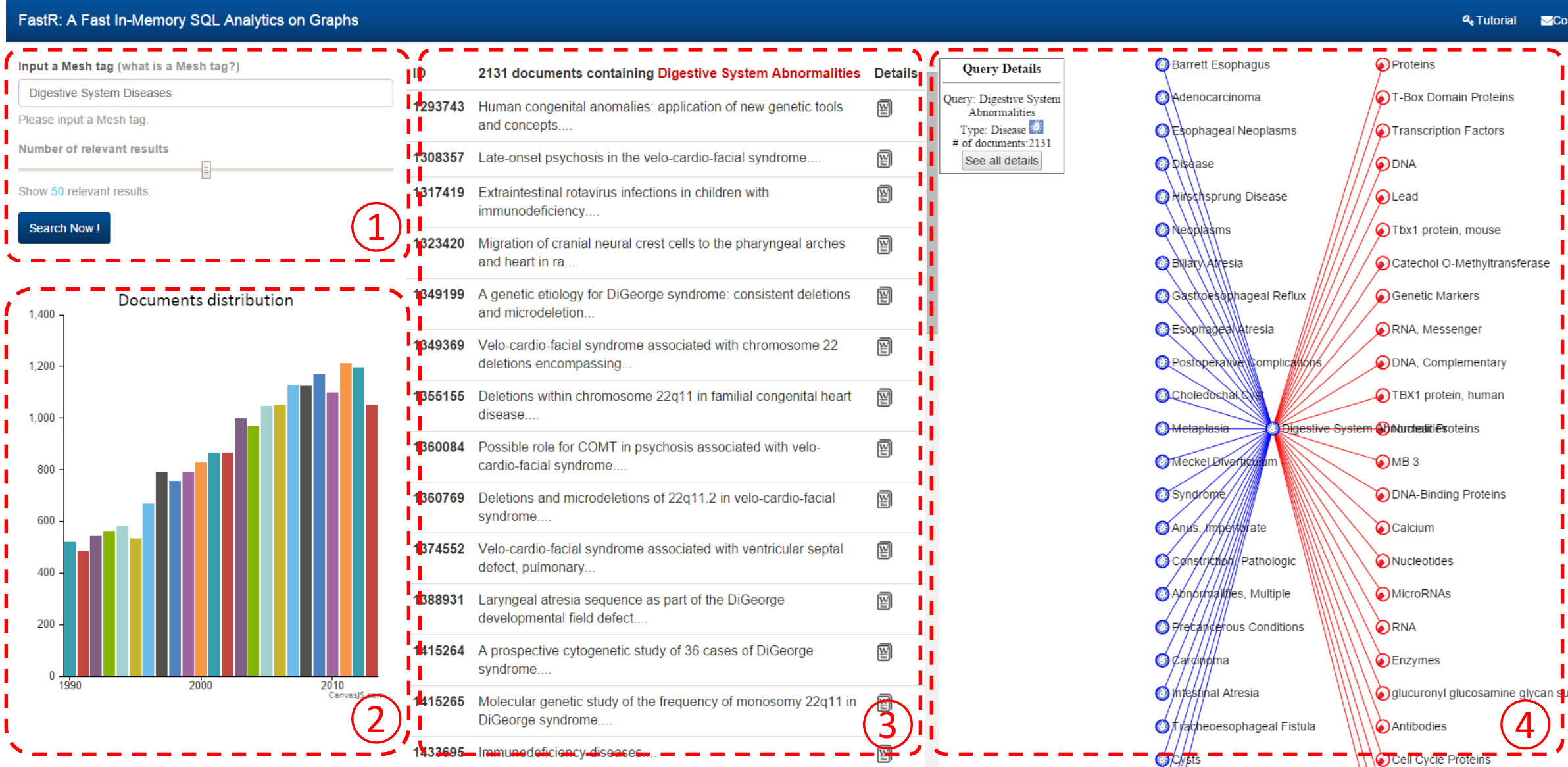}
\caption{\textbf{A screenshot of \fastr demo system}.  \textcircled{1} is the query generator with two components: I. Mesh tag input text box; and II. a slider to limit the number of relevant tags.
\textcircled{2}  is the exhibitor showing the publication statistics of the query tag. \textcircled{3} shows a list of documents containing the query tag. \textcircled{4} is the visual space explorer demonstrating the relevant tags with zoom-in/out, drag and rotation features. Demo link: {\small\url{http://137.110.160.52:8080/demo}}}
\label{fig:demo_fastr}
\end{figure*}
\end{gotofull}

\vspace{-2mm}
\begin{example}
\noindent\textbf{[Query SD: Similar Documents]} The Query~SD and the corresponding RQNA algebra expression in Figure~\ref{fig:xmpl:RQNA}(a) finds documents that are similar to a given document $d_0$ with ID $d_0^{ID}$, where similarity is defined as cosine similarity: Each document $d$ is associated with a vector $t^d = [t_1^d, \ldots, t_n^d]$, where $n$ is the number of terms across all documents. In a frequency-unaware definition, $t_i^d = 1	$ if the document $d$ contains the $i$-th term and $t_i^d = 0$ otherwise. The cosine similarity between two documents $x$ and $y$ is defined as $\sum_{i=1,\ldots,n} t_i^x t_i^y$.%
\smallfootnote{In practice, the queries also normalize for the sizes of $t^x$ and $t^y$ and, in later examples, the sizes of measures. The examples exclude the normalizations since they do not present any important additional aspect to the exhibited query pattern.}
%\vspace{-1mm}
{\small
\begin{tabbing}
Query SD:\\
\=\gt{SELECT dt2.Doc, COUNT(*)}\\
 \>\gt{FROM }\=\gt{DT dt1 JOIN DT dt2 }\gt{ON dt1.Term = dt2.Term}\\
% \>\>\\
 \>\gt{WHERE dt1.Doc = }$d_0^{ID}$\\
 \>\gt{GROUP BY dt2.Doc}
\end{tabbing}}
%\vspace{-1mm}
\noindent\textbf{[Query FSD: Frequency-and-Time-aware Document Similarity]} The following Query~FSD and the corresponding RQNA expression of Figure~\ref{fig:xmpl:RQNA}(b) computes time-aware and frequency-aware cosine similarity. In contrast to the frequency-unaware definition of Query~SD, $t_i^d$ is the number of occurrences of the $i$th term in document $d_0$. Furthermore, the Query~FSD raises the similarity degree of documents that are chronologically close.%

{\small
\begin{tabbing}
Query FSD:\\
\=\gt{SELECT dt2.Doc, $\frac{\gt{SUM(dt1.Fre * dt2.Fre)}}{\gt{abs(d1.Year-d2.Year)+1}}$}\\
\>\gt{FROM }\=\gt{(}\=\gt{(}\=\gt{(Document d1 JOIN DT dt1 ON d1.ID = dt1.Doc)}\\
\>                     \>      \>     \>\gt{JOIN DT dt2 ON dt1.Term = dt2.Term)}\\
\>                     \>      \>\gt{JOIN Document d2 ON d2.ID = dt2.Doc)}\\
\>\gt{WHERE d1.ID =}$d_0^{ID}$\\
\>\gt{GROUP BY dt2.Doc}
\end{tabbing}}

\begin{gotofull}
\noindent\textbf{[Query AS: Author Similarity]}
The Query~AS and the respective RQNA expression of Figure~\ref{fig:xmpl:RQNA}(e) specify an author with ID $a^{ID}$ and they find other authors that have similar publications, weighing higher those with publications in recent years. The similarity is measured by the frequency-aware cosine of common terms between the publications. It also favors documents that are chronologically close.

{\small
\begin{tabbing}
Query AS:\\
\=\gt{SELECT da2.Author, SUM(dt1.Fre$\times$dt2.Fre)/(2017-d.Year)}\\
\>\gt{FROM }\=\gt{(}\=\gt{(}\=\gt{(DA da1 JOIN DT dt1 ON da1.Doc=dt1.Doc)}\\
\>\>\>\>\gt{JOIN DT dt2 ON dt1.Term = dt2.Term)}\\
\>\>\>\gt{JOIN Document d ON dt2.Doc=d.ID)}\\
\>\>\gt{JOIN DA da2 ON dt2.Doc=da2.Doc}\\
\>\gt{WHERE da1.Author = }$a^{ID}$ \\
\>\gt{GROUP BY da2.ID}
\end{tabbing}}
\end{gotofull}
%\vspace{-2mm}
\noindent\textbf{[Query AD: Authors' Discovery]} Query~AD finds the authors who published papers that pertain to the terms identified by $t_1^{ID},\ldots,t_n^{ID}$ (e.g., authors that published papers related to ``neoplasms" and ``statins").  The also displays the number of papers per author.  Figure~\ref{fig:xmpl:RQNA}(c) shows the RQNA expression.
%\vspace{-1mm}
{\small
\begin{tabbing}
Query AD:\\
\=\gt{SELECT da.Author, COUNT(*)}\\
\>\gt{FROM DA da}\\
\>\gt{WHE}\=\gt{RE da.Doc IN}\\
\> \>\gt{(}\=\gt{SELECT dt.Doc FROM DT dt }\gt{WHERE dt.Term = }$t_1^{ID}$\gt{)}\\
%\> \> \>\\
\> \>\gt{INTERSECT}\\
\> \> \> $\vdots$ \\
\> \>\gt{INTERSECT}\\
\> \>\gt{(}\=\gt{SELECT dt.Doc FROM DT dt }\gt{WHERE dt.Term = }$t_n^{ID}$\gt{)}\\
%\> \> \>\\
\>\gt{GROUP BY da.Author}
\end{tabbing}}

\yannis{Keep the following in the long version, along with the query I placed in \gt{eat\{\}.}  chunbin:yes}

\begin{gotofull}
Note that relationship queries do not require that all subqueries have identical structure. Furthermore, the aggregation is not necessary. For example, the following query finds authors who have recently (after 2012) published a paper on ``statins'' (term id $583352$) and at least one of the paper's authors had also published on ``lung neoplasms" (term id $384053$).

\begin{tabbing}
\=\gt{SELECT da.Author}\\
\>\gt{FROM DA da}\\
\>\gt{WHE}\=\gt{RE da.Doc IN}\\
\> \>\gt{(}\=\gt{SELECT dt.Doc FROM DT dt}\\
\> \> \>\gt{WHERE dt.Term = 583352)}\\
\> \>\gt{INTERSECT}\\
\> \>\gt{(}\=\gt{SELECT d.ID FROM Document d}\\
\> \> \>\gt{WHERE d.Year > 2012)}\\
\> \>\gt{INTERSECT}\\
\> \>\gt{(}\=\gt{SELECT da.Doc}\\
\> \> \>\gt{FROM DA da JOIN DT dt ON da.Doc = dt.Doc}\\
\> \> \>\gt{WHERE dt.Term = 384053)}\\
\end{tabbing}
\end{gotofull}

\noindent\textbf{[Query FAD: Co-Occuring Terms Discovery]} Very similarly to the Query AD, the Query FAD finds what (other) terms occur in documents about ``neoplasms" and ``statins" and how often. Figure~\ref{fig:xmpl:RQNA}(d) shows its RQNA expression.
\end{example}

\vspace{-6mm}
\begin{gotofull}
\begin{example}
The Scripps Research Institute implemented the {\small\url{Knowledge.Bio}}\smallfootnote{http://knowledge.bio/}~\cite{bainbridge2011whole}system for exploring, learning, and hypothesizing relationships among concepts of the SemMedDB database\smallfootnote{http://skr3.nlm.nih.gov/SemMedDB/dbinfo.html}~\cite{kilicoglu2012semmeddb}, a repository of semantic predications (subject-predicate-object triples).
It currently contains information for approximately 70 million predications from all of PubMed citations. One attractive example in knowledge.bio is that given a patient with, say, a mutation in the Sepiaterin Reductase (SPR) gene, a collection of Movement Disorders and significantly disrupted sleep, knowledge.bio returns ``Serotonin'' as a possible treatment by exploring the relevance of those concepts. The biggest challenge of knowledge.bio is the low performance due to the large data size. By applying our \fastr algorithm, the running time of discovering the top-k relevant concepts is dramatically reduced.

\begin{figure}[h!]
\includegraphics[width=0.5\textwidth]{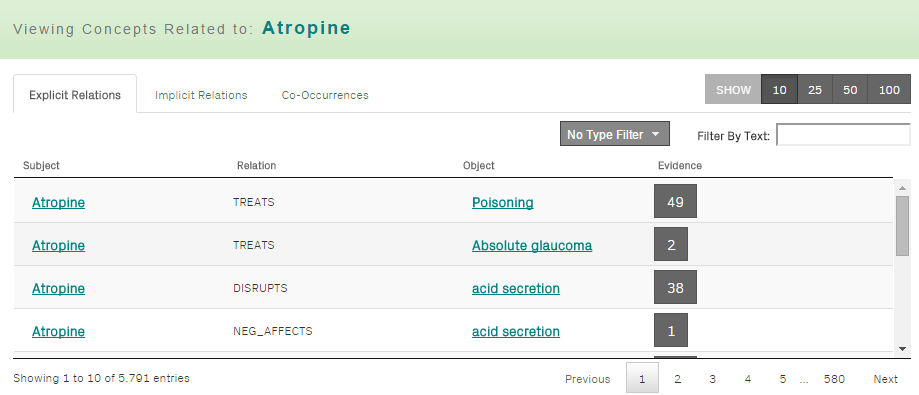}
\caption{\textbf{Example application of \fastr in biomedical area}.}
\label{figure:bio}
\end{figure}

\begin{figure}[h!]
\center
\includegraphics[width=0.4\textwidth]{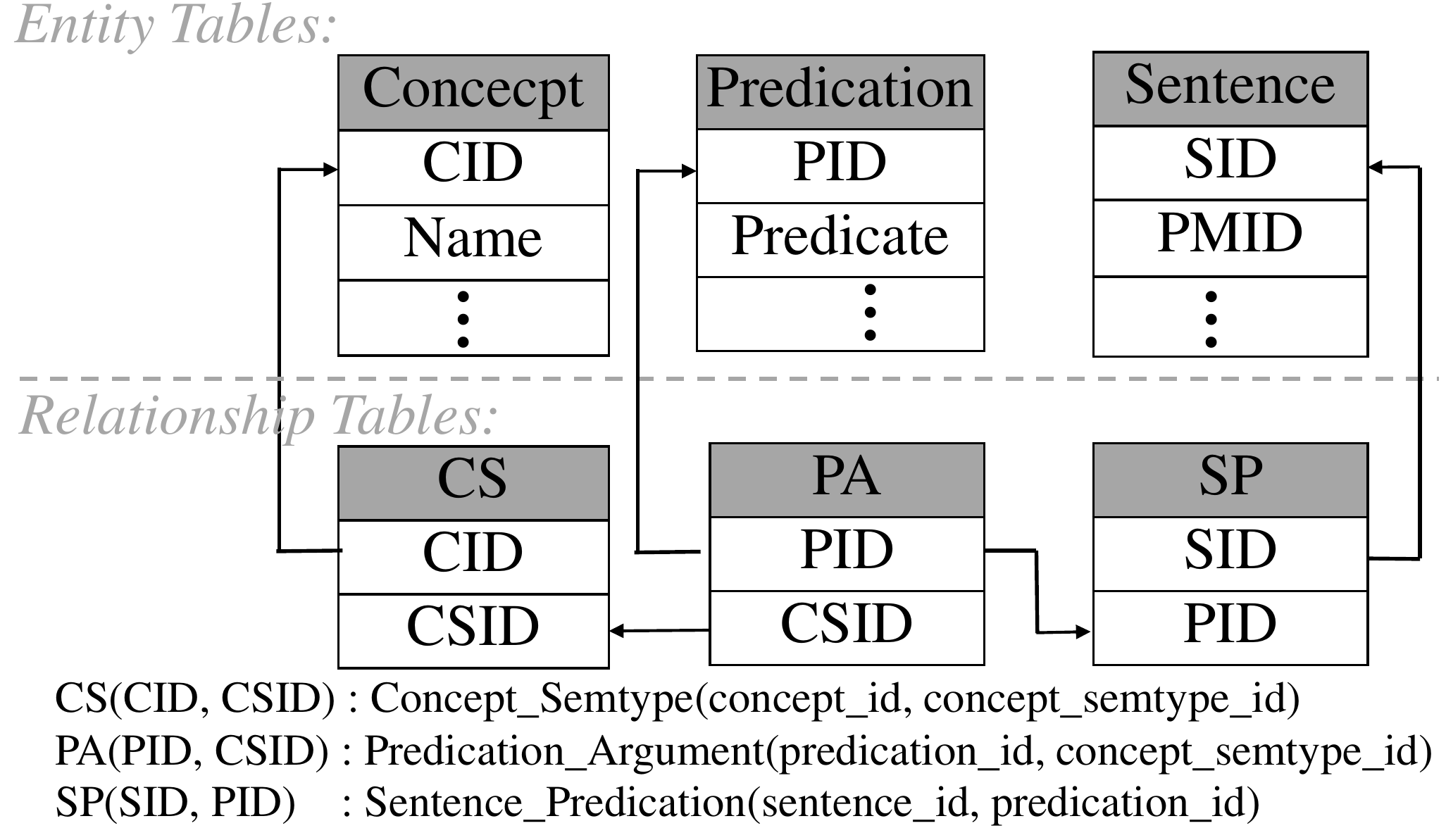}
%\vspace{-2mm}
\caption{\textbf{SemmedDB database schema}. \gt{CS}, \gt{PA} and \gt{SP} are relationship tables while the others are entity tables.}
\label{figure:semmedDB_Schema}
\end{figure}

Figure \ref{figure:bio} shows a screenshot of the \textit{Knowledge.Bio} system. The first result shows that ``Atropine'' treats ``Rattus norvegicus'', according to 319 evidence points. An evidence point is a sentence (in some document) that describes this relation.

Figure~\ref{figure:semmedDB_Schema} shows the schema of SemMedDB. As a use case from \textit{Knowledge.Bio}, consider the task of finding the concepts that are most relevant to ``Atropine''. To answer this query, \textit{Knowledge.Bio} issues the ``concept similarity (CS)'' SQL query in, which returns relevant concepts to ``Atropine''
\smallfootnote{The $c^{ID}$ is the concept id of ``Atropine''}.
%I removed the following until you answer: The results obtained by FastR are highlighted with red color in \textit{Knowledge.Bio}.}

{\small
\begin{tabbing}
CS query:\\
\=\gt{SELECT c2.CID, COUNT($\star$)}\\
\>\gt{FROM CS c2, PA p2, SP s2}\\
\>\gt{WHERE}\=\gt{s2.PID = p2.PID}\\
\>\>\gt{AND p2.CSID = c2.CSID AND s2.SID}\\
\>\>\gt{IN}\=\gt{(}\\
\>\>\>\gt{SELECT s1.SID}\\
\>\>\>\gt{FROM CS c1, PA p1, Sp s1}\\
\>\>\>\gt{WHERE}\=\gt{ s1.PID = p1.PID }\gt{AND p1.CSID = c1.CSID}\\
%\>\>\>\> \\
\>\>\>\>\gt{AND c1.CID =$c^{ID}$)} \\
\>\gt{GROUP BY CID}
\end{tabbing}}
The running time of this query on an Amazon Relational Database Service (Amazon RDS) with MySQL was 25 minutes. \fastr reduced the running time for that query to less than 1 second.
\end{example}
\end{gotofull}

\begin{gotobasic}
%\vspace{-7mm}
\begin{example}
The Scripps Research Institute implemented the {\small\url{Knowledge.Bio}}\smallfootnote{http://knowledge.bio/} system for exploring, learning, and hypothesizing relationships among concepts of the SemMedDB database\smallfootnote{http://skr3.nlm.nih.gov/SemMedDB/dbinfo.html}~\cite{kilicoglu2012semmeddb}, a repository of semantic predications (subject-predicate-object triples).
\yannis{to Chunbin, add in extended: done}
\begin{figure}[h!]
\center
\includegraphics[width=0.4\textwidth]{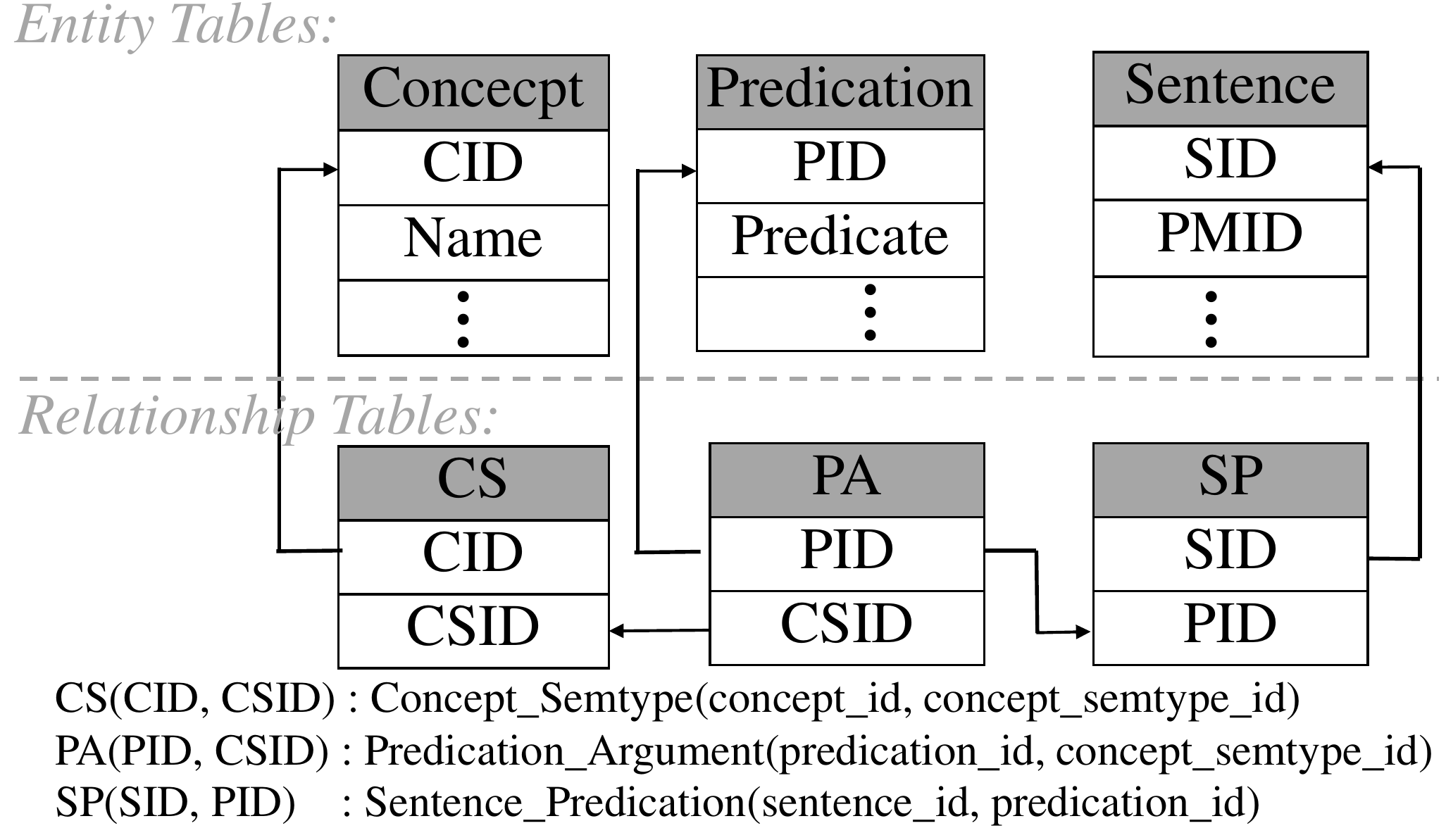}
%\vspace{-2mm}
\caption{\textbf{SemmedDB database schema}. \gt{CS}, \gt{PA} and \gt{SP} are relationship tables while the others are entity tables.}
\label{figure:semmedDB_Schema}
\end{figure}
Figure~\ref{figure:semmedDB_Schema} shows the schema of SemMedDB.
As a use case from \textit{Knowledge.Bio}, consider the task of finding the concepts that are most relevant to a given concept, e.g., ``Atropine''. To answer this query,  the ``concept similarity (CS)'' SQL query\smallfootnote{The $c^{ID}$ is the concept id of ``Atropine''} is issued. See the extended version \cite{lin2016fastr} for a discussion of use cases.
%I removed the following until you answer: The results obtained by FastR are highlighted with red color in \textit{Knowledge.Bio}.}

%\vspace{-2mm}
{\small
\begin{tabbing}
CS query:\\
\=\gt{SELECT c2.CID, COUNT($\star$)}\\
\>\gt{FROM CS c2, PA p2, SP s2}\\
\>\gt{WHERE}\=\gt{s2.PID = p2.PID}\\
\>\>\gt{AND p2.CSID = c2.CSID AND s2.SID}\\
\>\>\gt{IN}\=\gt{(}\\
\>\>\>\gt{SELECT s1.SID}\\
\>\>\>\gt{FROM CS c1, PA p1, Sp s1}\\
\>\>\>\gt{WHERE}\=\gt{ s1.PID = p1.PID }\gt{AND p1.CSID = c1.CSID}\\
%\>\>\>\> \\
\>\>\>\>\gt{AND c1.CID =$c^{ID}$)} \\
\>\gt{GROUP BY CID}
\end{tabbing}}
%\vspace{-3mm}

The running time of this query on an Amazon Relational Database Service (Amazon RDS) with MySQL was 25 minutes. \fastr reduced the running time for that query to less than 1 second.
\end{example}
\end{gotobasic}

% on a computer with 128GB memory
\yannis{to Chunbin: why is only the first tuple of the result highlighted as \fastr? Aren't all the lines result of \fastr? Chunbin: I fixed, they are all \fastrs results}
\yannis{to Chunbin: Rename the primary keys of the entity tables to \gt{ID}, per the convention assumed in the data.tex. Chunbin:ok}
\yannis{to Chunbin: what does it mean ``Amazon RDS" for MySQL? Do they provide many RDS's? Also, when one runs RDS is the query executed on the RDS or the computer? Chunbin: on the RDS.  Amazon RDS provides you six familiar database engines to choose from, including Amazon Aurora, Oracle, Microsoft SQL Server, PostgreSQL, MySQL and MariaDB}
\yannis{to Chunbin: bring back the further examples in the \gt{eat\{\}} in the extended. Chunbin :done}

\begin{gotofull}
\subsection{Further Examples}
Relationship queries can be found in a variety of applications. Some examples are outlined in the following list.

{\em Potential Virus Discovery in Network Security}~ \cite{franczek1999computer,knight2004method}: Consider a database documenting virus infections in a computer network with tables for ``virus'' entities, ``host IP'' entities, and virus instance - host IP relationships. To discover potential virus infections for a host who has reported a virus $s$, a relationship query first selects the set of hosts associated with $s$, and then retrieves and aggregates all the virus infections know for these hosts. The viruses with the high scores might also hide in the host computer.

{\em Friend Suggestion in Social Networks}~\cite{roth2010suggesting}: Consider a database with ``user'' entities, ``tweet'' entities, and a relationship associating tweets with users. For example, the relationship captures the information that a user read or shared a tweet. To provide friend suggestions for a given user $u$, a relationship query first discovers his/her tweets, then returns a sorted list of users based on their association with the discovered tweets.
\end{gotofull}

%\vspace{-2mm}
\begin{gotobasic}
The extended version \cite{lin2016fastr} provides additional examples, including cases on virus discovery in networks and analytics in social networks.
\end{gotobasic}

\section{\fastr Data Structure}
\label{sec:data-structure}
\normalsize
\begin{gotobasic}
The introduction presented the essentials of \fastr index structure. In this section, we illustrate more details of the lookup data structure and the fragment encodings.
\end{gotobasic}

\begin{gotofull}
Consider a relationship table $R(D_1, D_2, M_1, \ldots, M_m)$. The \fastr data structure is optimized towards two goals: (i) rapidly evaluating $\pi_{A}\sigma_{D_i=c}(R)$, where $A$ may be any attribute $D_j, j \neq i$ or $M_j$ and (ii) minimizing space by using compressed data structures. The most important mechanism towards compression is \emph{fragments} that encode each $\pi_{A}\sigma_{D_i=c}(R)$ using techniques such as compressed bitmaps and Huffman encoding.

For each table of a graph database, \fastr stores two \emph{indices} $\mathcal{I}_{R.D_1}$ and $\mathcal{I}_{R.D_2}$.%
\footnote{The administrator may elect to store only one of the two indices but then some queries will be not be amenable to \fastrs efficient processing.}
The only storage pertaining to $R$ are these indices.
The $\mathcal{I}_{R.D_1}$ index consists of a \emph{lookup data structure} on $D_1$ and \emph{fragments} corresponding to the other attributes. Given an ID $c$ and an attribute $A$, a \emph{lookup algorithm} uses the data structure to return (i) a pointer to a byte array that encodes the fragment $\pi_{A}\sigma_{D_i=c}(R)$ and (ii) the size of the byte array, which is required by the algorithms that decode fragments.
\end{gotofull}
\vspace{-3mm}
\paragraph*{Lookup Data Structure}
\begin{gotobasic}
Recall that \fastr stores two \emph{indices} $\mathcal{I}_{R.F_1}$ and $\mathcal{I}_{R.F_2}$ for a relationship table $R(F_1, F_2, M_1, \ldots, M_m)$, where $F_1$, $F_2$ are foreign keys and each $M_i$ is a measure.
\end{gotobasic}
Since $F_1$ is a foreign key, its values are IDs of an entity $E$. Therefore, they are integers with a range $[0, h-1]$, where $h$ is $|E|$. Based on this observation, the index $\mathcal{I}_{R.F_1}$ has one \emph{attribute byte array} for the foreign key attribute $F_2$ and one \emph{attribute byte array} for each one of the measure attributes $M_1, \ldots, M_m$. The attribute byte array of an attribute $A$ stores the fragments $\pi_{A}\sigma_{F_1=0}(R), \pi_{A}\sigma_{F_1=1}(R),$ $\ldots, \pi_{A}\sigma_{F_1=h-1}(R)$ consecutively. Notice that some fragments will be empty.

\begin{figure}[!htp]
\centering
\includegraphics[width=0.47\textwidth]{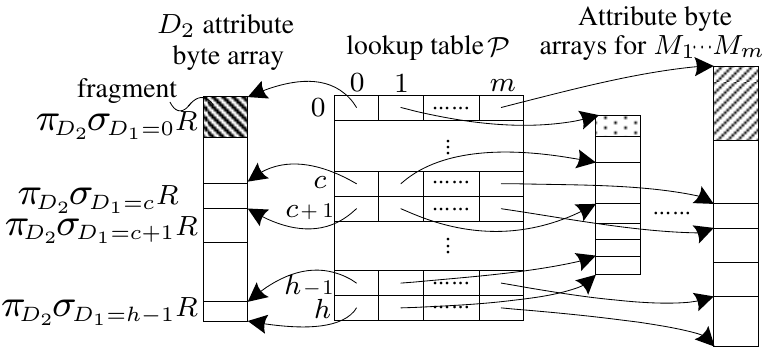}
\vspace{-3mm}
\caption{\textbf{\fastr Index $\mathcal{I}_{R.F_1}$}. The lookup table stores offsets of fragments. The fragments of different columns have different encodings.}
\label{fig:data-structures}
\end{figure}

The lookup data structure is a {\em lookup table} $\mathcal{P}$, as illustrated in Figure~\ref{fig:data-structures}. Each attribute $F_2, M_1, \ldots, M_m$ is associated with a column of the table; e.g., $F_2$ is associated with the $0$-th column, $M_1$ with the $1$st column, etc. Notice that no column is associated with $F_1$. The lookup table has $h+1$ rows, with array indices $0$ to $h$. Cell $\mathcal{P}[c][a]$ contains a pointer to fragment $\pi_{A}\sigma_{F_1=c}(R)$, where $a$ is the column number for the attribute $A$. The size of the fragment is $\mathcal{P}[c+1][a] - \mathcal{P}[c][a]$. There must be a row $\mathcal{P}[h]$ for computing the size of the last fragment $\pi_{A}\sigma_{F_1=h}(R)$. Notice that by computing the size of fragments from consecutive offsets, \fastr does not need to store fragment sizes explicitly.

For further space savings, the pointers are actually offsets that are represented with the minimum necessary number of bytes. If the fragment byte array of $A$ has size $b_A$, then the offsets pointing to fragments of the byte array, will be integers of size $\lceil \log_{256} b_A \rceil$ bytes.
For example, assume the $F_2$ attribute byte array (Figure~\ref{fig:data-structures}) is 3GB and is located at the 64-bit memory address \texttt{0x000000086} \texttt{0A23800}. Offset values pointing to it will be 4-byte integers, since $\lceil \log_{256} 3G \rceil =4$. Also, assume the offset to the $c$-th fragment $\pi_{A}\sigma_{F_1=c}(R)$ is \texttt{0x00B09000} and the offset to the $c+1$-st fragment is \texttt{0x00B09100}. Then, given a request for the fragment $\pi_{A}\sigma_{F_1=c}(R)$, the lookup algorithm returns the start pointer $\mathcal{F}_{R.A}$=$\mathtt{0x0000000860A23800}+\mathtt{0x00B09000}$ and the size $l$=$\mathtt{0x00B}$ $\mathtt{09100} - \mathtt{0x00B09000}$. If this fragment is encoded by an encoding method $\textbf{E}$, \fastr decodes it by using a macro $decode\textbf{E}(\mathcal{F}_{R.A}, l: \mathcal{A}_{R.A}, n)$ to produce a decoded array $\mathcal{A}_{R.A}$ and  the number of elements $n$ within it. In the following, we introduce encodings for fragments utilized in this paper.

\smallskip
\begin{gotofull}
\noindent\textbf{Retrieve a fragment $\pi_A\sigma_{F_i=c}(R)$.}
The key feature of \fastr lookup data structure is it can retrieve a fragment efficiently. More precisely,
\fastr first obtains an offset-arrays $\mathcal{P}=\mathcal{I}_{R.F_i}[c]$ by a random access, and its neighbour
$\mathcal{P}_{next}=\mathcal{I}_{R.F_i}[c+1]$. Then \fastr gets the start address of the fragment $\mathcal{F}_{R.A}= \mathcal{P}[A]$ and its length $l$=$\mathit{P}_{next}[A] - \mathcal{F}_{R.A}$. If this fragment is encoded by an encoding method $\textbf{E}$, \fastr decodes it by using a macro $decode\textbf{E}(\mathcal{F}_{R.A}, l: \mathcal{A}_{R.A}, n)$ to produce a decoded array $\mathcal{A}_{R.A}$ and  the number of elements $n$ within it. In the following, we introduce encodings for fragments utilized in this paper.
\end{gotofull}

\yannis{to chunbin: Change the height of the $D_2$, $M_1$, $M_m$ columns in Figure~\ref{fig:data-structures} to visualize that they do not have to be same size. The last index is wrong. It should be $h$ and the one before the end $h-1$. Chunbin: done}

\yannis{removed: A fragment of a foreign key column is called a \textit{F-Fragment} and one from a measurement column is called a \textit{M-Fragment}.Chunbin:OK}
\vspace{-1mm}
\paragraph*{Fragment Encodings}
\label{sec:fragments}
\yannis{The size of each fragment is rounded up to bytes.}
\eat{As we stated in the introduction, unlike column databases, \fastr does not require random access within fragments, because \fastrs query processing either uses all values in a fragment or none of them at all. This feature provides opportunity to use compressions that do not support random accesses.}
\begin{gotobasic}
\fastr currently uses the following four representative encodings.
%: Uncompressed Array (UA), Bit-aligned Compressed Array (BCA), Byte-aligned Compressed Bitmaps (BB) and Huffman-encoded arrays (Huffman).  Recall that we only discuss numerical values, for string columns, \fastr applies a dictionary encoding that maps strings to integers upon load time.
\eat{The extended version explains the reason why we choose these four encodings. Here we only discuss numerical values, for string columns, \fastr applies a dictionary encoding that maps strings to integers upon load time.}

\noindent \textbf{\textit{U}ncompressed \textit{A}rray (UA)}: An uncompressed array stores the original numerical values in their declared type.
%, which may be an 8-bit, 16-bit, 32-bit or 64-bit type.

\noindent \textbf{\textit{B}it-aligned \textit{C}ompressed \textit{A}rray (BCA)}: Assume a foreign key attribute points to the IDs of an entity, which range from $0$ to $h-1$. Then each foreign key value needs $\lceil\log_2 h\rceil$ bits. Consequently, a fragment $\pi_A \sigma_{F=c} R$ with size $n_c$ takes $\lceil \frac{n_c \cdot\lceil\log_2 h\rceil}{8}\rceil$ bytes. Notice that the fragment is padded to an integer number of bytes (hence the use of $\lceil . \rceil$) because fragments always take whole bytes, since \fastr needs efficient random access to the start of fragments.

\noindent \textbf{\textit{B}yte-aligned compressed \textit{B}itmap (BB)}: Given an array $[v_1,\ldots,v_n]$,
%, where each $v_i$ is a non-negative integer
the equivalent uncompressed bitmap is a sequence of $n$ bits, such that the bits at the positions $v_1,\ldots, v_n$ are 1 and all other bits are 0. \fastr uses the byte-aligned method to compress the bitmap~\cite{antoshenkov1995byte,wu2006optimizing}. The first bit of a byte is a flag that declares whether (i) the next seven bits are part of a number that also uses consequent bytes or (ii) the remaining seven bits actually represent the length number by themselves.

\noindent \textbf{Huffman-encoded Array (Huffman)}: \fastr employs the standard Huffman encoding scheme~\cite{huffman1952method,storer1988data,ChungW99}. It uses an efficient decoding algorithm proposed in ~\cite{ChungW99}, which avoids tree traversals (i.e., avoids random access on the heap). This gives performance advantages due to CPU caching (L1 and L2) effects, since an array is a consecutive block of data.

Note that, \fastr does not have any restriction on encoding methods, thus it allows wide range of encoding methods even they do not support random accesses. The extended version provides more details on describing the encoding methods.
\end{gotobasic}

\begin{gotofull}
Fragments are generally encoded by different methods. Unlike column databases, \fastr does not require random access in the compressed data, because \fastrs query processing either uses all values in a fragment or none of them at all. \fastr currently uses four representative encodings: Uncompressed arrays, bit-aligned compressed array, byte-aligned compressed bitmaps and huffman-encoded arrays. The latter three have been chosen because (i) each one of them compresses well (and better than the rest) in particular, easy-to-recognize scenarios, and (ii) carefully implemented decoding algorithms have acceptable CPU overhead in the cases of ``specialty" of each encoding. Nevertheless, other encodings can also be applied, such as PforDelta~\cite{zukowski2006super} that combines aspects of run length encoding and delta encoding.

All fragments of an attribute's byte array use the same encoding. However, it is allowed (and it is sometimes beneficial) \matthias{Can we find an example?} \yannis{to Chunbin: Has it happened in our optimum encoding of either PubMed or SemMedDB? Chunbin:unfortunately, no.} to encode a measure attribute $M_i$ of a table $R(D_1, D_2, \ldots, M_i,\ldots)$ in $\mathcal{I}_{R.D_1}$ with an encoding method different from the one used in $\mathcal{I}_{R.D_2}$. Finally, note that the four encoding methods are used only for numerical values. For string columns, \fastr applies a dictionary encoding that maps strings to integers upon load time, similarly to column databases. In our applications, the dictionary tables that represent the string-to-integer mappings are stored outside main memory, in a conventional database.
\end{gotofull}

\begin{gotofull}
\fastr currently uses the following representative encodings:

\smallskip
\noindent \textbf{\textit{U}ncompressed \textit{A}rray (UA)}: An uncompressed array stores the original numerical values in their declared type, which may be an 8-bit, 16-bit, 32-bit or 64-bit type. Notice that if the index $\mathcal{I}_{R.D_1}$ encodes an attribute $A$ as uncompressed array, then the byte array of attribute $A$ is identical to what the column for $A$ would be in a column database that stores $R$ sorted by $D_1$.

\yannis{to Chunbin: The name Bit Aligned Dictionary is a remnant of the time that we were using these as if they were dictionaries. However, they are simply compressed arrays, since everything is arrays. Correct? Hence the name should change to Bit-Aligned Compressed Arrays.Chunbin: correct, and I changed all the BD to BCA accordingly.}

\smallskip
\noindent \textbf{\textit{B}it-aligned \textit{C}ompressed \textit{A}rray (BCA)}: Since \fastr does not require random access within the fragments, the BCA encoding uses fewer than 32 (or 64) bits to store each number of an array. In particular, let us assume that a foreign key attribute $D$ points to the IDs of an entity $E$, which range from $0$ to $h-1$. Then each foreign key value needs $\lceil\log_2 h\rceil$ bits. Consequently, a fragment $\pi_A \sigma_{D=c} R$ with size $n_c$ takes $\lceil \frac{n_c \cdot\lceil\log_2 h\rceil}{8}\rceil$ bytes. Notice that the fragment is padded to an integer number of bytes (hence the use of $\lceil . \rceil$) because fragments always take whole bytes, since \fastr needs efficient random access to the start of fragments.

\yannis{to self: mention earlier that the dictionary tables do mapping. mention earlier that \fastr conditions may only be equalities when a field is encoded. mention how we deal with non-equality conditions}
When a measure attribute is dictionary-encoded upon loading, it takes values from $0$ to $m-1$, where $m$ is the number of unique values of $M$. Hence, each value is encoded in $\lceil\log_2 m\rceil$ bits. Finally, if an unsigned integer measure attribute is not dictionary-encoded, its values are encoded in $\lceil\log_2 r \rceil$, where $r$ is the largest value.

\yannis{to Chunbin, we can keep this piece in the extended version, after cleaning up its flow:Chunbin:yes}

\vspace{2mm}
\noindent \textbf{\textit{U}ncompressed \textit{B}itmap (UB)}: For each fragment, \fastr maintains a bit array with size $\mathcal{D}+\Delta'$, where $\Delta'\in[0,7]$ to ensure that fragments are byte-aligned for efficiently accessing a random fragment.  Note that the size of an uncompressed bitmap only relies on the domain size and is independent from the fragment size. For example, assume that $\mathcal{D}=12$ and a fragment with values \{0,2,5,8,11\} should be encoded as 10100100100100000. Note that the size of this bit-array is $12+4=16$ rather than $12$ since \fastr requires an additional $4$ bits. One important and attractive feature of bit-arrays is that intersection operations can be transformed to a logical (bitwise) AND operation.

\vspace{2mm}
\noindent \textbf{\textit{B}yte-aligned compressed \textit{B}itmap (BB)}: Bitmaps have been heavily used in analytical processing and provide yet another alternative to uncompressed arrays. Conceptually, given an array $[v_1,\ldots,v_n]$, where each $v_i$ is a non-negative integer, the equivalent uncompressed bitmap is a sequence of $n$ bits, such that the bits at the positions $v_1,\ldots, v_n$ are 1 and all other bits are 0. Bitmap compression methods are based on encoding the length of the (typically long) sequences of zeros that appear between the ones \cite{wu2002compressing,guzun2014tunable}. Compression algorithms differ from each other on the specifics of encoding the lengths. \fastr uses the byte-aligned method to represent a length number \cite{antoshenkov1995byte,wu2006optimizing}. The first bit of a byte is a flag that declares whether (i) the next seven bits are part of a number that also uses consequent bytes or (ii) the remaining seven bits actually represent the length number by themselves. For example, the length numbers for the  sequences of zeros in the uncompressed bitmap (UB) are $100$, $3000$ and $95$ respectively.  \yannis{to chunbin: spell out ``UB" as Uncompressed Bitmap, since we no longer have UB. Chunbin:fixed.}

\vspace{1mm}
\includegraphics[width=0.46\textwidth]{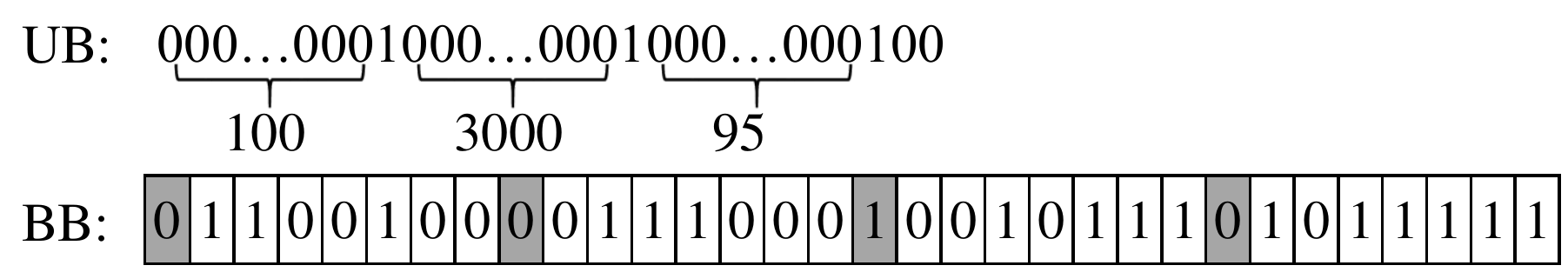}
\vspace{1mm}

\fastr uses the first byte to represent the length number $100$ and the next two bytes for the number $3000$ (since $2^7<3000<2^{14}$). Note that the first bit of the first byte is 1, which means the seven bits in the following byte is still part of the number. Notice that \fastr uses the little endian format, when it represents multi-byte numbers.

\vspace{2mm}
\noindent \textbf{Huffman Encoding}: For attributes with many duplicate values and an uneven frequency distribution, e.g. Zipf distribution, \fastr employs Huffman encoding \cite{vinokur1992huffman,vinokur1986huffman}. \fastr maintains a global Huffman tree, which reflects frequencies in the entire column, but encodes each fragment separately. In addition, \fastr uses an efficient decoding algorithm that avoids tree traversals (i.e., avoids random access on the heap). In lieu of the usual encoding tree, it uses an array as in~\cite{ChungW99}. This gives performance advantages due to CPU caching (L1 and L2) effects, since an array is a consecutive block of data.
\end{gotofull}

\smallskip
We compare the performance/storage tradeoff of the various encodings analytically and experimentally.

%\section{Compression Quality Analysis}
%\label{sec:analysis}
The following table summarizes the space needed by each fragment. Assume each fragment contains $N$ elements and the domain size of the column containing this fragment is $D$.
\begin{gotobasic}
\vspace{-2mm}
\begin{table}[htbp]
\small
\centering
\renewcommand{\tabcolsep}{1.5mm}
\begin{tabular}{|l|c|}
\hline
Uncompressed Array (UA) &  $32\cdot N\cdot\lceil\log_{2^{32}}D\rceil$ \\
\hline
Bit-aligned Compressed Array (BCA) &  $8\cdot \big\lceil \frac{N\cdot\lceil\log_{2}D\rceil}{8}\big\rceil$ \\
\hline
Byte-aligned Compressed Bitmap (BB) & $N \cdot(8\cdot\lceil\log_{128}\frac{D-N}{N}\rceil)$\\
\hline
Huffman & $8\cdot\lceil\frac{N\cdot E_D+D}{8}\rceil$\\
\hline
\end{tabular}
\end{table}
\vspace{-2mm}
\end{gotobasic}
%$8\cdot\lceil \big( N\cdot( \frac{s}{H_{D,s}}\sum_{i=1}^D\frac{ln(i)}{i^s}$+$ln(H_{D,s}))\big)/8 \rceil$

\begin{gotofull}
\begin{table}[!htp]
\small
\centering
\renewcommand{\tabcolsep}{1.5mm}
\begin{tabular}{|l|c|}
\hline
Uncompressed Array (UA) &  $32\cdot N\cdot\lceil\log_{2^{32}}D\rceil$ \\
\hline
Uncompressed bitmap (UB) & $8\cdot \big\lceil \frac{D}{8}\big\rceil$ \\
\hline
Bit-aligned Compressed Array (BCA) &  $8\cdot \big\lceil \frac{N\cdot\lceil\log_{2}D\rceil}{8}\big\rceil$ \\
\hline
Byte-aligned Compressed Bitmap (BB) & $N \cdot(8\cdot\lceil\log_{128}\frac{D-N}{N}\rceil)$\\
\hline
Huffman& $8\cdot\lceil\frac{N\cdot E_D+D}{8}\rceil$\\
\hline
\end{tabular}
\end{table}
\end{gotofull}

where $E_D=-\sum_{i=1}^Dp_i\log p_i$  is the entropy of the column, $p_i$ is the probability of occurrences of element $i$. \footnote{\small Here we report the lower bound of the space needed by Huffman. The space needed by Huffman is bounded by  [$8\lceil\frac{N\cdot E_D+D}{8}\rceil$, $8\lceil\frac{N\cdot E_D+N+D}{8}\rceil$)~\cite{navarro2005new}.}

\yannis{to Chunbin: add Hoffman with necessary explanations in text as need be.Chunbin:Yes}
%where $H_{D,s}$ is the $D$-th generalized harmonic number $H_{D,s}$=$\sum_{k=1}^D\frac{1}{k^s}$.

\begin{figure}[h!]
\centering
\includegraphics[width=0.5\textwidth]{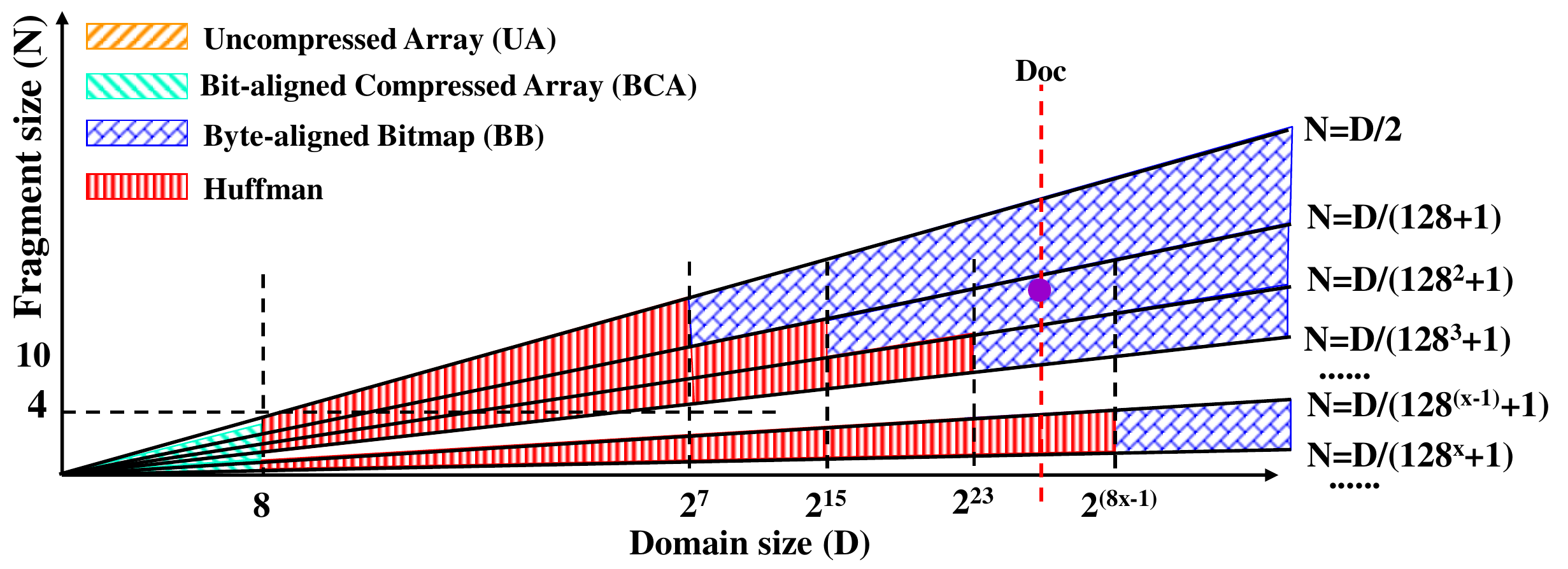}
\vspace{-5mm}
\caption{\textbf{Space cost comparison of different encodings}. Each area is colored by the compression method with minimal space cost. For example, in the blue area BB compresses the most. The dotted red line with purple circle shows the domain size of \gt{Doc} in the PubMed dataset, which indicates that BB is the best method for  \gt{Doc} fragments}
\label{fig:compression_analysis_1}
\end{figure}

Figure~\ref{fig:compression_analysis_1} shows the most compact way to encode a fragment in key/foreign key columns, as a function of (a) number $N$ of elements in the fragment (vertical axis), and (b) the size of the underlying domain $D$ (horizontal axis).\smallfootnote{Since BCA only applies for fragments containing only unique values.} % and (c) the parameter $s$ of a zipf distribution
Not surprisingly, uncompressed arrays are never the most compact method.
\begin{gotofull}
(For one, they always take more space than bit-aligned compressed arrays.)
\end{gotofull}
In the common cases, where $D>2^7$ and $N \leq D/8$, compressed bitmaps are more compact than bit-aligned compressed arrays when $\frac{D}{128^x+1}\leq N<\frac{D}{8}$, for an $x\geq 1$.
\yannis{to Chunbin: You can add a few more if you think there are more interesting cases.Chunbin:sure}

\begin{gotobasic}
Ideally, each fragment should be encoded by the one with minimal space cost indicated by Figure~\ref{fig:compression_analysis_1}, but in this way, it costs more space to record the encoding method for each fragment. We discuss this trade-off in the extended version and observe that applying the same encoding for all the fragments of a column is reasonably optimal or close to optimal in the datasets.
\end{gotobasic}

\begin{gotofull}
Figure~\ref{fig:compression_analysis_1} implies that different fragments of the same column may be most compactly encoded with different methods. For example, a fragment of more documents id's of a term should be encoded with BB, otherwise, it is less suitable to apply BCA. Though applying different encodings for different fragments can achieve minimal space cost, the penalty is that we need to remember the encoding method for each fragment, which increases the space cost. To balance this trade-off, in this paper, we apply the same encoding (the one with minimal space cost for the fragment with average size) for fragments in the same column. Note that, fragments in different columns still benefit from applying different encodings. And only one encoding type is required to store for each column, which can be stored in the metadata.
\end{gotofull}

% \begin{figure}[h!]
% \centering
% \includegraphics[width=0.5\textwidth]{compression_comparison7}
% \caption{\textbf{Space cost comparison of different encodings}. Each area is colored by the compression method with minimal space cost. For example, in the red area BCA compresses the most. The dotted red lines show the domain sizes of \gt{Term}, \gt{Author} and \gt{Doc} in the PubMed dataset. In addition, the red triangle and circle indicates that BB is the best method for \gt{Term} fragments and \gt{Doc} fragments, while BCA is the best for \gt{Author} fragments.}
% \label{fig:compression_analysis_1}
% \end{figure}

\section{\fastr Query Processing}
\label{sec:qp}

\yannis{to self, for RQNA section: mention that, for brevity, we omit operators and enhancements that are used in practice, yet they have no repercussions on performance. Example, projection and group-by that ranem their inputs, execute scalar functions.}

\yannis{to Chunbin: Adjust RQNA and examples so that projections are pushed down. As things stand now the translation from RQNA to physical operators cannot be accomplished. }

\yannis{to Chunbin: Change the RQNA algebraic plans to include pushing projections above the joins, where the projections show which attributes will need to be retrieved from the right hand side. As you see in the logical-to-physical translation below, such projections are needed. Chunbin:Already push down, correct?}

The \textit{\fastr Query Processor} (Figure~\ref{fig:architecture}) produces executable C++ source code, as it was illustrated in the introduction. The processor first transforms the given query into an RQNA expression. Then, it translates the RQNA expression into a plan consisting of physical operators. E.g., the plans in Figure~\ref{fig:xmpl-physical-plans} correspond to the RQNA expressions in Figure~\ref{fig:xmpl:RQNA}. Then the code generator translates the plan into code, using also the metadata provided by the \textit{\fastr Loader}.

\yannis{to self, check later whether we have space and need to repeat the following:
The generated \fastr code reveals the following benefits: (1) \fastr uses a bottom-up pipelining method to avoid intermediate results and it also get rids of function call overload by using tight for-loops to iterate fragments, such simple loops are amenable to compiler optimization, and CPU out-of-order speculation~\cite{nes2012monetdb}; (2) \fastr uses proper encoding for each fragment to minimize space cost, since no random accesses are required within the accessed fragments, compressions do not support random accesses are allowed; and (3) \fastr is further optimized by tuning to the dense IDs, it uses arrays for aggregation and duplicates checking.
}

\chunbin{Why we mention the next sentence? Want to motivate why we need code generator?
In an actual FastR deployment, most SQL queries are parametrized (i.e., they may contain SQL's ``\gt{?}" in lieu of constants) and are prepared in advance of running.}

\yannis{to self: Conclusions and Future Work promise on non-RQNA. This section focuses on relationship queries. The conclusion and future work sections discuss algebraic expressions that contain RQNA subexpressions. They also discuss future extensions of the query processor. }

Section~\ref{sec:physical-operators} presents the physical operators and the key intuitions in the translation of RQNA expressions into plans. \begin{gotobasic}\cite{lin2016fastr} provides the complete translation algorithm.\end{gotobasic} \begin{gotofull}Appendix~\ref{sec:appendix} provides the complete translation algorithm.\yannis{to Chunbin: Write a translation pseudocode in the extended. Chunbin:yes}\end{gotofull} Then Section~\ref{sec:code_generator} describes how the \fastr code generator translates plans into code, essentially by mapping each physical operator into an efficient code snippet and stitching these snippets together.

\subsection{\fastr Physical Operators}
\label{sec:physical-operators}
\fastrs physical operators are designed to (i) utilize the \fastr data structure, and (ii) enable code generation to produce bottom-up pipelined execution code.  In the following, we explain the physical operators' syntax and semantics, neglecting for now the bottom-up pipelined execution aspects.

\noindent \textbf{Fragment-based Join} The operator $\rijoin_{B;R\mapsto r}^{r.A_1,\ldots,r.A_n} \mathcal{I}_{R.B'}L$ inputs the result of an expression $L$ that produces a column $B$, generally among others. For each value $b \in B$, the operator uses the index $\mathcal{I}_{R.B'}$ to retrieve (and decompress) the fragments $\pi_{A_i} \sigma_{r.B'=b} (R\mapsto r),$ $i=1,\ldots,n$. Intuitively $L$ would be the left operand of a conventional join and $R \mapsto r$ would be the right side. Conceptually, one may think that the fragments are combined into a result table whose schema has the attributes $A_1,\ldots,A_n$ (and also the attributes of $L$). However, in reality, the decompressed fragments are not combined into rows. In adherence to the late binding technique \cite{journals/ftdb/AbadiBHIM13,abadi2009column} of column-oriented processing, the ordering of the items in the fragments dictates how they can be combined into tuples.

The $\rijoin$ operator is useful for executing both selections and joins of the RQNA expressions:

\begin{compact_item}
\item A projection/join combination $\pi_{\mathit{attrs}(L), r.A_1,\ldots,r.A_n} (L \Join_{B=r.B'} R\mapsto r)$ where $B$ is an attribute of $L$ and $B'$ is a foreign key of a relationship table $R$ or $B'$ is the ID of an entity table $R$, translates to $L \rijoin_{B; R\mapsto r}^{r.A_1,\ldots,r.A_n} \mathcal{I}_{R.B'}$.

\item A projection/selection combination $\pi_{r.A_1,\ldots,r.A_n} \sigma_{r.B' = c} (R\mapsto r)$, where $c$ is a constant and $B'$ is a foreign key of a relationship table $R$ or $B'$ is the ID of an entity table $R$, translates to $\{ [B:c] \} \rijoin_{B; R\mapsto r}^{r.A_1,\ldots,r.A_n} \mathcal{I}_{R.B'}$. Essentially, \fastr reduces the selection into a join, by considering the left-hand-side argument to be a table with a single tuple and a single attribute $B$, whose value is $c$.
\end{compact_item}
\vspace{-3mm}
\noindent \textbf{Fragment-based Semijoin}
The operator $\risemijoin_{B; R\mapsto r}^{r.A_1,\ldots,r.A_n} \mathcal{I}_{R.B'} L$ operates similarly to the fragment-based join but returns only attributes from $(R\mapsto r)$ if there is a matching tuple in $L$. It is introduced in the plan when the RQNA expression has an expression $\pi_{r.A_1,\ldots,r.A_n}((R \mapsto r) \semijoin_{B = r.B'} L)$.

The operator maintains a lookup structure for values from the $B$ column of $L$; for each value $b \in B$, the operator checks the lookup structure to find out whether that particular value $b$ was already received earlier. If it were, then it is dismissed. If this is the first time that $b$ is received, then the operator marks in the lookup structure that this $b$ has been received and proceeds to use the index $\mathcal{I}_{R.B'}$ to retrieve (and decompress) the fragments $\pi_{r.A_i} \sigma_{r.B'=b} (R\mapsto r), i=1,\ldots,n$, as the join would.
\begin{gotobasic}
If $L$ is relatively large, it is best to use a boolean array, despite the fact that the query needs to initialize all array elements to false. Otherwise, a hash set or a tree is preferable.
\end{gotobasic}
\begin{gotofull}
The lookup structure can be a hash set, a tree-based set or an array of booleans whose size is the domain of $r.B'$. If $L$ is relatively large, it is best to use a boolean array, despite the fact that the query needs to initialize all array elements to false. If $L$ is relatively small, a hash set or a tree is preferable. In the absence of a size estimator, which would estimate the size of $L$, \fastr chooses the boolean array approach. The rationale is that the initialization of the array penalizes both fast queries (such as the AD query) and slow queries (such as the CS query) with a few milliseconds (the initialization time in AD and CS are 0.057ms and 121.72ms respectively), which depend exclusively on the size of the underlying domain.
Such initialization penalty is unimportant in absolute terms (i.e., the online user experience is not affected by a few milliseconds) but it saves significant lookup cost in the case of the relatively slow queries, where $L$ is usually large.
\end{gotofull}

\vspace{-4mm}
\begin{figure}[h!]
\centering
\includegraphics[width=1\columnwidth]{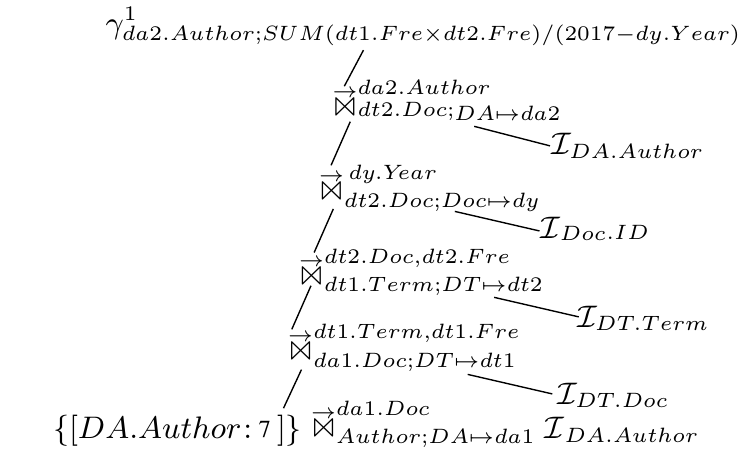}
\vspace{-5mm}
\caption{\textbf{Example Physical algebraic plan}. This is the physical algebra plan of queries AS.}
\label{fig:xmpl-physical-plans}
\end{figure}

\begin{gotobasic}
While joins and semijoins are sufficient, the extended version also describes the occassional replacement of semijoins with a \emph{merge intersection} operator that merges sorted one-attribute relations. In the example queries, it is useful for the Query~AD.
\end{gotobasic}

\begin{gotofull}

\noindent \textbf{Merge Intersection}
The operator $\riinter^{\theta}_{L_1,\ldots,L_m}$ computes the result of $L_1\cap\ldots\cap L_n$, when each argument $L_i$ consists of a single sorted attribute. In the case of RQNA queries, an $L_i$ meets the necessary conditions if it is an expression of the form $\pi_{l_i.B_i} \sigma_{l_i.K = c} (L_i \mapsto l_i)$, where $K$ is a (foreign or primary) key attribute.

The superscript $\theta=0$ indicates that the merging should happen directly on the encoded fragments.  In a common example, if all the fragments are encoded with compressed bitmap then the merge intersection can be directly applied on the compressed bitmaps. Otherwise, $\theta=1$ means fragments $L_1\ldots L_n$ are encoded with different encodings. Then each fragment needs to be decoded before intersecting by merging. More precisely, a (preallocated, since we know the worst case scenario in advance) array with size $\min\{|L_1|,\ldots,|L_m|\}$ is initialized to store the final intersection results. The operator has pointers pointing to the beginning of each fragment. The pointed-to values are compared: If all the values are the same, then this value is added to the array and all pointers are incremented. Otherwise, all pointers except the one with maximal value are moved forward
until the pointed-to values are greater or equal than the maximal one. Then it repeats the above step until all the values in a fragment have been accessed. The operator $\riinter^{\theta}_{L_1,\ldots,L_m}$ is introduced in the plan when the RQNA expression has a series of semijoins and the right arguments of the semijoins are a single sorted column.
\end{gotofull}
\begin{gotofull}
In particular, an RQNA subexpression $\pi_{r.A_1,\ldots,r.A_n} ((R \mapsto r) \semijoin_{r.B' = l_1.B_1} \pi_{l_1.B_1} L_1) \semijoin \ldots \semijoin_{r.B' = l_m.B_m} (\pi_{l_m.B_m} L_m)$ translates into the following expression:

$\risemijoin_{B; R\mapsto r}^{r.A_1,\ldots,r.A_n} \mathcal{I}_{R.B'} \riinter^{\theta}_{(\pi_{l_1.B_1} L_1),\ldots,(\pi_{l_m.B_m} L_m)}$

\eat{For example, see the RQNA expression of query AS in Figure~\ref{fig:xmpl:RQNA}(e) and the corresponding plan in Figure~\ref{fig:xmpl-physical-plans}. }
\yannis{to Chunbin: complete. Chunbin: done}
\end{gotofull}

\yannis{to Chunbin: Check that writeup corresponds to figures and the actual algorithm. I did not have the chance. There may be a cleaner translation algorithm than the one I suggest. Chunbin: Yes, I will.}
\chunbin{I am confused with the following example.}

\eat{ of selection results $F_i=\{ [B:c] \} \rijoin_{B; R\mapsto r}^{A} \mathcal{I}_{R.B'}$.
 has a series of fragment-based semijoins could achieve the same result. However, the merge intersection operator exploits the ordering of $L_1, \ldots, L_m$.}

\matthias{I think this only works if the values are sorted. Why can we assume that? Chunbin: we do not assume the values are sorted, since they are already sorted.}

\noindent \textbf{Aggregation Operator}
\matthias{Why do we write $\alpha$ twice in the notation? Chunbin: I removed the one in the superscript, correct me if I made mistake here.}
The aggregation operator $\gamma^{1}_{r.D; \alpha(s(A_1, \ldots, A_n))}$ groups its input according to the single group-by attribute $r.D$ and aggregates the results of the scalar function $s(A_1,\ldots,A_n)$ using the associative aggregation function $\alpha$ (e.g., \gt{min}, \gt{max}, \gt{count}, \gt{sum}).
\begin{gotobasic}
Recall, in relationship queries the single group-by attribute $r.D$ is the foreign key of a relationship table or the ID of an entity. In either case, the range of $r.D$ is the same as the range of the underlying entity ID. Consequently, the $\gamma^{1}$ operator's superscript $1$ signifies the assumption that the domain of $G$ is small enough to allow for the allocation of an array, whose size is the domain of $r.D$ and each entry is a number, initialized to zero. Every time a``tuple'' from $r$ is processed, this array is updated at $r.D$ accordingly. In addition, an array of booleans registers which values of $r.D$ were actually found. For example, the aggregation $\gamma^1_{\mathit{Author}; \frac{\mathit{SUM}(\mathit{DT}_2.\mathit{Fre}\times \mathit{DT}_1.\mathit{Fre})}{(2017-\mathit{DY}.\mathit{Year})}}$ of the Query AS, as shown in Figure~\ref{fig:xmpl-physical-plans}, initializes an array and a boolean  register array with sizes of the domain \gt{Author}.
\end{gotobasic}
\begin{gotofull}
Recall, in relationship queries the single group-by attribute $r.D$ is the foreign key of a relationship table or the ID of an entity. In either case, the range of $r.D$ is the same as the range of the underlying entity ID. Consequently, the $\gamma^{1}$ operator's superscript $1$ signifies the assumption that the domain of $G$ is small enough to allow for the allocation of an array, whose size is the domain of $r.D$ and each entry is a number, initialized to zero. Every time a ``tuple'' from $r$ is processed, this array is updated at $r.D$ accordingly. In addition, an array of booleans registers which values of $r.D$ were actually found. As was also the case with the lookup structure of the semijoin, it is preferable to incur the penalty of initializing such arrays, instead of using hash sets or tree-based sets that have more expensive lookup times. For example, the aggregation $\gamma^1_{\mathit{Author}; \frac{\mathit{SUM}(\mathit{DT}_2.\mathit{Fre}\times \mathit{DT}_1.\mathit{Fre})}{(2017-\mathit{DY}.\mathit{Year})}}$ of the Query AS, as shown in Figure~\ref{fig:xmpl-physical-plans}, initializes an array with size of the domain \gt{Author} to store the aggregated score for each author and also a boolean register array to check which authors are actually accessed.
\end{gotofull}

\yannis{Example would be good here: For example, the aggregation $\gamma^{1}_{\gt{da2.Author};\ldots}$ of the Query AS ...Chunbin: added}

\yannis{May have to move to future Work. Chunbin: I moved the following paragraph to full version.}
\matthias{I don't understand the following sentence. What are ``other'' non-relationship queries? Chunbin: the queries beyond relationship queries.}

\begin{gotofull}
In non-relationship queries the group-by list may involve more than one group-by attributes, in which case it uses a hash table~\cite{garcia2008database} instead of an array.  In other non-relationship queries, the aggregation function $f$ is non-associative (e.g., the median) in which case it is not enough to allocate an aggregate values' array with just one number per domain entry.
\end{gotofull}

\begin{figure}[h!]
\centering
\includegraphics[width=1\columnwidth]{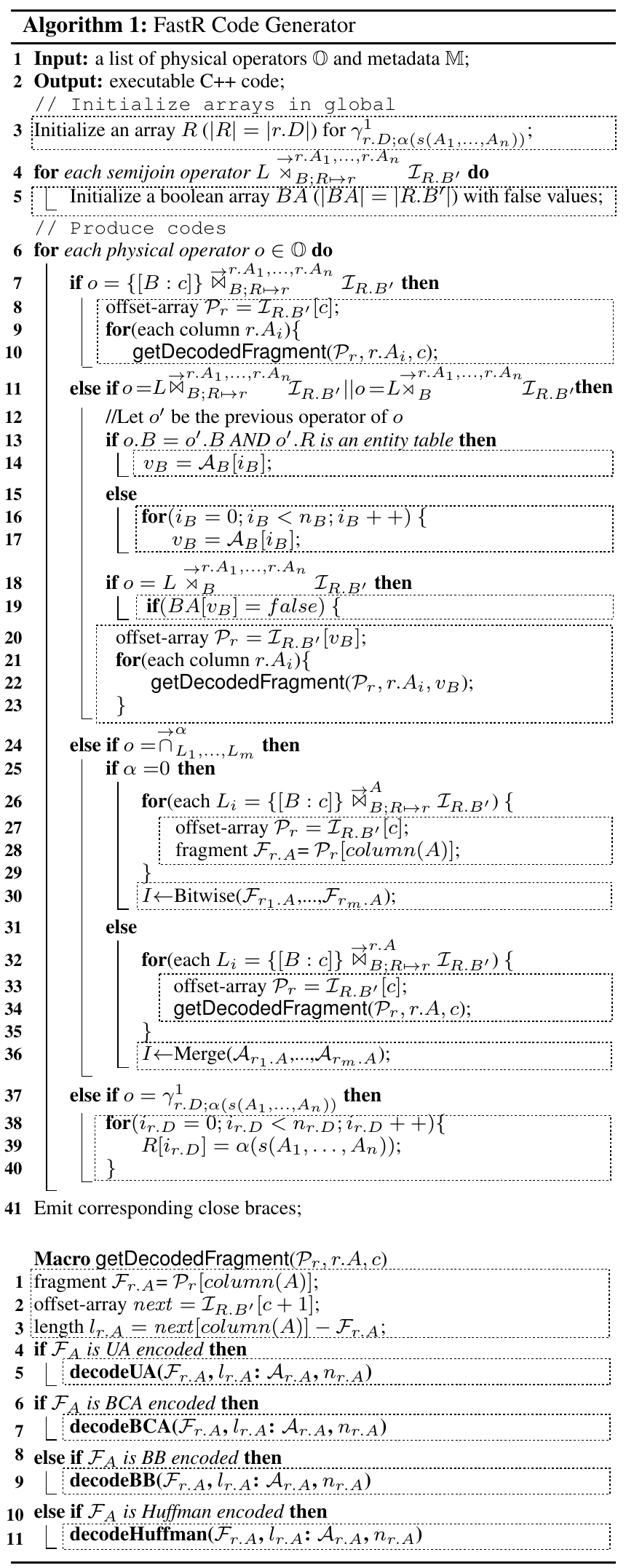}
\end{figure}

\subsection{\fastr Code Generator}
\label{sec:code_generator}
% Recall that the \textit{FastR Code Generator} receives the physical plan of relationship queries and FastR metadata and statistics (see Figure~\ref{fig:architecture}) to generate executable C++ codes like the example algorithm for AS query.

The \fastr code generator has two main components: to-be-emitted codes boxed by dotted lines in the pseudocode; and control commands that determine which code pieces should be emitted. The inputs of the code generator are (1) physical plan and (2) \fastr metadata and  statistics, which specifies the encodings of each fragment. The code generator has two phases: (1) initialize necessary buffers (lines 2-5); and (2) emit code pieces for each physical operator (6-44). More precisely, in the first phase, the \fastr code generator initializes an array $R$ to store final aggregation results (line 3) and several boolean arrays for duplicate-checking in semijoin operations (lines 4-5). In the second phase, it emits selection codes (lines $8-10$) for each  selection operator $\{ [B:c] \} \rijoin_{B; R\mapsto r}^{r.A_1,\ldots,r.A_n} \mathcal{I}_{R.B'}$. The  \textsf{getDecodeFragment} macro emits codes for (1) retrieving a fragment (lines 1-3) and (2) calling corresponding decode macros for the fragments (lines 4-11). The encoding information is obtained from metadata. For each join  $L \rijoin_{B;R\mapsto r}^{r.A_1,\ldots,r.A_n} \mathcal{I}_{R.B'}$  and semijoin $L \risemijoin_{B; R\mapsto r}^{r.A_1,\ldots,r.A_n} \mathcal{I}_{R.B'}$ operators (lines 11$-$26), the generator first checks whether
the previous operator operates on an entity table and the operated columns are the same. If yes, then one for-loop can be avoided (lines 13-17). If it is a semijoin operator, then one more duplicate-checking step should be added (line 19). The remaining steps (lines 20-23) of join/semijoin are the same with the selection operator.
For each intersection operator $\riinter^{\alpha}_{F_1,\ldots,F_m}$, the generator first identifies whether all the fragments are encoded with the same bitmap encodings by checking the metadata. If yes ($\alpha=0$),  the generator emits codes to perform intersection directly on encoded fragments (lines 27-30). Otherwise, it emits code to perform intersection on decoded fragments (lines 33-36). Then \fastr code generator emit aggregation code pieces (lines 38-40) for the aggregation operator $\gamma^{1\alpha}_{r.D; \alpha(s(A_1, \ldots, A_n))}$. Finally, it emits corresponding close braces in order to produce \fastr algorithm with correct syntax  (line 41).

\noindent\textbf{Memory Requirement.} The required memory size for \fastr is $4\cdot|r.D|+\sum_{i=1}^k\cdot |r_i.B'|$ bytes, where $|r.D|$ is the domain size of $r.D$ for the aggregation operator $\gamma^{1}_{r.D; \alpha(s(A_1, \ldots, A_n))}$, $k$ is the number of semijoin operators, and $|r_i.B'|$ is the domain size of $r_i.B'$ for the $i-$th semijoin operator  $L \risemijoin_{B;R\mapsto r}^{r.A_1,\ldots,r.A_n} \mathcal{I}_{R_i.B'}$.

\eat{
\begin{gotobasic}
\noindent\textbf{Parallel Computing.}  \fastr can use multiple cores/threads to perform parallel computation. The fact that in \fastr (1) each fragment is independent of others, so \fastr can assign fragments to different threads; and (2) the query processing is more CPU-bounded than memory-bounded especially when decompressing encoded fragments.
%The full version also discusses the extension to GPU environment.
\end{gotobasic}
}

\begin{gotofull}

\noindent\textbf{Parallel Computing.}  \fastr can use multiple cores/threads to perform parallel computation. The fact that in \fastr (1) each fragment is independent of others, so \fastr can assign fragments to different threads; and (2) the query processing is more CPU-bounded than memory-bounded especially when decompressing encoded fragments. We would like to mention one important technical detail of how to handle two kinds of global arrays, i.e., boolean arrays for each semijoin operation, and a numerical array for aggregation operator. To guarantee the correctness of \fastr, it applies spinlock~\cite{catozzi2001operating} in each array slot, which is experimentally verified to be more efficient than just using one spinlock on the entire boolean array. For the numerical array, \fastr utilizes the same strategy, which is experimentally verified to be generally faster than maintaining one independent array in each thread than aggregate them together in the main thread (see the experiments in Appendix).

\noindent\textbf{Discussion.} Relationship queries are mostly CPU-bounded. The bottleneck to answering queries is CPU computation. Some research work~\cite{esmaeilzadeh2011dark,chung2010single} indicates that CPU performance is not likely to improve significantly in the future. Thus heterogeneous approaches are introduced to cope with the scalability issue of CPU by using additional computation units.  GPU (graphics processor units) is the most representative one due to its commercial availability, full blown programmability and better backward compatibility~\cite{yuan2013yin,he2011high}. We claim that \fastr can be easily applied to GPU environment to further speed up the performance. In particular,  each GPU core maintains a complete index. Then elements can be evenly assigned to each core to perform their own computations.
\end{gotofull}

\begin{gotobasic}
\section{Main-memory column database}
\label{sec:main-memory}
%\yannis{OUT OF ORDER SECTION}
%\yannis{EXPLAIN THE GOAL OF CREATING THE OMC. EXPLAIN THAT THE OMC KEY ASPECT IS THE TWO COPIES.}

\yannis{This setion needs serious editing. Chunbin: I rewrote it.}
\chunbin{I moved the details of PMC and OMC to appendix in the full version. For the basic version, I only keep very short description.}
\chunbin{To Yannis: which one is better, dissect or isolate?}

To dissect the benefit from compiled codes of \fastr and provide a fair comparison, we implement two main-memory column databases served as main-memory baselines.
One is a {\em plain main-memory column database (PMC)} without optimizations. The other one is {\em fully optimized main-memory column database (OMC)}. Note that, they are both generated by code generators and compiled into executable C++ plans. In addition, the logical query plans in PMC and OMC are the same with that in \fastr, i.e., the same RQNA expressions. Both PMC and OMC use operator-at-a-time execution model as  MonetDB~\cite{stonebraker2005c,abadi2008column}, which means both PMC and OMC materialize complete intermediate results produced by each operator.

More precisely, PMC maintains one copy of each unsorted table, and use whole column scan when operating each operator.  OMC maintains two copies of each table, such that each copy is sorted based on one foreign key column. OMC applies all the optimizations that can improve the performance of processing  relationship queries. The optimizations are  (1) applying run-length encoding for sorted column in order to both improve the lookup performance and reduce the space cost; (2) utilizing binary search for sorted columns instead of whole column scan. The detailed codes of PMC and OMC can be found in our full version \cite{lin2016fastr}.

\chunbin{I moved PMC and OMC to appendix section and they will be in the full version only.}
% The results are stored in maps.
% The generated codes for the operators $\sigma$, $\Join$, $\rtimes$, $\cap$ and $\gamma$ in PMC and OMC are shown in Algorithm 1 - Algorithm 4 respectively.

\end{gotobasic}

\section{Experiments}\label{sec:experiments}
We evaluate \fastrs novelties by running relationship queries on three real-life datasets.
%The reported execution times are the averages over five out of seven runs by excluding the minimal and maximal run time.
In all experiments, the full data are located in RAM.

\begin{gotofull}
\subsection{Experimental Setting}
\label{exp:setting}
\end{gotofull}
\yannis{to Chunbin: Fix machine. It is still the old 4-core one. Chunbin: done}The experiments were conducted on a computer with a \textit{4th} generation Intel \textit{i7-4770} processor ($8M$ Cache, $8$ cores, $3.6$ GHz, single socket) running Ubuntu $14.04.1$ with $16GB$ RAM. The size of $L_1$, $L_2$ and $L_3$ caches are $256KB$, $1MB$ and $8MB$ respectively. All
the algorithms are coded in C++. The C++ plans were compiled with g++ 4.8.4, using the \textit{-O3} optimization option.  \yannis{to self: the spinlock should have been mentioned in the query processing, not here. Chunbin: I moved it back}

\begin{table}[t]
\small
\renewcommand{\tabcolsep}{0.8mm}
\begin{tabular}{lcr}
    \begin{minipage}{.55\linewidth}
        \begin{tabular}{l|r}\hline\hline
        Table name & 	\#  rows	\\\hline
        \gt{DT(Doc,Term,Fre)} 	&207,092,075 \\\hline
        \rowcolor{Gray}
        \gt{DT(Doc,Term,Fre)}	&901,388,401 \\\hline
        \gt{DA(Doc,Author)}	&61,329,130 \\\hline
        \gt{Document(ID,Year)}	&23,176,635 \\\hline \hline%14,293,636
        \end{tabular}
    \end{minipage} &
    \ &
    \begin{minipage}{.3\linewidth}
        \begin{tabular}{l|r}\hline\hline
        Entity ID & 	Domain Size	\\\hline
        \gt{Doc}(ument) &23,326,299 \\\hline
        \gt{Term}	&27,883 \\\hline
        \rowcolor{Gray}
        \gt{Term}	&259,728 \\\hline
        \gt{Author}	& 6,301,521 \\\hline \hline
        \end{tabular}
    \end{minipage}
\end{tabular}

\smallskip

\begin{tabular}{l|l|r|r|r} \hline\hline
    Table & Fragment & Average size & Maximal size & Standard deviation\\ \hline
    \multirow{2}{*}{\gt{DT}} & \gt{Doc} & 7427.18 & 8192342 & 197.56\\\cline{2-5}
                        & \gt{Term} & 14.48 & 667 & 17.52\\ \hline

    \multirow{2}{*}{ \cellcolor{gray!25}\gt{DT}} &  \cellcolor{gray!25}\gt{Doc} &  \cellcolor{gray!25}3470.50 &  \cellcolor{gray!25}8192342 &  \cellcolor{gray!25}318.72\\\cline{2-5}
    &  \cellcolor{gray!25}\gt{Term} &  \cellcolor{gray!25}63.06 &  \cellcolor{gray!25}753 &  \cellcolor{gray!25}39.06\\ \hline
    \multirow{2}{*}{\gt{DA}} & \gt{Doc} & 5.99 & 5712 & 21.93\\\cline{2-5}
                        & \gt{Author} & 4.35 & 3163 & 8.04\\ \hline
                     \gt{Document} & \gt{Year} & 1.00 & 1 & 0.00\\ \hline\hline
\end{tabular}
\vspace{-3mm}
 \caption{\textbf{Data characteristics of PubMed-M and PubMed-MS}. PubMed-MS has larger size of terms than PubMed-M, which results in larger size of \gt{DT} table in PubMed-MS. The gray cells indicates the difference between PubMed-M and PubMed-MS.}
 \label{table:pubmed_dataset}
\end{table}

\begin{table}
%[htbp]
\small
\renewcommand{\tabcolsep}{3.8mm}
%\begin{tabular}{c|c|c|c} \hline\hline
%    Table name & 	\# of rows  & & 	Domain Size \\\hline
%    \gt{Concept\_Semtype (CS)}	&1550482  &\gt{Concept} &1339227\\\hline
%        \gt{Predication\_Argument (PA)}	&37508726 & \gt{Sentence}	&146055876 \\\hline
%        \gt{Sentence\_Predication (SP)}	&81929321 &\gt{Predication}	&17359895 \\\hline \hline
% \hline\hline %\pbox{20cm}{sentence fragment\\ for predication}
%\end{tabular}
\begin{tabular}{c|c|c|c} \hline\hline
    Table & 	\#  rows  &  Table & 	\#  rows \\\hline
    \gt{CS}	&1550482  &\gt{Concept} &1339227\\\hline
        \gt{PA}	&37508726 & \gt{Sentence}	&146055876 \\\hline
        \gt{SP}	&81929321 &\gt{Predication}	&17359895 \\\hline
        %\pbox{20cm}{sentence fragment\\ for predication}
\end{tabular}
\smallskip
\renewcommand{\tabcolsep}{0.3mm}
\begin{tabular}{c|c|c|c|c} \hline
    Table & Fragment & Ave size & Max size & Standard deviation\\ \hline
    \multirow{2}{*}{\gt{CS}} & \gt{concept\_semtype\_id} & 1.16& 5.00 &0.39\\ \cline{2-5}
                        & \gt{concept\_id} & 1.00& 1.00& 0.00\\ \hline
    \multirow{2}{*}{\gt{PA}} & \gt{predication\_id} &122.00 &109532& 845.15\\\cline{2-5}
                        & \gt{concept\_semtype\_id} & 2.15& 38& 0.53\\ \hline
    \multirow{2}{*}{\gt{SP}} & \gt{sentence\_id} & 4.65& 125367 &112.36\\ \cline{2-5}
                        & \gt{predication\_id}  & 1.61& 140 &1.07\\ \hline\hline %\pbox{20cm}{sentence fragment\\ for predication}
\end{tabular}
\vspace{-3mm}
 \caption{\textbf{Data characteristics of SemmedDB dataset}. SemmedDB has slower fan-out than PubMed.}
 \label{table:semmeddb_dataset}
\end{table}

\begin{gotobasic}
\noindent \textbf{Dataset.} We evaluated all the systems and design choices in three datasets: PubMed-M, PubMed-MS and SemmedDB. Table~\ref{table:pubmed_dataset} and Table~\ref{table:semmeddb_dataset}  summarize their data characteristics respectively. The common schema of PubMed-M and PubMed-MS was introduced in Section~\ref{sec: introduction} and the schema of SemmeDB in Section~\ref{sec:queries}. PubMed-M contains $27,883$ terms, which correspond to PubMed's MeSH terms. In contrast, PubMed-MS has $259.728$ terms because it includes $~230,000$ PubMed Supplemental terms, in addition to the MeSH terms, As a result PubMed-MS has a larger \gt{DT} table. However, PubMed-MS has fewer documents per term ratio (also called, ``lower fanout of \gt{Term} in \gt{DT}) compared with PubMed-M, since Mesh terms are (on average) associated with many documents while supplemental terms are (on average) associated with few documents. The \textit{fanout} of a column $R.F$ is defined by $\frac{|R|}{|E|}$, where $F$ is a foreign of table $R$ pointing to the ID of an entity table $E$. The fanout is the expected number of tuples of $R$ that have a given value of $E.\textrm{ID}$. It affects performance significantly as revealed by comparing the results of the very high fanout PubMed-M, with the medium fanout PubMed-MS and the low fanout SemmedDB.
\end{gotobasic}

\begin{gotofull}
\noindent \textbf{PubMed Dataset} We use the subset of PubMed publications from 1990 to 2015, which has about $23$ million citations for biomedical literature from the National Library of Medicine bibliographic database, life science journals, and online books.
The citations in PubMed are labelled by descriptors from MeSH (National Library of Medicine's controlled vocabulary thesaurus). MeSH contains about $220,000$ descriptors (terms), organized in an hierarchy. In addition, PubMed also provides additional descriptors (called \textit{Supplemental}) that also label citations. The number of supplemental terms is around $380,000$. Mesh terms are (on average) associated with large number of citations while supplemental terms are (on average) associated with few citations. That is, Mesh term and Supplemental term have different fanout  (see formal definition below). Since both space cost and running time are fanout-sensitive, we use two versions of PubMed in our experiments. In one series of experiments, we use PubMed with only Mesh terms, called \emph{PubMed-M}. The second series of experiments, we use PubMed with both Mesh terms and supplemental terms, called \emph{PubMed-MS}.

\yannis{there should be a ``recall from Section X" that \gt{anonymous1} only loads...}

\yannis{do we comment later on fanout and fragment sizes. Chunbin:yes}

\begin{definition}
\textbf{Fanout}. Given a relationship table $R$ and a foreign key $D$ of it, pointing to the ID of an entity table $E$, we define as \emph{fanout of $D$ in $R$}, also called \emph{fanout of $E.\mathrm{ID}$ in $R$} the ratio of the number of tuples of $R$, divided by the number of tuples in $E$ (which is also the number of unique values of the ID of $E$). The fanout is the expected number of tuples of $R$ that have a given value of $E.\textrm{ID}$.
\end{definition}

Table \ref{table:pubmed_dataset} presents the schema and statistics of Pubmed-M and Pubmed-MS: two relationship tables (\gt{DT(Doc,Term,Fre)} and \gt{DA(Doc, Author)}) and one entity table \gt{Document(Doc, Year)}.
%\footnote{Note, only query-relevant tables are loaded into RAM in our experiments. Other tables/attributes, such as \gt{Abstract} attribute in \gt{Document} table, are left in the conventional database.}
%\footnote{Note, the \gt{anonymous1} dashboard actually loads on FastR the particular tables only. Other document attributes, such as title and abstract, are left in the conventional database.}
The tables \gt{DA} and \gt{Document} are identical in Pubmed-M and Pubmed-MS. The statistics of \gt{DT} differ, depending on whether it is PubMed-M (white lines) or Pubmed-MS (gray lines). Note that, PubMed-MS has lower \gt{Term} fanout than PubMed-M, while PubMed-MS has larger table sizes than PubMed-M.

\noindent \textbf{SemMedDB Dataset.} The Semantic MEDLINE Database (SemMedDB) is a repository of semantic predications (subject-predicate-object triples) extracted by SemRep \cite{rindflesch2003interaction}, which is a semantic interpreter of biomedical text. SemMedDB currently contains information about approximately $82.2$ million predications from all of PubMed citations (around $25$ million citations from 1809 to 2015 June 30th).
% In our experiments, the following projection tables are utilized: \textit{Concept\_Semtype (concept\_id, concept\_semtype\_id)},
%\textit{Predication\_Argument (predication\_id, concept\_semtype\_id)} and
%\textit{Sentence\_Predication (sentence\_id, predication\_id)}.
Table \ref{table:semmeddb_dataset} summarizes the data characteristics of SemmedDB dataset.
%: three relationship tables (\gt{CS(CID,CSID)}, \gt{PA(PID,CSID)} and \gt{SP(SID,PID)})

In the interest of reducing the main memory requirements, the FastR databases for PubMed and SemMedDB do not include strings such as document titles, author names and term names.  Rather, they only load data that capture the graph structure of the database and certain measures. (The Postgres and MonetDB databases, which we measure below, also do not store such strings.) In practice, the applications issue SQL queries on FastR to obtain entity IDs and associated measures(see Section~\ref{sec:queries}). Then a conventional database system is used to translate the IDs to printable names. In our experiments, we only consider the FastR part of a query.

\noindent \textbf{Dataset Summary.} Comparing the fanout and table sizes among these three datasets, we notice that
PubMed-M has the highest fanout, while SemmedDB has the lowest fanout. PubMed-MS is the biggest dataset, while SemmedDB is the smallest one.
\end{gotofull}

\smallskip
\noindent \textbf{Compared Systems.}
We compared \fastr with the graph database Neo4j 2.3.2 (enterprise edition),
the row-oriented database PostgreSQL $9.4.0$% \smallfootnote{http://www.postgresql.org/}
and the column database MonetDB%\smallfootnote{https://www.monetdb.org/Home}. The reported deployment of \fastr minimizes space by choosing the most compressed encoding method for each attribute. For example, we use Huffman encoding for the \gt{DT.Fre} column.
%In particular, it uses the byte-aligned compressed bitmap encoding, since it minimizes space.
\yannis{to Chunbin and Ben: Is it bitmap for everything? I thought we use Huffman on the frequencies. Chunbin: No bitmap is not for everything, we use Huffman on the frequencies.}

%\vspace{-3mm}
%\begin{compact_item}
%\item
\begin{gotobasic}
%Neo4j is an open-source NoSQL graph database, where vertices and edges are modeled natively and they are allowed to have attributes.
Since Neo4j does not support the SQL syntax, we translated the experiments' queries into Cypher\smallfootnote{http://neo4j.com/docs/stable/cypher-query-lang.html}, Neo4j's query language. See the Cypher statements in~\cite{lin2016fastr}.
\end{gotobasic}

\begin{gotofull}
Neo4j is an open-source NoSQL graph database implemented in Java and Scala. Each entity in entity tables like \gt{Term} is modeled as a vertex, and the attributes of the entity become the attributes of the vertex. Each tuple $(d_1,d_2,m_1,\cdots,m_n)$ in the relationship table, e.g., \gt{DT}, is modeled as an edge between $d_1$ and $d_2$, where $d_1$ and $d_2$ are in foreign key columns and $m_1,\cdots,m_n$ are in measure attributes. $m_1,\cdots,m_n$ become the attributes of this edge. Neo4j does not support SQL queries. Section~\ref{sec:appendix_graph} shows the translation of the experiments' queries into Cypher\smallfootnote{http://neo4j.com/docs/stable/cypher-query-lang.html}, Neo4j's query language.
\end{gotofull}

In the Postgres and MonetDB experiments, entity tables are sorted and indexed on their primary keys. We boost performance by having relationship tables be stored sorted on their first attribute and have indices on all foreign key attributes. We measure the \textit{warm} running time for queries, i.e., each query is run twice and we only report the second measurement. The first one is used just to bring all the necessary data (for the evaluation of the query) into the RAM buffers. Furthermore, we run and average five warm runs.

\yannis{to Chunbin: Is it a fair simplification of what happened with MonetDB indices? Chunbin: I think so.}
\eat{In the MonetDB experiments each table is also sorted based on its first column. We also built indices. Note that MonetDB  often makes its own decision to create and maintain indices whose size cannot be measured with MonetDB commands. Therefore, when reporting the space cost in MonetDB, we monitor and report the actual memory usage via the OS, rather than MonetDB commands. As was the case for Postgres also, we report only the \textit{warm} running times.}

\yannis{to self: make clear that it is understandable/redundant in light of PMC and OMC sections. Chunbin:sure}

In addition, PMC and OMC are employed as baselines that isolate the effect of compiled code from the other contributions of \fastr.
\eat{In a variant of OMC, experiments also isolate the effects of small domain adjustments in \fastr.}
\begin{gotofull}
% PMC does not use any optimizations.
% Since we do not have an end-to-end functioning PMC or OMC, we handcrafted the C++ code for the PMC and OMC plans.
The PMC maintains one copy of each table, while the OMC maintains two copies of each relationship table sorted by $F_1$ and $F_2$ respectively. We manually rewrite the experiments' queries to make optimal use of the sorting.
% For example, OMC loads two copies of the \gt{DT} table. The first one is sorted based on \gt{Doc} and the second is sorted on \gt{Term}. It encodes the sorted column by run-length-encoding and lookups the encoded column by using binary search.
Each column is maintained in an individual array in PMC and OMC. The logical query plans of PMC/OMC are the same with that of \fastr, i.e., the same RQNA expression. In addition, both PMC and OMC use code generators to produce executable C++ codes.
\end{gotofull}

\begin{table*}[htbp]
\small
\centering
\renewcommand{\tabcolsep}{1.5mm}
\begin{tabular}{l|r|r|r|rrrrrrrr} \hline\hline
                              & & &      &(a)    &(b)      & (c)     & (d)     & (e)      & (f)                 & \multirow{2}{*}{$\frac{\mathrm{MonetDB}}{\mathrm{\fastr}}$} & \multirow{2}{*}{$\frac{\mathrm{OMC}}{\mathrm{\fastr}}$}   \\
                            & Query & Join \# tuples &Result \# tuples &  Neo4j  &Postgres   & MonetDB   & PMC     & OMC      & \fastr & &\\\hline
\multirow{5}{*}{Pubmed-M}	&	SD	& 22,401,361 & 6,409,707&	14.7	&	211.2	&	10.8	&	23.36	&	2.47	&	\textbf{	0.230	}	&	47.0	&	10.7	\\\cline{2-12}	
                        	&	FSD	& 22,401,361 & 6,409,707&	86.6	&	567.5	&	23.9	&	33.98	&	5.93	&	\textbf{	0.821	}	&	29.1	&	7.2	\\\cline{2-12}	
                        	&	AD	& 99,734 & 57,584&	7.4	    &	158.1	&	3.2	&	4.82	&	0.73	&	\textbf{	0.037	}	&	86.5	&	19.7	\\\cline{2-12}	
                        	&	FAD	& 717,487 & 5,643&	18.2	&	198.6	&	4.5	&	7.16	&	0.88	&	\textbf{	0.064	}	&	70.3	&	13.8	\\\cline{2-12}	
                        	&	AS	& 147,273,421 & 6,393,107 &	5546.5	&	29520.5	&	4474.8	&	5832.30	&	194.77	&	\textbf{	5.662	}	&	790.3	&	34.4	\\\hline	
 \multirow{5}{*}{Pubmed-MS}	&	SD	& 136,151,592 & 12,466,510&	61.6	&	741.2	&	49.3	&	400.24	&	10.25	&	\textbf{	1.068	}	&	46.2	&	9.6	\\\cline{2-12}	
                        	&	FSD	& 136,151,592 & 12,466,510&	146.8	&	2148.7	&	112.8	&	1892.30	&	32.80	&	\textbf{	4.376	}	&	25.8	&	7.5	\\\cline{2-12}	
                        	&	AD	& 85,982 & 64,765&	6.9   	&	112.9	&	2.7	&	19.67	&	0.67	&	\textbf{	0.035	}	&	77.1	&	19.1	\\\cline{2-12}	
                        	&	FAD	& 1,503,368 & 9,556&	11.1	&	119.6	&	3.5	&	24.99	&	0.71	&	\textbf{	0.062	}	&	56.5	&	11.5	\\\cline{2-12}	
                        	&	AS	& 1,391,434,113 & 9,803,226&	9604.8	&	180164.1&	28918.8	&	33321.74	&	3083.30	&	\textbf{	54.720	}	&	528.5	&	56.3	\\\hline	
                 SemmedDB	&	CS	& 207,191 & 5,057&	21.0	&	53.1	&	4.7	&	23.58	&	2.12	&	\textbf{	0.031	}	&	151.6	&	68.4	\\\hline\hline	
\end{tabular}
\vspace{-3mm}
 \caption{\textbf{End-to-end runtime performance tests (in seconds)}. Numbers in bold are the fastest ones.}
 \label{table:performance}
\end{table*}
% the fastr-UA is 4.510 for AS
\begin{table*}[htbp]
\small
\centering
\renewcommand{\tabcolsep}{4.7mm}
\begin{tabular}{l|rrrrrrrr} \hline\hline
             & (a) & (b)      & (c)     & (d)   & (e)   & (f)            & \multirow{2}{*}{$\frac{\mathrm{MonetDB}}{\mathrm{\fastr}}$} & \multirow{2}{*}{$\frac{\mathrm{OMC}}{\mathrm{\fastr}}$}\\
             & Neo4j & Postgres & MonetDB & PMC   & OMC   & \fastr                                                               \\ \hline
   PubMed-M  & 34.36 & 20.92    & 3.69    & 3.09  & 3.49  & \textbf{1.47}  & 2.51   & 2.37  \\ \hline
   PubMed-MS & 112.15 & 78.90    & 13.27   & 11.42 & 11.82 & \textbf{3.51}  & 3.78    & 3.37 \\ \hline
   SemmedDB  & 10.39 & 6.84     & 1.23    & \textbf{0.97}  & 2.05  & 1.36  & 0.90 & 1.51 \\ \hline\hline
\end{tabular}
\vspace{-3mm}
 \caption{\textbf{Space cost for each system (in GB)}. Numbers in bold are the smallest ones.}
 \label{table:space_cost}
\end{table*}

\noindent \textbf{Experimental Results}
%\label{exp:results}
We ran the queries SD, FSD, AD, FAD and AS on PubMed (PubMed-M and PubMed-MS) and the query CS on SemmedDB (see queries in Section~\ref{sec:queries}).
\eat{Each reported running time is the average time over $5$ runs, excluding the first run on Neo4j, Postgres and MonetDB, which is used just to warm the memory.} \begin{gotobasic}The extended version provides additional information of how the selection constants in the queries are chosen. It also provides additional commentary on the results.\end{gotobasic}
\yannis{to Chunbin: What is the process? Random choice? Is it the same constant for all 5 runs? If not, the warming is not really happening. If yes, what do we mean below when we say ``ON AVERAGE 6"? If describing the constants' selection can be done in a few lines, we may leave it in basic. Here is a piece about selection from an old version. Is it still applicable? The AS query, which is a six-way join query, ends with a database crash on Postgres and MonetDB if the given author has many publications.
Therefore, to ensure that all the compared systems get results for the AS query, we chose the given author ID to have 6 (on average) publications. Note that the join size and the result size are still high, since the fanout of all the tables is high in the PubMed dataset. Chunbin: We use the same constant for all 5 runs. I did not say "on average 6". "we chose the given author ID to have 6 (on average) publications" means we choose author ids that has around 6 publications, not run this query 6 times.}
\yannis{to Chunbin and Ben: Add two columns between Query and Neo4j: (1) Final Join \#tuples and (2) Result \#tuples. No comments about it in the basic. Do not change the measured code for this. Rather make a separate version that has counters. That is, counting should not change the performance. Chunbin: Yes, we will add the two columns}

\begin{gotobasic}
We first measure the overall running time performance and the overall space cost for each algorithm/database. The results show that \fastr outperforms MonetDB and OMC by 10-$10^3$ and 7-70 times respectively, and uses generally less space cost. The benefits of \fastr are gained from the combination of the following effects: (1) the use of compiled code, (ii) bottom-up pipelining execution strategy, (iii) data structure optimized by tuning to the dense IDs, and (iv) applying aggressive data compression schemes. The gap between the speedups of \fastr and OMC over MonetDB indicate the power of compiled code. In order to isolate the effect of each of the other three optimizations in \fastr, we implemented the following variants of \fastr and OMC.

\vspace{-1mm}
\begin{compact_item}
\item \textbf{\fastrUA}: \fastr with uncompressed array encodings.
\item \textbf{\fastrUAB}: It is like \fastrUA but instead of locating by array lookup $a[v]$ the position offset for a given search value $v$, \fastrUAB uses binary search to find the position offset. Therefore, \fastrUAB does not use the dense ID assumption.
\item \textbf{\fastrUAM}: It is like \fastrUA but instead of using an array to store the final aggregation results, it uses a hashmap. It uses a hashmap to replace the boolean array used by semi-joins.
\item \textbf{OMC-denseID}: It is like OMC but uses arrays instead of hashmaps in both lookup and aggregation. The lookup data structure in OMC-denseID is the same with that in \fastr.
\end{compact_item}

Then we conducted further experiments in order to isolate the effect of each optimization.

\begin{compact_item}
\item Measured the effect of \fastr using the dense IDs assumption against \fastr without using it by comparing (i) \fastrUA vs.  \fastrUAB, and (ii) \fastrUA vs.  \fastrUAM .
\item Measured  the effect of bottom-up pipelining against materializing intermediate results by comparing \fastrUA vs. OMC-denseID.
\item Analyzed the performance of different compressions and evaluated the effect of different number of threads on \fastr performance.
\end{compact_item}
\end{gotobasic}

\begin{gotofull}
Section~\ref{exp:performance} measures the overall running time performance of each system. Section~\ref{exp:space_cost} measures the overall space cost for each algorithm/database. The results show that \fastr outperforms MonetDB and OMC by 10-$10^3$ and 7-70 times respectively, and uses generally less space cost. The benefits of \fastr are gained from the combination of the following effects: (1) the use of compiled code, (ii) bottom-up pipelining execution strategy, (iii) data structure optimized by tuning to the dense IDs, and (iv) applying aggressive data compression schemes. The gap between the speedups of \fastr from MonetDB and that from OMC indicates the power of the use of  compiled code. In order to isolate the  effect of the other three optimizations in \fastr, we implemented the following variants of \fastr and OMC.

\begin{compact_item}
\item \textbf{\fastrUA}. It is \fastr with uncompressed array encodings.
\item \textbf{\fastrUAB}. It is \fastrUA but instead of locating directly to corresponding positions for search values in the lookup structure, \fastrUAB uses binary search to find the positions.
\item \textbf{\fastrUAM}. It is \fastrUA but instead of using an array to store the final aggregation results, it uses a hashmap. And it uses a hashmap to replace the boolean array for semi-joins.
\item \textbf{OMC-denseID} . It is OMC but using arrays instead of hashmaps in both lookup and aggregation. The lookup data structure in OMC-denseID is the same with that in \fastr.
\end{compact_item}

Then we conducted the further experiments in order to isolate the  effect of the other three optimizations.

\begin{compact_item}
\item Measured the effect of \fastr using the dense IDs assumption against \fastr without using it by comparing (i) \fastrUA vs.  \fastrUAB in Section~\ref{exp:binary}, and (ii) \fastrUA vs.  \fastrUAM in Section~\ref{exp:hash}.
\item Measured  the effect of bottom-up pipelining against materializing intermediate results by comparing \fastrUA vs. OMC-denseID in Section~\ref{exp:intermediate_results}.
\item Analyzed the performance of different compressions in Section~\ref{exp:different_encoding} and evaluated the effect of different number of threads on \fastr performance in Section~\ref{exp:multithreads}.
\end{compact_item}
\end{gotofull}

\begin{gotofull}

\vspace{-3mm}
\subsection{Overall runtime performance}
\label{exp:performance}
\end{gotofull}
\begin{gotobasic}
\noindent\textbf{Overall runtime performance.}
\end{gotobasic}
Table~\ref{table:performance} reports the average running time of each query for each system, using 8 threads\smallfootnote{We applied spinlock~\cite{catozzi2001operating} to protect the correctness of sharing the arrays for of the semijoin and aggregate operators.} \fastr outperforms the others for all the queries. We further observed that:
\yannis{to Chunbin: A bit annoying that the first ratios are MonetDB/FastR but the next ones are FastR/MonetDB. Any reason? Chunbin: To be honest, no, we can the second one to MonetDB/FastR. Yannis: Then change the space ratios to also become MonetDB/FastR and OMC/FastR. Chunbin: I did}

\vspace{-3mm}
\begin{compact_item}
\item \fastr outperforms MonetDB and OMC by about $170$ times and $20$ times on the average (see ratio columns). If \fastr only apply UA compression, it will achieve better performance, e.g., the running time of  AS query on PubMed-M is 4.45s.
%For example, FastR is $790.3$ times faster than MonetDB for AS query in PubMed-M dataset. The high performance of FastR verifies the advantages of its new data organization and the efficiency of query-aware compiled pipelined codes.

\begin{gotofull}
\item MonetDB outperforms Postgres, which is explicable by the analytic nature of relationship queries. MonetDB also  outperforms PMC, which means that the performance boost by MonetDB's indexing is more important than the performance boost by PMC's compiled code.

\item Neo4j beats Postgres in all the queries, which is explicable given the many graph path navigations in relationship queries. Interestingly, MonetDB outperforms Neo4j in almost all cases, which indicates that for such OLAP queries, column-databases are probably favorable to graph databases. The exception is the AS on PubMed-MS, where long relationship paths and relatively low fanout give the edge to Neo4j.
    \yannis{to Chunbin: could it be that Neo4j queries lose because they go from node to node, while some queries on SQL do not use the entity tables and instead jump from relationship tuple to relationship tuple. Answer here. we do not necessarily have to write it in the paper. Chunbin: Yes, I think so.}
\end{gotofull}

\item \begin{gotobasic}MonetDB outperforms PMC: the indexed plans advantage of MonetDB beats the compiled code advantage of PMC.\end{gotobasic} OMC outperforms MonetDB, since (a) OMC uses code generation and (b) has two copies of each relationship table. For example, OMC uses two copies of the \gt{DT} table in the SD query. Therefore each OMC lookup is a binary search on the sorted column (hence essentially tieing the index-based lookups of MonetDB) and the lookups' results are run-length encoded on the sorted column, hence reducing the size of intermediate results.

\item High fanout is favorable to GQ-Fast: The improvement over the competing systems is usually higher in the queries SD, FSD, FAD and AS,
    \yannis{to Chunbin: complete which are the queries that use DT. Chunbin: done} when they use the \gt{DT} of Pubmed-M, than it is in the queries that use the \gt{DT} of Pubmed-MS. The difference is the higher fanout of \gt{Term} in the \gt{DT} of PubMed-M.
% For example, FastR is $19.7$ times faster than OMC for AD query in PubMed-M, which is higher than $19.1$ (the improvement over OMC for AS in PubMed-MS).
We conjecture that high fanouts ammortize over larger fragments the fixed costs of the decompression routines, therefore extending \fastrs advantages.
\end{compact_item}

%\subsection{The effect of avoiding row ids}
%\label{exp:row_ids}
\yannis{I ate the effect or row id-s. The teaching of this section is not clear. One cannot tell whether the effect is due to having chunks of related values together (as both OMC and FastR do) Vs being due on some difference between row-id's and offsets. If I had just a little more time, I would have eliminate this section in its current form. Chunbin:got it.}

\eat{When executing the AS query, there are around one million requests for obtaining fragments from \gt{DA.Author}. The way to obtain a fragment in FastR is different from PMC: FastR uses offsets in the $\mathcal{I}_\gt{DA.Author}$ index, which means FastR avoids row ids. Here we measure the effect of avoiding row ids by comparing the running time of obtaining a variable number of fragments for both FastR and PMC. In Figure \ref{exp:fig_avoid_row_id}(a), we change the number of requests on the \gt{DA.Doc} attribute from $100$ to $10^7$. Figure \ref{exp:fig_avoid_row_id}(a) shows that FastR outperforms PMC by ten times. In Figure \ref{exp:fig_avoid_row_id}(b), we fix the number of requests to be 300, which is the number of requests issued by the second join in the AS query. Then we change the fragment size from $1000$ to $10^7$. As shown, FastR scales well with different sizes of fragments, while the running time increases linearly with the enlarge of fragment sizes. Note that, the running time of FastR with different fragment sizes are the same, since FastR maintains a fragment structure to directly obtain fragments without materializing row ids. The running time of obtaining fragments for FastR is independent from the fragment size. In particular, FastR outperforms PMC by four order of magnitudes on the average.
}

\begin{gotofull}
\subsection{Overall space cost}
\label{exp:space_cost}
\end{gotofull}
\begin{gotobasic}
\noindent\textbf{Overall space cost.}
\end{gotobasic}
Table~\ref{table:space_cost} presents the overall space costs.
\fastr has the lowest space cost in PubMed-M and PubMed-MS. Interestingly, \fastr also uses much less space than PMC even though PMC stores only one copy of each table while \fastr stores two ``copies" (i.e., two indices); this indicates the importance of the dense compressions. In SemmedDB, \fastr still uses less space than OMC, but more space than PMC. The reason is the fanout of the SemmedDB (averaging at $1.16$), which dilutes the effect of fragment compression since fragments are very small and space is spent on padding them to full bytes. Though PMC uses marginally smaller space than \fastr in SemmedDB, \fastr is still the best overall choice as it is $760$ (i.e., 23.58/0.031) times faster (Table~\ref{table:performance}).
\begin{gotobasic}
\cite{lin2016fastr} provides additional commentary.
\end{gotobasic}

\begin{gotofull}
Both MonetDB and OMC use more space than PMC, since PMC only maintains one copy of data and no indices, while OMC stores two copies of data and MonetDB builds indices. One would expect the space cost of OMC to always be less than twice that of PMC, since OMC utilizes run-length encoding to compress the sorted column of each table. This is the case in PubMed-M and PubMed-MS; indeed OMC has only marginally higher space cost than PMC in the PubMed datasets. However, many tables in SemmedDB have very small fanout. For example, in \gt{CS}, each concept has only $1.16$ concept\_semtypes associated with it on the average (see Table \ref{table:semmeddb_dataset}). Therefore, in the case of SemmedDB, the run length encoding, which involves one additional counter/integer, ends up wasting space.

Both Neo4j and Postgres have large space cost, since they build large indices and also spend extra space on maintaining particular data format.
For example, Postgres stores additional ``header" data for each row. For example, the pure data size for SemmedDB is $0.97$ GB, while Postgres spends $4.22 $GB to store it. Furthermore, the Postgres index size is large. For example, the index size is $9.57$ GB in the PubMed-M dataset, while the data is $11.35$ GB. Neo4j stores
data in nodes and relationships of a multi-graph with pairs of key-value  properties.
\end{gotofull}

\yannis{to self, where does this go?  For example, the space costs of byte-aligned compressed bitmap encoding, bit-aligned dictionary encoding, and uncompressed array encoding of \gt{DT.Term} are $1.43GB$, $2.09GB$ and $3.66GB$ respectively.  Therefore, choosing the right encoding method is important (recall also the formal analysis of Section \ref{sec:analysis}).
}

\begin{gotofull}
\subsection{Effects of the dense IDs}
\label{exp:dense}
\end{gotofull}
\begin{gotobasic}
\noindent\textbf{Effects of the dense IDs.}
\end{gotobasic}
The dense IDs assumption was used to allow \fastr to use arrays for semijoins and aggregations instead of using other data structures like hash index. The two following experiments check its importance.

\begin{gotofull}
\subsubsection{\fastrUA~vs. \fastrUAB}
\label{exp:binary}
\end{gotofull}
\begin{gotobasic}
\noindent\textbf{\fastrUA~vs. \fastrUAB.}
\end{gotobasic}
\yannis{to self: mention that it only accounts for small savings. Chunbin: around 15\%.}
We conducted experiments to evaluate the performance of retrieving fragments in \fastr. \eat{For any given lookup value $t$, \fastr obtains fragments by directly locating to the $t$-th position in the lookup table instead of using binary searches, which are widely utilized in existing databases. We implemented another version of \fastr, called \textbf{\fastr-Binary}, that uses binary search to obtain fragments. }
Table~\ref{table:binary} shows the running time of different queries for \fastrUAB and \fastrUA on PubMed-M and SemMedDB~\smallfootnote{We achieve similar improvements in PubMed-MS. Due to space limitation, we omit them.}. As shown, \fastrUA outperforms \fastrUAB for all the queries. For example, \fastrUA saves around $12\%$ running time over \fastrUAB for AS query.  In addition, we also observed that, queries (FSD, AS and CS) with larger number of lookup requests benefits more compared with queries with smaller number of lookup requests, e.g., SD, AD and FAD.
%In addition, we also conducted experiments to systematically test the performance of FastR and FastR-Binary with different number of lookup requests. As shown in Figure~\ref{fig:binary}(b) FastR gains more benefits against FastR-Binary with more number of requests.

\begin{table}[htbp]
\small
\centering
\renewcommand{\tabcolsep}{1.5mm}
\begin{tabular}{l|rrrr} \hline\hline
    & Ave \# lookups & \fastrUAB & \fastrUA & $\theta$\\\hline
SD  &  22       & 247.94  & \textbf{177.08} & 28.58\%  \\ \hline
FSD &  21748262 & 1129.72 & \textbf{435.60} & 61.44\%  \\ \hline
AD  &  23609    & 38.67   & \textbf{30.33} & 21.57\%  \\ \hline
FAD &  23609    & 27.84   & \textbf{25.95} & 6.79\%  \\ \hline
AS  &  58589421 & 7364.92 & \textbf{4510.11} & 38.76\%  \\ \hline
CS  &  132975  & 16.21   & \textbf{8.62} & 46.82\%  \\ \hline\hline
\end{tabular}
\vspace{-3mm}
 \caption{\textbf{\fastrUA vs. \fastrUAB (in milliseconds)}. The last column shows the improvements, where $\theta$= 1-$\frac{\mathrm{\fastrUA}}{\mathrm{\fastrUAB}}$.}
 \label{table:binary}
\end{table}

%\begin{figure}[htbp]
%\centering
%\includegraphics[width=0.48\columnwidth]{exp_fastr_binary_1}
%\includegraphics[width=0.48\columnwidth]{exp_fastr_binary_2}
%\caption{\textbf{Effect of using positions to replace binary searches}.}\label{fig:binary}
%\end{figure}

\begin{gotofull}
\vspace{-2mm}
\subsubsection{\fastrUA vs. \fastrUAM}
\label{exp:hash}
\end{gotofull}
\begin{gotobasic}
\noindent\textbf{\fastrUA vs. \fastrUAM.}
\end{gotobasic}
\yannis{to self: it changes the ratio significantly .}
\yannis{to Chunbin: The differences in the AD, FAD and CS case are invisible. Either use log or write tables with numbers. When it is so little data, the tables are better. Include the ratio also. In next papers take care yourself of such obvious things. We spend too many revision rounds on exchanging notes on easy fixes that we shouldn't even be discussing. Chunbin: Yes, I changed the figures to tables.}
We measured the benefit of choosing an array for aggregation in \fastr over a hashmap by comparing \fastrUA with \fastrUAM.
\eat{ We implemented two versions of \fastr. One is \textbf{\fastr-Hash} that uses hashmap for aggregation and semijoin.
\yannis{to Chunbin and Ben: I added ``semijoin". Is it true? Chunbin: yes}
The other one is \textbf{\fastr} that uses a fix size array for aggregation and semijoin.}
As shown in Table~\ref{table:hash}, \fastrUA outperforms \fastrUAM for all the queries. \fastrUA gains more improvement for the queries with larger output like AS query (\fastrUA saves about $33\%$ running time) than queries with smaller output like CS query.
\yannis{to Chunbin and Ben: why do you say significant improvement on large output? looking at the ratios at Figure~\ref{fig:hash}(a) the ratio is higher on SD and FSD. (I know what you want to write but you really have to work to write it, potentially presenting additional data or relating to the data of other sections. The reader is not in your mind and is not a detective.)}
\yannis{to Chunbin: is it Pubmed-M or PubMed-MS? Why not both? Chunbin: in PubMed-M. In PubMed-MS, it has similar improvement.}

%In addition, in order to better analyze the effect of the output size, we vary domain sizes of the final output column from 100 to $10^7$. As shown in Figure~\ref{fig:hash}(b), with larger output size, \fastr obtains more benefits.
\yannis{to Chunbin and Ben: I do not understand what the Figure~\ref{fig:hash}(b) experiment does. What queries? Chunbin: I deleted it.}

\begin{table}[htbp]
\small
\centering
\renewcommand{\tabcolsep}{2mm}
\begin{tabular}{l|rrrr} \hline\hline
    & Ave \# results & \fastrUAM & \fastrUA & $\theta$\\\hline
SD  &  27,443,100 & 908.95  & \textbf{177.08} & 80.52\%  \\ \hline
FSD &  27,307,529 & 1342.82  & \textbf{435.60} & 67.56\%  \\ \hline
AD  &  200,679    & 34.84    & \textbf{30.33} & 12.94\%  \\ \hline
FAD &  56,518     & 31.63    & \textbf{25.95} & 17.96\%  \\ \hline
AS  &  20,019,297 &7766.83   & \textbf{4510.11} & 41.93\%  \\ \hline
CS  & 5,057       & 10.06    & \textbf{8.62} & 14.31\%  \\ \hline\hline
\end{tabular}
\vspace{-3mm}
 \caption{\textbf{\fastrUA vs. \fastrUAM (in milliseconds)}. The last column shows the improvements where $\theta$=1-$\frac{\mathrm{\fastrUA}}{\mathrm{\fastrUAM}}$.}
 \label{table:hash}
\end{table}

%\begin{figure}[h!]
%\centering
%\includegraphics[width=0.48\columnwidth]{exp_fastr_map_1}
%\includegraphics[width=0.48\columnwidth]{exp_fastr_map_2}
%\caption{\textbf{Effect of using arrays instead of hashmaps for aggregation}}\label{fig:hash}
%\end{figure}

\begin{gotofull}
\subsection{The effect of pipelining}
\label{exp:intermediate_results}
\end{gotofull}
\yannis{to self: We compared \fastr-UA (ie, no compression) with OMC-dense (i.e., array lookup, start-end, aggregation and semijoin arrays)}
\chunbin{I modified this section}
\begin{gotobasic}
\noindent\textbf{The effect of pipelining.}
In this experiment we compared OMC-denseID with \fastrUA in order to measure the benefit of pipelining over materializing intermediate results. Note that, OMC-denseID and \fastrUA have the same lookup data structure and same array for final aggregation.  Table~\ref{table:avoid_intermediate} reports the running time of \fastrUA and OMC-denseID on five instances of the AS query, where each instance queries for another author id, $A_1 - A_5$. The number of accessed fragments for those queries varies from  $~7M$ to $~585M$. As shown, \fastrUA outperforms OMC-denseID by around $15\times$. As the number of accessed elements increases, the running time of OMC-denseID increases significantly, since OMC-denseID materializes larger intermediate result columns.

\begin{table}[h!]
\small
\centering
\renewcommand{\tabcolsep}{1.5mm}
\begin{tabular}{l|r|r|r|r}\hline\hline
            &\# fragments &\# elements &OMC-denseID & \fastrUA 	\\\hline
   $A_1$	&7,484,532  & 51,730,682&4.12  & 1.02 \\\hline
   $A_2$	&9,287,804 & 65,687,183 &22.15 &2.24\\\hline
   $A_3$	&87,467,470 & 619,809,092&74.90 &5.38 \\\hline
   $A_4$	&184,219,134 & 1,305,764,797&171.33&13.76 \\\hline
   $A_5$	&585,932,678 & 4,153,322,719&297.47&49.01 \\\hline \hline
%   $D_1$	&78 & 7,701,798&0.32  &0.02 \\\hline
%   $D_2$	&107 & 15,002,676&0.83  &0.06 \\\hline
%   $D_3$	&392 & 61,490,754&2.80  &0.19 \\\hline
%   $D_4$	&476 & 131,690,364&8.52  &0.44 \\\hline
%   $D_5$	&660 & 405,002,636&17.43 &0.73 \\\hline \hline
\end{tabular}
\vspace{-3mm}
 \caption{\textbf{Avoiding intermediate results.}}
 \label{table:avoid_intermediate}
\end{table}
%Table~\ref{table:avoid_intermediate} also shows the performance of \fastrUA and OMC-denseID for the AD query, with different document ids ($D_1 \ldots D_5$) plugged in the \gt{WHERE} clause each time.
\end{gotobasic}

\begin{gotofull}
\begin{figure}[h!]
\centering
\includegraphics[width=0.48\columnwidth]{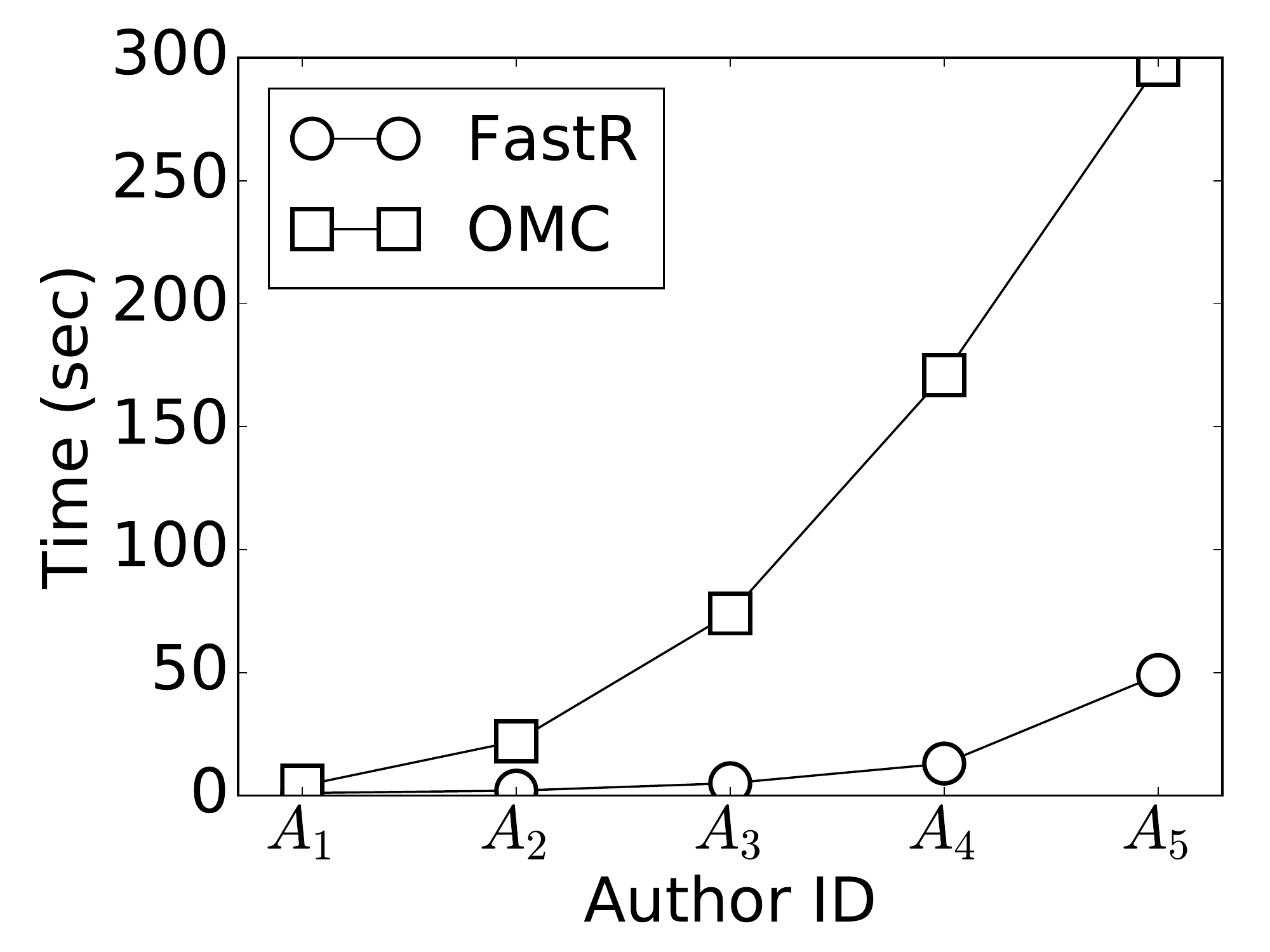}
\includegraphics[width=0.48\columnwidth]{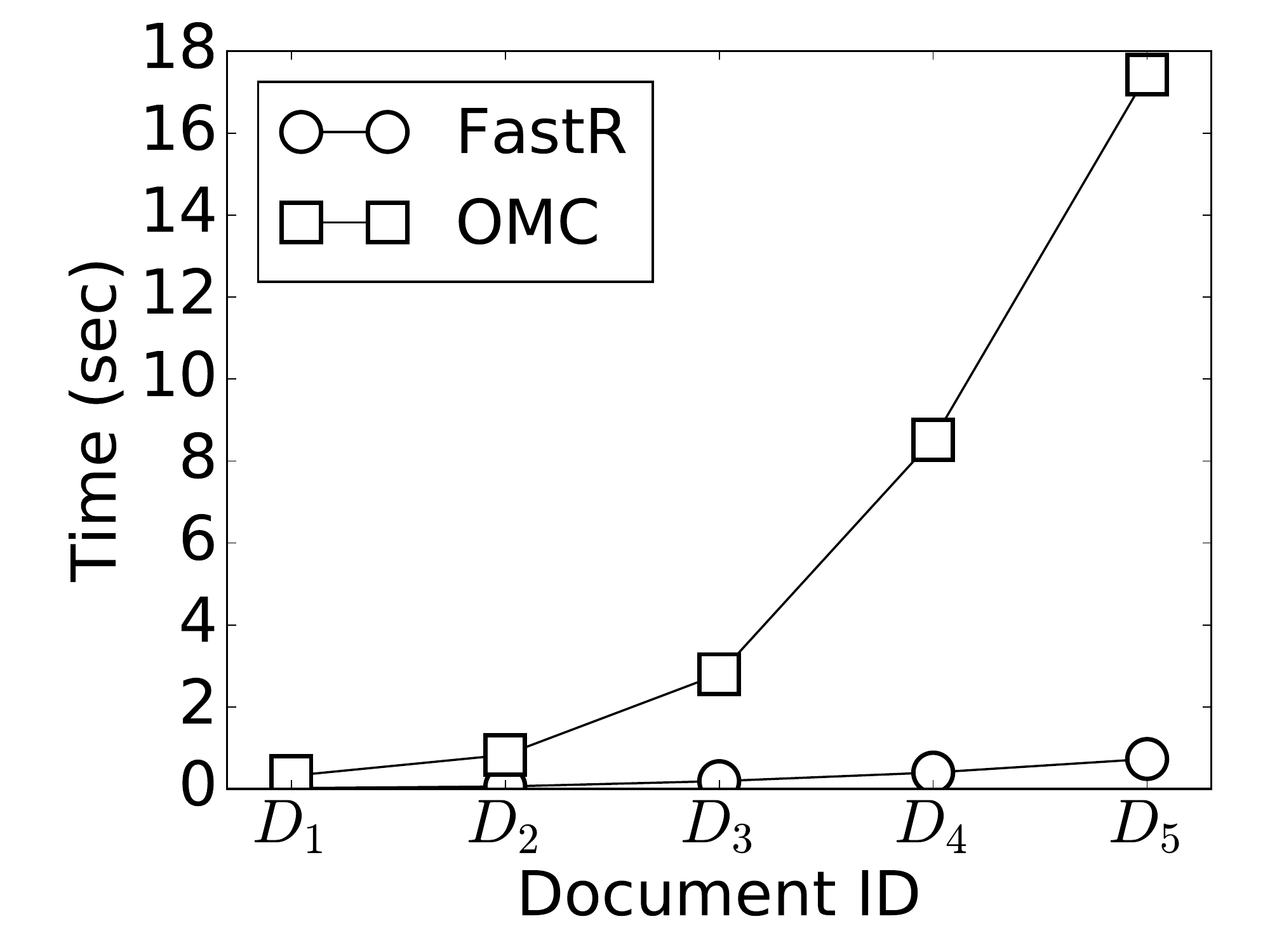}
\vspace{-3mm}
\caption{\textbf{Avoiding intermediate results}}\label{exp:fig_avoid_intermediate}
\end{figure}
In this experiment we compared OMC-denseID with \fastrUA in order to measure the benefit of pipelining over materializing intermediate results. Note that, OMC-denseID and \fastrUA have the same lookup data structure and same array for final aggregation.  Figure~\ref{exp:fig_avoid_intermediate} reports the running time of \fastrUA and OMC-denseID on five instances of the AS query, where each instance queries for another author id, $A_1 - A_5$. The number of accessed fragments for those queries varies from  $~7M$ to $~585M$ (see Table~\ref{table:query_statistics}). As shown, \fastrUA outperforms OMC-denseID by around $15\times$. As the number of accessed elements increases, the running time of OMC-denseID increases significantly, since OMC-denseID materializes larger intermediate result columns.

\begin{table}[h!]
\small
\centering
\renewcommand{\tabcolsep}{2.5mm}
\begin{tabular}{l|r|r}\hline\hline
            & 	\# of accessed  fragments & total \# of accessed elements	\\\hline
   $A_1$	&7,484,532  & 51,730,682 \\\hline
   $A_2$	&9,287,804 & 65,687,183 \\\hline
   $A_3$	&87,467,470 & 619,809,092 \\\hline
   $A_4$	&184,219,134 & 1,305,764,797 \\\hline
   $A_5$	&585,932,678 & 4,153,322,719 \\\hline \hline
   $D_1$	&78 & 7,701,798 \\\hline
   $D_2$	&107 & 15,002,676 \\\hline
   $D_3$	&392 & 61,490,754 \\\hline
   $D_4$	&476 & 131,690,364 \\\hline
   $D_5$	&660 & 405,002,636 \\\hline \hline
\end{tabular}
\vspace{-3mm}
 \caption{\textbf{Query statistics}}
 \label{table:query_statistics}
\end{table}
 Figure~\ref{exp:fig_avoid_intermediate} also shows the performance of \fastrUA and OMC-denseID for the AD query, with different document ids ($D_1 \ldots D_5$) plugged in the \gt{WHERE} clause each time.
\end{gotofull}

\eat{For \fastr, we use uncompressed arrays so that \fastr does not need to pay extra time on decompression. For OMC we did two modifications: (1) We changed the aggregation method from hashmap to array in order to exclude the difference between hashmap and array. (2) We changed the run-length encoded array to an array with corresponding domain size. Instead of storing a triple (t, s, l), OMC only stores a tuple (s, l) in position $t$. This array has exactly the same size with that in \fastr. Then we changed the lookup method in OMC by locating positions instead of binary searches in order to provide the same lookup method with \fastr.}

%Though it is still not an accurate comparison due to the different lookup structures in \fastr and OMC, it is close enough to argue that the difference between this \fastr and this modified OMC is from pipelining.
\yannis{to Chunbin and Ben: I am lost: How is the lookup structures different? The previous sentence said that they both locate positions rather than do binary search. So, what is different? Chunbin: In the paper, "Instead of storing a triple (t, s, l), OMC only stores a tuple (s, l)" means OMC does not store the value "t" explicitly, but it still have start and length. That is slightly different from \fastr. The same thing is that, now both OMC and \fastr can directly get start offset from t-th position. The difference is that, \fastr should get next non-zero start-offset to compute length, but OMC can directly get the length. Of cause, OMC needs to spend more space here compared with \fastr.}

\yannis{to Chunbin: This comes across as wrong figure number. Check them on the pdf not only the labels.Chunbin: this figure number is correct, i.e., Figure 15.}

\yannis{to Chunbin and Ben: Something seems inconsistent here. According to overall runtime comparisons, \fastr is 34 (in Pubmed-M) and 56 (in Pubmed-MS) times faster than OMC. In this subsestion's experiment we made \fastr faster by using UA and we also made OMC faster by using the dense optimizations (lookup instead of binary + arrays instead of hash). According to the previous sections the lookup Vs binary is a mere 10\% improvement for AS. The array Vs has is a 25\% improvement. Then how has the OMC/\fastr ratio fallen to 15 from 34? If I multiply $34 \times 90\% \times 75\%$ I get 23. How do we explain the gap between 15 and 23? Chunbin: In this experiment, we used different author ids. Since we want to show the trend of the performance changes with different \# fragments accessed.}

\yannis{to Chunbin and Ben: The lines in the \fastr/OMC figure are misleading because the ratios between the A points do not follow any usual reporting principle (log or linear). Chunbin: correct, but we choose the author ids from the real data, and we can not make sure that the \# accessed fragments increase linear. It is a five-way join query, what we can control is the author ids. So we choose author ids with different publications in order to get different \# of accessed fragments. Since the \# of accessed fragments are not linear increased, so the final running time is not linear. May be a better way is not to plot them in a line-chart, but using a table. That is, combining Figure 15 and Table 5 into one table.}

\yannis{to Ben and Chunbin: What does it mean ``increases significantly"? I want to see data and a figure that explains the correlation of \#accessed fragments and running times. Is it superlinear? Is it linear?Chunbin: again, we can not control the \# of accessed fragments to be linear.}
\yannis{to Ben and Chunbin: Same issues.}

\yannis{to Ben and Chunbin: What does it mean ``total \#accessed elements" and why does it matter? Does it even matter?  What is an element anyway? Don't introduce undefined terms. Chunbin: First ``total \# accessed fragments" means the number of fragments that accessed in the query. I think this one is clear to you. Since each fragment has different size, i.e., different number of element (value or integer) inside. So we report ``total \# accessed elements" here. Without this one, reviewers may say: ok, the \# of accessed fragments is large, but how about the size of each accessed fragment? are they small or large.}

\begin{gotofull}
\subsection{Analysis on different encodings}
\label{exp:different_encoding}
\end{gotofull}
\begin{gotobasic}
\noindent\textbf{Analysis on different encodings.}
\end{gotobasic}
\yannis{LATER: The terminology and notation of this section needs to align with the terminology of the data structure section better. For now, I removed the subscripts from $\gt{DT}_1$. I don't see where such notation is defined. Chunbin: agree. Yannis: And now you switched from one undefined notation to an irrelevant and undefined notation. These numbers have nothing to do with queries. Think what they really have to do with and use/define the relevant notation.}
We investigate the performance including compression quality and decompression time of the encoding methods that are employed by \fastr, i.e., uncompressed array (UA), bit-aligned compressed array (BCA), byte-aligned bitmap (BB) and Huffman encoding.  Table~\ref{table:compression} reports the encoded size of each column in the PubMed-MS dataset. As shown, no one encoding is the best for all the columns. Adopting a suitable compression can significantly save space. For example, by using BB, the space cost of \gt{dt1.Term} reduced from 3660.29 MB to 1431.12 MB. The selection of suitable compressions are based on our analysis results in Section~\ref{sec:data-structure}. For example, as we discussed UA always uses most space. In addition, BB achieves the minimal space on \gt{dt2.Doc}, which indicates the correctness of our analysis.
\yannis{to Chunbin: The data do not indicate ``the correctness of our analysis" because there are no data on what our analysis had predicted for these cases. Include them if you want to make such a point. At least in the extended version. Chunbin: I will add it in the extended version.}
\yannis{to Ben and Chunbin: I am surprised by the frequency compression numbers. Given how few frequencies there are I would expect BCA and Huffman to achieve a significant compression ratio over the UAs 8-bit numbers. What's going on? Chunbin: Ben will explain this one.}

%  Based on the analysis results in Section~\ref{sec:analysis}, the best encodings for \gt{DT.Term}, \gt{DT.Doc} and \gt{DA.Doc} are byte-aligned compressed bitmaps (BB), while the best one for \gt{DA.Author} is bit-aligned dictionary (BD). For example, the domain size of \gt{DT.Term} is $223705<2^{23}$. As shown in Table~\ref{table:pubmed_dataset}, the average size of fragment is $22.51$. Therefore, for \gt{DT.Term}, BB should be the best one theoretically.

\begin{table}[h!]
\small
\centering
\renewcommand{\tabcolsep}{2.5mm}
\begin{tabular}{l|r|r|r|r} \hline\hline
                         & UA & BCA & BB & Huffman\\ \hline
\gt{dt1.Term}&	3605.55	&	2033.25	&	\textbf{1376.39}	&	1565.60	\\ \hline
\gt{dt1.Fre}&	901.39	&	454.12	&	N/A	&	\textbf{142.46}	\\ \hline
\gt{dt2.Doc}&	3605.55	&	2816.93	&	\textbf{1047.71}	&	2779.37	\\ \hline
\gt{dt2.Fre}	&901.39	&	450.74	&	N/A	&	\textbf{134.84}	\\ \hline
\gt{da1.Doc}	&245.26	&	198.75	&	\textbf{187.54}	&	325.70	\\ \hline
\gt{da2.Author}&	245.26	&	\textbf{183.95}	&	205.10	&	275.56	\\ \hline
\gt{dy.Year}	&57.17	&	\textbf{14.20}	&	N/A	&	14.29	\\ \hline\hline
\end{tabular}
\vspace{-3mm}
 \caption{\textbf{Size of encoded columns (MB)}. The bold fonts show the minimal space for each column. BB only applies for fragments with unique values, so \gt{dt1.Fre} and \gt{dt2.Fre} can not be encoded by BB.}
 \label{table:compression}
\end{table}

\yannis{LATER, to Chunbin: I cannot see why ``According to the analysis, if a fragment size is greater than $14$, then BB is the best encoding method."}
\chunbin{I will modify the figure in Analysis Section to reflect this one}

\begin{table*}[htbp]
\small
\centering
\renewcommand{\tabcolsep}{3.2mm}
\begin{tabular}{l|l|l||c||r|r|r|r} \hline\hline
  	                    &	\# elements per fragment	&	\# fragments	&	compression ratio	&	1 thread	&	2 threads	&	4 threads	&	8 threads		\\\hline
                BCA 	&	100000$_{\pm1000}$              &	8000    	    &	76.23$\%$        	&	1535.506	&	864.594  	&	450.890	    &	378.227		\\\hline
                BB   	&	100000$_{\pm1000}$	            &	8000         	&	31.75$\%$        	&	1501.806	&	835.154  	&	428.818	    &	371.442		\\\hline
            Huffman 	&	100000$_{\pm1000}$              &	8000         	&	73.08$\%$        	&	52198.713	&	29688.662  	&	14934.780	&	7925.446		\\\hline\hline
\end{tabular}
\vspace{-3mm}
 \caption{\textbf{Space cost and decompression time of BCA, BB and Huffman}. Domain size is 1 billion, data follows zipf distribution with factor $s=1.5$. Fragments only contain unique values, which simulates fragments in foreign-key columns.}
 \label{table:decompression_unique}
\end{table*}

\begin{gotobasic}
\begin{table*}[htbp]
\small
\centering
\renewcommand{\tabcolsep}{3.6mm}
\begin{tabular}{l|l|l||c||r|r|r|r} \hline\hline
  	                    &	\# elements per fragment	&	\# fragments	&	compression ratio	&	1 thread	&	2 threads	&	4 threads	&	8 threads		\\\hline
 \multirow{3}{*}{BCA} 	&	100                     	&	8000000	        &	21.88$\%$       	&	1581.167	&	801.313	    &	410.039	    &	348.422		\\\cline{2-8}
                        %&	100000                  	&	8000	        &	21.88$\%$       	&	1397.182	&	708.072	    &	362.327 	&	307.879		\\\cline{2-8}
                    	&	10000000                   	&	80	            &	21.88$\%$       	&	1286.579	&	652.020	    &	333.645	    &	283.507		\\\hline
\multirow{3}{*}{Huffman}&	100	                        &	8000000	        &	12.28$\%$       	&	5055.162	&	2543.277	&	1280.826	&	668.050		\\\cline{2-8}
                        %&	100000                  	&	8000        	&	11.41$\%$       	&	4635.229	&	2332.007	&	1174.428	&	612.555		\\\cline{2-8}
                    	&	10000000                   	&	80	            &	11.39$\%$         	&	4374.838	&	2201.003	&	1108.453	&	578.143		\\\hline\hline
\end{tabular}
\vspace{-3mm}
 \caption{\textbf{Space cost and decompression time of BCA and Huffman}. Domain size is 100, data follows zipf distribution with factor $s=1.5$. Fragments contains duplicates, which simulates fragments in measure attributes.}
 \label{table:decompression_duplicates}
\end{table*}
\end{gotobasic}

\begin{gotofull}
\begin{table*}[htbp]
\small
\centering
\renewcommand{\tabcolsep}{3.4mm}
\begin{tabular}{l|l|l||c||r|r|r|r} \hline\hline
  	                    &	\# elements per fragment	&	\# fragments	&	compression ratio	&	1 thread	&	2 threads	&	4 threads	&	8 threads		\\\hline
 \multirow{8}{*}{BCA} 	&	10	                        &	80000000    	&	21.88$\%$        	&	1647.730	&	835.046  	&	427.301	    &	363.090		\\\cline{2-8}
                    	&	100                     	&	8000000	        &	21.88$\%$       	&	1581.167	&	801.313	    &	410.039	    &	348.422		\\\cline{2-8}
                    	&	1000                    	&	800000	        &	21.88$\%$	        &	1517.293	&	768.942	    &	393.475 	&	334.347		\\\cline{2-8}
                    	&	10000                   	&	80000	        &	21.88$\%$	        &	1456.000	&	737.880	    &	377.580	    &	320.840		\\\cline{2-8}
                    	&	100000                  	&	8000	        &	21.88$\%$       	&	1397.182	&	708.072	    &	362.327 	&	307.879		\\\cline{2-8}
                    	&	1000000	                    &	800	            &	21.88$\%$       	&	1340.741	&	679.468	    &	347.690 	&	295.442		\\\cline{2-8}
                    	&	10000000                   	&	80	            &	21.88$\%$       	&	1286.579	&	652.020	    &	333.645	    &	283.507		\\\cline{2-8}
                    	&	100000000	                &	8	            &	21.88$\%$       	&	1234.606	&	625.680	    &	320.167	    &	272.055		\\\hline
\multirow{8}{*}{Huffman}&	10	                        &	80000000    	&	18.75$\%$       	&	5203.430	&	2617.872	&	1318.393	&	687.644		\\\cline{2-8}
                    	&	100	                        &	8000000	        &	15.38$\%$       	&	5055.162	&	2543.277	&	1280.826	&	668.050		\\\cline{2-8}
                    	&	1000                       	&	800000      	&	12.34$\%$       	&	4911.119	&	2470.809	&	1244.330	&	649.014		\\\cline{2-8}
                    	&	10000                      	&	80000	        &	11.56$\%$       	&	4771.180	&	2400.405	&	1208.874	&	630.521		\\\cline{2-8}
                    	&	100000                  	&	8000        	&	11.41$\%$       	&	4635.229	&	2332.007	&	1174.428	&	612.555		\\\cline{2-8}
                    	&	1000000	                    &	800	            &	11.39$\%$       	&	4503.152	&	2265.558	&	1140.964	&	595.100		\\\cline{2-8}
                    	&	10000000                   	&	80	            &	11.39$\%$         	&	4374.838	&	2201.003	&	1108.453	&	578.143		\\\cline{2-8}
                    	&	100000000               	&	8           	&	11.39$\%$       	&	4250.180	&	2138.287	&	1076.868	&	561.670		\\\hline\hline
\end{tabular}
 \caption{\textbf{Space cost and decompression time of BCA and Huffman}. Domain size is 100, data follows zipf distribution with $s=1.5$. Exist duplicates in each fragment, which simulates fragments in measure columns.}
 \label{table:decompression_duplicates_2}
\end{table*}
\end{gotofull}

% \begin{figure}[h!]
% \centering
% \includegraphics[width=0.48\columnwidth]{compression_1}
% \includegraphics[width=0.48\columnwidth]{compression_2}
% \caption{\textbf{Effect of using positions to replace binary searches}.}\label{fig:pure_decoding}
% \end{figure}

\vspace{-6mm}

In addition, we also conducted experiments to evaluate the decompression performance of these encodings for two kinds of fragments: one is fragments on foreign key columns containing only unique values. The other one is fragments on measure attributes, which have many duplicates. For the first case, we generated a \gt{DT} table with \gt{DT.Doc} in  zipf distribution with factor $s=1.5$ and the domain size is 1 billion. Then we randomly choose 8000 fragments whose sizes are located in [100000-1000, 100000+1000].  We observed from Table~\ref{table:decompression_unique}  that BB achieves the highest compression quality (it saves $69,25\%$ space) and the highest decompression performance (it is about $30$ times fater than Huffman). Huffman has the worst performance, since the domain size is large which requires Huffman to maintain a large decoding table that can not be fitted in L1 or L2 caches.

Further, we tested the decompression performance for Huffman and BCA for fragments on measure attributes. We created  a \gt{DT} table with \gt{DT.Fre} in zipf distribution with factor $s=1.5$ and the domain size is 100. Table~\ref{table:decompression_duplicates_2} reports the decomprssion time and compression quality for Huffman and BCA with different number of fragments and different size of fragments. As shown, Huffman achieves the highest compression quality and has comparable decompression performance against BCA. Compared with the results in Table~\ref{table:decompression_unique}, we noticed that the decompression performance of Huffman is significantly improved. This is because the domain size is small (say 100) here, which can be fitted into the L1 cache. This result also indicates that Huffman is perfect for measure attributes.

Finally, we observed that multiple threads improve the decompression performance for all the encoding methods. For Huffman, the performance is continuously improved with increasing number of threads, while the improvements of BB and BCA become slower with more threads. This is because Huffman is more CPU bounded than memory bounded.

\begin{figure}[h!]
\centering
\includegraphics[width=0.31\columnwidth]{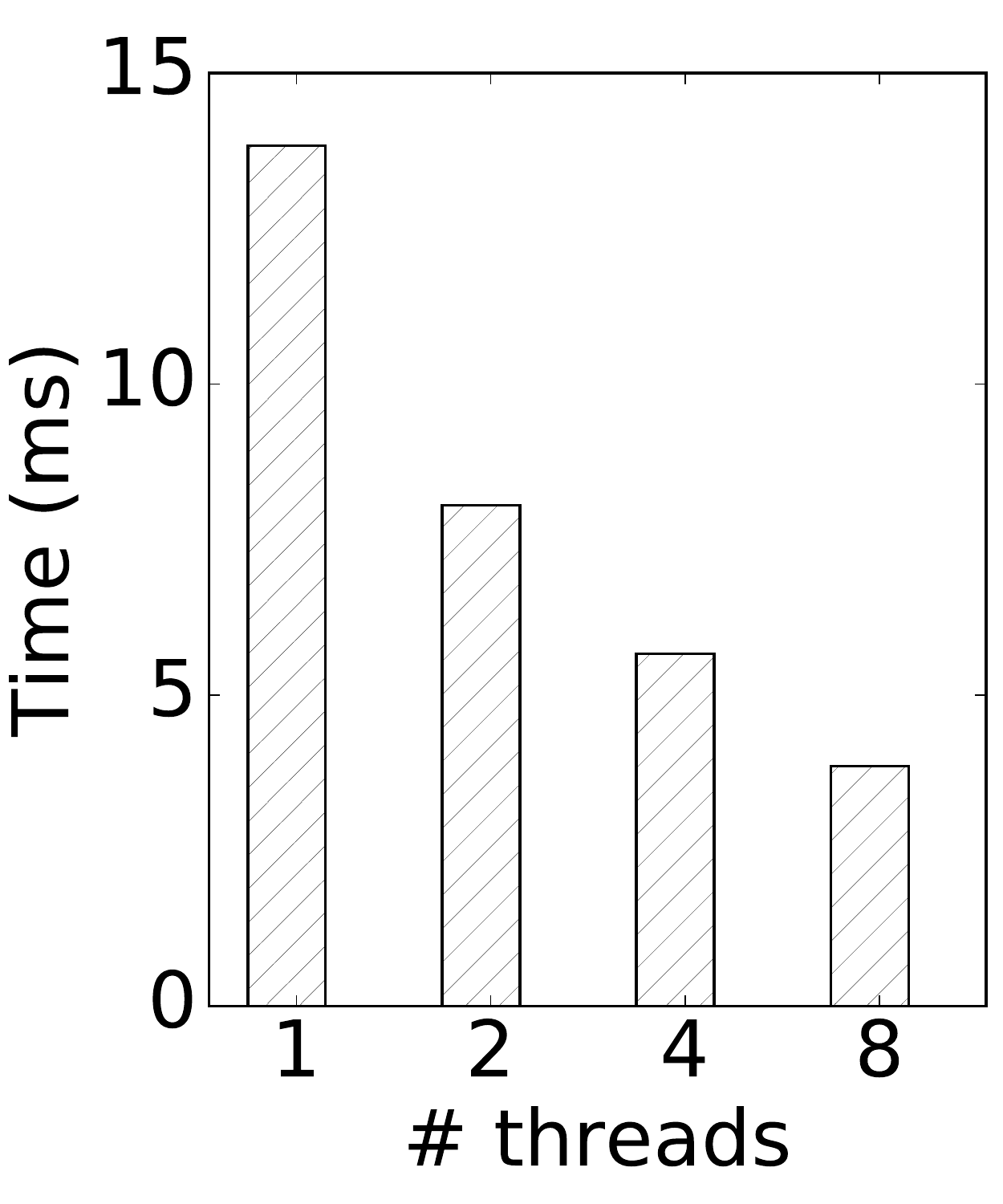}
\includegraphics[width=0.31\columnwidth]{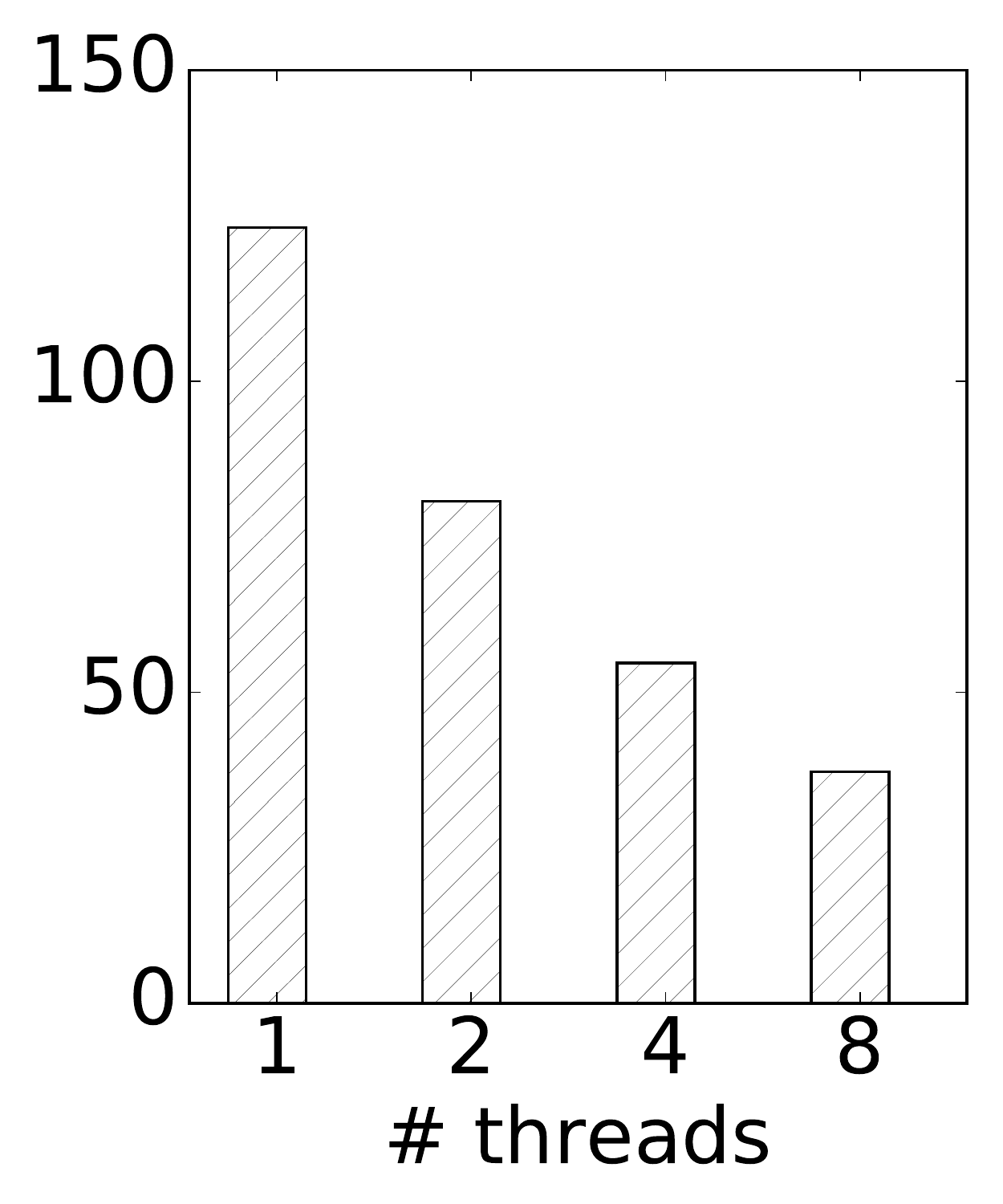}
\includegraphics[width=0.31\columnwidth]{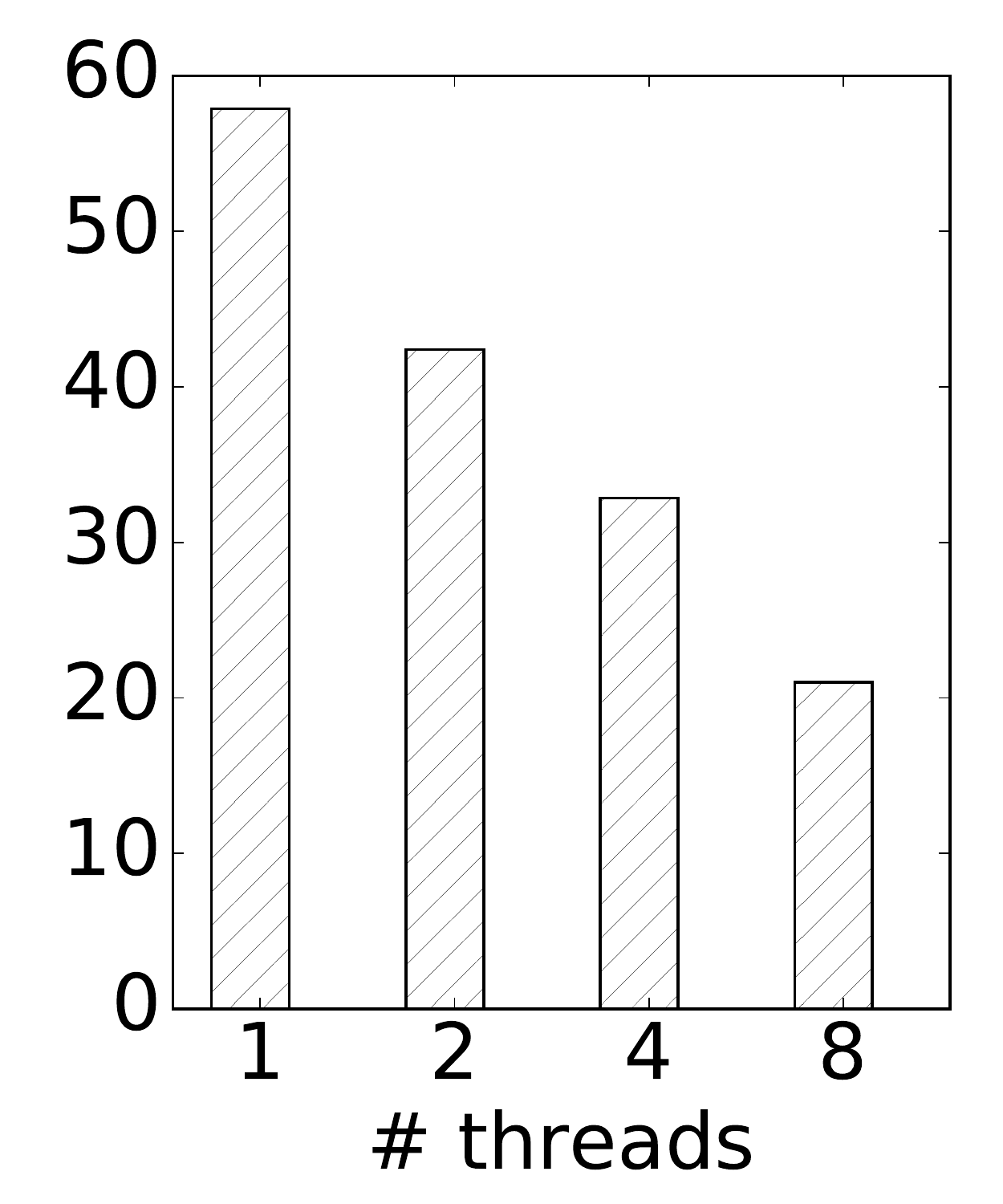}
\vspace{-2mm}
\caption{\textbf{Running time of AS query on PubMed-M and PubMed-MS, and running time of CS on SemmedDB with threads from 1 to 8}.}\label{fig:multi-cores}
\end{figure}

\begin{gotofull}
\subsection{Effect of multiple threads}
\label{exp:multithreads}
\end{gotofull}
\begin{gotobasic}
\noindent\textbf{Effect of multiple threads.}
\end{gotobasic}
At last, we evaluate the effect of multiple threads for the overall performance. Figure~\ref{fig:multi-cores} shows the running time of AS query on PubMed-M and PubMed-MS and that of CS query on SemmedDB with threads from 1 to 8. We observed that: (1) multiple threads improve the performance; and (2) the improvements are not as significant as expected. This is because there exits skew problem in multiple threads. For example, the difference between the minimal and maximal number of processed fragments in different threads is around $2$ million for AS query. The skew problem can be solved by employing load-balance algorithms.
\vspace{-3mm}

\section{Conclusions and Future Work}
\label{sec:conclusion}
In this paper, we defined and studied relationship queries. \fastr uses a new fragment-based data structure and a new, coordinated bottom-up pipelining execution strategy to answer relationship queries. We itemized the sources of \fastrs superior performance and measured how much each one of them contributes to the overall performance speedup.

In the future we will study and prototype how \fastrs data structures and strategy can be incorporated in a general SQL processor, whereas \fastr will be executing relationship subqueries and conventional query processing techniques will be used to further combine and process the results of the relationship subqueries. Furthermore, we will incorprorate strategies for skew avoidance and best utilization of the multiple cores.

\begin{gotofull}
\section{Appendix}
\label{sec:appendix}
\subsection{Proof of analytical results}
For byte-aligned bitmap, in order to analyze the space cost, we need to know the distribution of elements in the fragment, since the space cost varies with different positions of 1's in the bitmap. Here we assume the elements are uniformly distributed. Therefore, the expected length of runs is $\frac{D-N}{N}$. According to Chernoff Bound~\cite{Mitzenmacher2005}, the actual length of runs is close to the expected value within $\epsilon$ with high probability $1-e^{-\delta}$.   Therefore, for byte-aligned bitmap, the space cost with high probability is:
here, we assume $N<\frac{D}{2}$ (i.e., $\frac{D-N}{N}>1$), which means the number of 1's is less than half of the domain size. If the number of 1's is greater than half of the domain size, then FastR could reverse the bitmap by switching 0 and 1.

In the following, we will compare the space costs of these four encoding methods comprehensively in order to choose the one with minimal space costs. Figure \ref{fig:compression_analysis_1} shows the comparison results. The x-axis is the domain size $D$ and the y-axis is the fragment size $N$. The different colors indicate the best encoding methods in terms of space cost for different $D$ and $N$.
%For example, the \gt{DT.Term} in PubMed dataaset, the domain size is 223705 and the average size of fragments is 23. As shown in Figure~\ref{fig:compression_analysis_1} the red point $(223705, 23)$ is located in the blue area, so \textsf{BB} is the encoding method with minimal space cost for \gt{DT.Term}, which can be verified by the accurate space costs in Table~\ref{table:compression}.

Figure \ref{fig:compression_analysis_1} is plotted based on the analytical results for these cases:

\begin{compact_item}
\item Case 1: For any $D$ and $N$, $\mathcal{S}_{\textsf{UA}}\geq\mathcal{S}_{\textsf{BD}}$ holds.
\item Case 2: If $D \leq 8$, then $\mathcal{S}_{\textsf{UB}}$ = min\{$\mathcal{S}_{\textsf{BD}}$ , $\mathcal{S}_{\textsf{BB}}$ , $\mathcal{S}_{\textsf{UA}}$\}.
    \yannis{to Chunbin: you just wrote that $S_{UA} \geq S_{BD}$ always. What's the point of keeping UA around? Chunbin: I want to make it clear that the one with smallest space is smaller than all the others.}
\item Case 3: If $\frac{D}{128^x+1}\leq N<\frac{D}{128^{x-1}+1}(x\geq2)$ and $D> 2^{8x-1}$, then $\mathcal{S}_{\textsf{BB}}\leq$  min\{$\mathcal{S}_{\textsf{UB}}$ , $\mathcal{S}_{\textsf{BD}}$ , $\mathcal{S}_{\textsf{UA}}$\}.
\yannis{to Chunbin: Can i replace this with $\frac{D}{128^x+1}\leq N<\frac{D}{8}(x\geq2)$?}
\item Case 4: If $\frac{D}{128^x+1}\leq N<\frac{D}{128^{x-1}+1}(x\geq2)$ and $D\leq 2^{8x-1}$, then $\mathcal{S}_{\textsf{BD}}\leq$  min\{$\mathcal{S}_{\textsf{UB}}$ , $\mathcal{S}_{\textsf{BB}}$ , $\mathcal{S}_{\textsf{UA}}$\}.
\item Case 5: If $\frac{D}{128^1+1}\leq N\leq\frac{D}{8}$ and $D> 2^{8-1}$, then $\mathcal{S}_{\textsf{BB}}\leq$  min\{$\mathcal{S}_{\textsf{UB}}$ , $\mathcal{S}_{\textsf{BD}}$ , $\mathcal{S}_{\textsf{UA}}$\} .
\yannis{to Chunbin: If we fix Case~3 the way I suggest then Case 5 is simply Case 3 for $x=1$. Correct? Chunbin: yes}
\item Case 63: If $\frac{D}{128^1+1}\leq N\leq\frac{D}{8}$ and $D\leq 2^{8-1}$, then $\mathcal{S}_{\textsf{BD}}\leq$  min\{$\mathcal{S}_{\textsf{UB}}$ , $\mathcal{S}_{\textsf{BB}}$ , $\mathcal{S}_{\textsf{UA}}$\}.
\item Case 7: If $\frac{D}{8}\leq N<\frac{D}{2}$ and $D> 2^7$, then $\mathcal{S}_{\textsf{UB}}$ = min\{$\mathcal{S}_{\textsf{BD}}$ , $\mathcal{S}_{\textsf{BB}}$ , $\mathcal{S}_{\textsf{UA}}$\}.
%\item Investigating the effect of the multicores (Section~\ref{exp:multicores}).
\end{compact_item}
For the rest cases, i.e., $\frac{D}{8}\leq N<\frac{D}{2}$ and $D\leq 2^7$, Table~\ref{table:space_compare} shows the comparison details.

\yannis{to Chunbin: I removed the table that presented comparisons within $\frac{D}{8}\leq N<\frac{D}{2}$ and $D\leq 2^7$.Chunbin:OK, I see. But we can keep it in the full version}

\begin{table}[htbp]
\small
\centering
\begin{tabular}{c|l} \hline
   & Conditions\\\hline
   \multirow{2}{*}{$\mathcal{S}_{\textsf{BD}}$ $\leq$ min\{$\mathcal{S}_{\textsf{UB}}$ , $\mathcal{S}_{\textsf{BB}}$ , $\mathcal{S}_{\textsf{UA}}$\}}& $N\leq4$ \\\cline{2-2}
                 &  $N\leq5$ AND $32<D<64$ \\\cline{2-2}
                 &  $N\leq6$ AND $40<D<64$ \\\cline{2-2}
                 &  $N\leq7$ AND $48<D<64$ \\\cline{2-2}
                 &  $N\leq8$ AND $48<D<64$ \\\cline{2-2}
                 &  $N\leq9$ AND $56<D<64$ \\\hline
    \multirow{2}{*}{$\mathcal{S}_{\textsf{UB}}$ $\leq$ min\{$\mathcal{S}_{\textsf{BD}}$ , $\mathcal{S}_{\textsf{BB}}$ , $\mathcal{S}_{\textsf{UA}}$\}}
                & $N\leq5$ AND $D\leq32$\\\cline{2-2}
                & $N\leq6$ AND $D\leq40$\\\cline{2-2}
                & $N\leq7$ AND $D\leq48$\\\cline{2-2}
                & $N\leq8$ AND $D\leq48$\\\cline{2-2}
                & $N\leq9$ AND $D\leq56$\\\cline{2-2}
                & $N\geq10$\\\hline
\end{tabular}
\caption{Space cost comparisons within $\frac{D}{8}\leq N<\frac{D}{2}$ and $D\leq 2^7$.}
 \label{table:space_compare}
\end{table}

\yannis{to Chunbin: the proofs are in \gt{eat\{\}}. Put them back in the extended version.Chunbin:Yes}

FastR maintains a fragment-array (byte-array) to continuously store fragments for each column, which requires the size of each encoded fragment is a  multiplication of eight in bits. Therefore, for the UB and BD, we need to add an additional $\Delta$ ($\Delta \in [1,7]$) bits to make sure the size meets the requirement of the fragment-array structure. Since FastR applies UB and BD only on fragments with distinct values. Thus, in this section, we focus on discussing the fragment with distinct values, which is a the case for foreign key columns. Let $D$ be the domain size, $N$ ($N\leq D$) be the number of elements in a fragment. Let $\mathcal{S}_A$ be the space cost of an encoding method $A$.

We provide next the detailed proofs for cases~1 to~7.

\begin{proof} (Case 1)
Recall that $\mathcal{S}_{\textsf{UA}}=32\cdot N\cdot\lceil\log_{2^{32}}D\rceil$ and  $\mathcal{S}_{\textsf{BD}}=N\cdot\lceil\log_{2}D\rceil+\Delta$. Consider $(2^{32})^{i}<D\leq(2^{32})^{(i+1)}$ ($i>0$), then $\mathcal{S}_{\textsf{UA}}=32\cdot(i+1)N$ while $\mathcal{S}_{\textsf{BD}}\in(32\cdot iN, 32\cdot(i+1)N]$. Therefore for $\mathcal{S}_{\textsf{UA}}\geq\mathcal{S}_{\textsf{UA}}$.
\end{proof}

\begin{proof} (Case 2)
There is at least one element in the fragment, thus $\mathcal{S}_{\textsf{BB}}\geq32$, since \textsf{UA} uses 32 bits for each value. Similarly, $\mathcal{S}_{\textsf{BB}}\geq8$ and $\mathcal{S}_{\textsf{BD}}\geq8$, since the size of \textsf{BB} and \textsf{BD} are required to be a multiplication of eight, i.e., $\mathcal{S}_{\textsf{BB}}\geq8$ and $\mathcal{S}_{\textsf{BD}}\geq8$.  If $D \leq 8$ holds, then we have $\mathcal{S}_{\textsf{UB}}$=$D \leq 8$. Therefore, $\mathcal{S}_{\textsf{UB}}$ $\leq$ min\{$\mathcal{S}_{\textsf{UA}}$ ,$\mathcal{S}_{\textsf{BB}}$, $\mathcal{S}_{\textsf{BD}}$\}.
\end{proof}

\begin{proof} (Case 3)
If $\frac{D}{128^{x}+1}\leq N<\frac{D}{128^{x-1}+1}(x\geq2)$ holds, then we have: $128^{x-1}<\frac{D-N}{N}\leq128^x$. Then $\mathcal{S}_{\textsf{BB}}=8\cdot x\cdot N$. Because $N<\frac{D}{128^{x-1}+1}<\frac{D}{8\cdot x}$, so $\mathcal{S}_{\textsf{BB}}<D\leq\mathcal{S}_{\textsf{UB}}$, since $\mathcal{S}_{\textsf{UB}}=D+\Delta$. Compare \textsf{BB} with \textsf{BD}. $\mathcal{S}_{\textsf{BB}}\leq \mathcal{S}_{\textsf{BD}}$ if $D>2^{8x-1}$ holds. Because  $\mathcal{S}_{\textsf{BD}}\geq8\cdot x\cdot N=\mathcal{S}_{\textsf{BB}}$. Therefore,  If $\frac{D}{128^x+1}\leq N<\frac{D}{128^{x-1}+1}(x\geq2)$ and $D> 2^{8x-1}$, then $\mathcal{S}_{\textsf{BB}}\leq$  min\{$\mathcal{S}_{\textsf{UB}}$ , $\mathcal{S}_{\textsf{BD}}$ , $\mathcal{S}_{\textsf{UA}}$\}.
\end{proof}

\begin{proof} (Case 4)
If $D>2^{8x-1}$ holds, then $\mathcal{S}_{\textsf{BD}}\leq8\cdot x\cdot N=\mathcal{S}_{\textsf{BB}}$. Recall the proof of Case 3, it is easy to get
\emph{If $\frac{D}{128^x+1}\leq N<\frac{D}{128^{x-1}+1}(x\geq2)$ and $D\leq 2^{8x-1}$, then $\mathcal{S}_{\textsf{BD}}\leq$  min\{$\mathcal{S}_{\textsf{UB}}$ , $\mathcal{S}_{\textsf{BB}}$ , $\mathcal{S}_{\textsf{UA}}$\}.}
\end{proof}

\begin{proof} (Case 5)
$\mathcal{S}_{\textsf{BB}}=8N$ and $\mathcal{S}_{\textsf{UB}}=D+\Delta$. Thus $\mathcal{S}_{\textsf{BB}}\leq\mathcal{S}_{\textsf{UB}}$ since $ D\geq8N$ holds.
If $D> 2^{8-1}$, then  $\mathcal{S}_{\textsf{BD}}>8N=\mathcal{S}_{\textsf{BB}}$. Thus \emph{If $\frac{D}{128^1+1}\leq N\leq\frac{D}{8}$ and $D> 2^{8-1}$, then $\mathcal{S}_{\textsf{BB}}\leq$  min\{$\mathcal{S}_{\textsf{UB}}$ , $\mathcal{S}_{\textsf{BD}}$ , $\mathcal{S}_{\textsf{UA}}$\}.}
\end{proof}

\begin{proof} (Case 6)
$\mathcal{S}_{\textsf{BB}}=8N$ and $\mathcal{S}_{\textsf{UB}}=D+\Delta$. Thus $\mathcal{S}_{\textsf{BB}}\leq\mathcal{S}_{\textsf{UB}}$ since $ D\geq8N$ holds.
If $D\leq 2^{8-1}$, then  $\mathcal{S}_{\textsf{BD}}\leq8N=\mathcal{S}_{\textsf{BB}}$, therefore we have:
\emph{If $\frac{D}{128^1+1}\leq N\leq\frac{D}{8}$ and $D\leq 2^{8-1}$, then $\mathcal{S}_{\textsf{BD}}\leq$  min\{$\mathcal{S}_{\textsf{UB}}$ , $\mathcal{S}_{\textsf{BB}}$ , $\mathcal{S}_{\textsf{UA}}$\}.}
\end{proof}

\begin{proof} (Case 7)
Recall that $\mathcal{S}_{\textsf{UB}}=D+\Delta$, $\mathcal{S}_{\textsf{BD}}=N\cdot\lceil\log_{2}D\rceil+\Delta'$ and $\mathcal{S}_{\textsf{BB}}=8N$. If $\frac{D}{8}\leq N<\frac{D}{2}$ holds,  $D\leq8N$, thus $D+\Delta\leq8N$. The reason is that $8N$ is a multiplication of eight, so if $D=8N$, $\Delta=0$. Therefore $\mathcal{S}_{\textsf{UB}}\leq\mathcal{S}_{\textsf{BB}}$.

Then we compare $\mathcal{S}_{\textsf{UB}}$ with $\mathcal{S}_{\textsf{BD}}$. If $D> 2^7$, $\mathcal{S}_{\textsf{BD}}\geq8N+\Delta'>\mathcal{S}_{\textsf{BB}}>\mathcal{S}_{\textsf{UB}}$.  Therefore, we have
\emph{If $\frac{D}{8}\leq N<\frac{D}{2}$ and $D> 2^7$, then $\mathcal{S}_{\textsf{UB}}$ = min\{$\mathcal{S}_{\textsf{BD}}$ , $\mathcal{S}_{\textsf{BB}}$ , $\mathcal{S}_{\textsf{UA}}$\}.}
\end{proof}

\subsection{Translation Algorithm}
In this section, we introduce the translation algorithm that translates RQNA expressions into physical plans.
\begin{algorithm}[htbp]\small
 Mapping each $\pi_{\mathit{attrs}(L), r.A_1,\ldots,r.A_n} (L \Join_{B=r.B'} R\mapsto r)\in \mathcal{Q}_R$ to $\mathcal{Q}_P\Leftarrow$$L \rijoin_{B; R\mapsto r}^{r.A_1,\ldots,r.A_n} \mathcal{I}_{R.B'}$;
 
 Mapping each $\pi_{r.A_1,\ldots,r.A_n} \sigma_{r.B' = c} (R\mapsto r)$ to $\{ [B:c] \} \rijoin_{B; R\mapsto r}^{r.A_1,\ldots,r.A_n} \mathcal{I}_{R.B'}$;
 
 Mapping each $\pi_{r.A_1,\ldots,r.A_n}((R \mapsto r) \semijoin_{B = r.B'} L)$ to $\risemijoin_{B; R\mapsto r}^{r.A_1,\ldots,r.A_n} \mathcal{I}_{R.B'} L$;
 
 Mapping each $\pi_{r.A_1,\ldots,r.A_n} ((R \mapsto r) \semijoin_{r.B' = l_1.B_1} \pi_{l_1.B_1} L_1) \semijoin \ldots \semijoin_{r.B' = l_m.B_m} (\pi_{l_m.B_m} L_m)$ to $\risemijoin_{B; R\mapsto r}^{r.A_1,\ldots,r.A_n} \mathcal{I}_{R.B'} \riinter^{\theta}_{(\pi_{l_1.B_1} L_1),\ldots,(\pi_{l_m.B_m} L_m)}$;
 
 Mapping each $\gamma_{r.D; \alpha(s(A_1, \ldots, A_n))}(L)$ to $\gamma^{1}_{r.D; \alpha(s(A_1, \ldots, A_n))}$;
 \caption{Translation Algorithm}\label{translation_algorithm}
 \end{algorithm}

\subsection{PMC Algorithm}
Here we introduce the detailed algorithms for PMC
Let $L$ be the intermediate result produced by previous operator.

(1) For $\pi_{A_1,\dots,A_n}\sigma_{B=c}(R)$, \textsf{PMC Selection} (Algorithm \ref{alg:PMC_selection}(lines $1-8$)) performs selection operation by scanning the whole column to get row ids (lines $3\sim 5$) and then executes random accesses to get values(lines $6\sim8$). Note that, PMC materializes row ids, i.e., map $A$.

(2) For $(\pi_{A_1,\ldots,A_n}L)\Join_{L.B_1=R_2.B_2}(\pi_{A_1,\ldots,A_m}R_2)$, \textsf{PMC Join} (Algorithm \ref{alg:PMC_join} (lines $2-11$)) is provided to get join results. It first scans column $R_2.B_2$ for each value in $L.B_1$ to get row ids (lines $3-6$), then uses row ids to find values in corresponding rows in other columns (lines $7-11$);

(3) For $(\pi_{A_1,\ldots,A_n}L)\rtimes_{L.B_1=R_2.B_2}(\pi_{A_1,\ldots,A_m}R_2)$, PMC applies the same algorithm as $\Join$. The join algorithm only produces results based on the values in  $L.B_1$, which insures the correctness of the semijoin.

(4) For $(\pi_{A}L)\cap (\pi_{A}R_2)$,  \textsf{PMC Intersection} (Algorithm \ref{alg:PMC_intersection} (lines $2-9$)) scans column $R_2.A$ for each value in $L.A$.

(5) For $\gamma_{\mathds{A},F}L$, Algorithm \textsf{Aggregation} (Algorithm \ref{alg:PMC_aggregation}) is called to compute the aggregated score for each value $t$ by a function $F$.

Note that, PMC does not apply optimizations.

\begin{algorithm}[h!]\small
Initialize a map $\mathcal{R}$;\\
\tcc{PMC Selection}
Initialize a temporary map $\mathcal{A}$;\\
\For{$i=0;i<|B|;i++$}
{
    \If{$B[i]==c$}
    {
        $\mathcal{A}\leftarrow i$;\\
    }
}
\For{each $i \in \mathcal{A}$}
{
    \For{$col=0;col<n; col++$}
    {
        $\mathcal{R} \leftarrow A_{col}[i]$; \\
    }
}

\tcc{OMC Selection}
Do binary search in column $B$ to find $(b,s,e)$ where $b=c$;\\
\For{$i=s;i<e;i++$}
{
    \For{$col=0;col<n; col++$}
    {
        $\mathcal{R}= A_{col}[i]$;\\
    }
}
\textbf{Return $\mathcal{R}$};\\
\caption{Selection ($\pi_{A_1,\dots,A_n}\sigma_{B=c}(R)$)}
\label{alg:PMC_selection}
\end{algorithm}

\begin{algorithm}[h!]\small
Initialize a map $\mathcal{R}$;\\
\tcc{PMC Join}
Initialize a temporary map $\mathcal{A}$;\\
\For{$i=0;i<|L.B_1|;i++$}
{
    \For{$j=0;j<|B_2|;j++$}
    {
        \If{$B_1[i]==B_2[j]$}
        {
            $\mathcal{A}\leftarrow j$;
        }
    }
    \For{each $j\in \mathcal{A}$}
    {
        \For{$col_1=0;col_1<m;col_1++$}
        {
            $\mathcal{R}$ = $R_2.A_{col_1}[j]$;
        }
        \For{$col_2=m;col_2<(m+n);col_2++$}
        {
            $\mathcal{R}$ = $L.A_{col_2-m}[j]$;
        }
    }

}

\tcc{OMC Join}
\For{$i=0;i<|L.B_1|;i++$}
{
    Do binary search in column $R_2.B_2$ to find (b,s,e) where $b=L.B_1[i]$;\\
    \For{$j=s;j<e;j++$}
    {
        \For{$col_1=0;col_1<m;col_1++$}
        {
            $\mathcal{R}$ = $R_2.A_{col_1}[j]$;\\
        }
        \For{$col_2=m;col_2<(m+n);col_2++$}
        {
            $\mathcal{R}$ = $L.A_{col_2-m}[j]$;\\
        }
    }
}
\textbf{Return $\mathcal{R}$};
\caption{Join ($\pi_{A_1,\ldots,A_n}L\Join_{L.B_1=R_2.B_2}\pi_{A_1,\ldots,A_m}R_2$)}
\label{alg:PMC_join}
\end{algorithm}

\begin{algorithm}[h!]\small
Initialize a map  $\mathcal{R}$;\\
\tcc{PMC Intersection}
flag = 0;\\
\For{$i=0;i<|L.A|;i++$}
{
    \For{$j=0;j<|R_2.A|;j++$}
    {
        \If{$R_1.A[i]==R_2.A[j]$}
        {
            $\mathcal{R}\leftarrow R_1.A[i]$;\\

        }
    }
}
\tcc{OMC Intersection}
i=0; j=0;\\
\While{$i<|L.A| AND j<|R_2.A|$ }
{
    \If{$L.A[i]<R_2.A[j]$}
    {
        i++;
    }
    \ElseIf{$L.A[i]>R_2.A[j]$}
    {
        j++;
    }
    \Else
    {
        $\mathcal{R}\leftarrow L.A[i]$;\\
        i++; j++;\\
    }
}
\textbf{Return $\mathcal{R}$};
\caption{Intersection($(\pi_{A}L)\cap (\pi_{A}R_2)$)}
\label{alg:PMC_intersection}
\end{algorithm}

\begin{algorithm}[h!]\small
\tcc{PMC and OMC Aggregation}
Initialize a map  $\mathcal{R}$;\\
Scan $L.A$ and get the aggregated score $F(t)$ for each $t\in A$ based on the function $F$;\\
$\mathcal{R}\leftarrow(t,F(t))$;\\
\textbf{Return $\mathcal{R}$};
\caption{Aggregation($\gamma_{A,F}L$)}
\label{alg:PMC_aggregation}
\end{algorithm}

\subsection{OMC Algorithm}
OMC utilizes optimizations -- sorting, clustering, maintaining copies, and applying compressions -- in column databases\cite{journals/ftdb/AbadiBHIM13,stonebraker2005c,abadi2006integrating}. Recall that, OMC does not use hash index, since an individual column in a relationship table is not a primary key. We introduce the optimizations in OMC in details:

\textbf{Dictionary Encoding}
In our experiments, we applied Dictionary encoding to all the columns. Dictionary encodings reduce the amount of data read from main-memory by replacing attribute values with shorter representations (e.g. integers).

\textbf{Clustering and sorting}
To speedup the lookup operation, one optimization is to sort the columns. Consequently, all the tuples with the same values are stored contiguously. By clustering, all the same values are grouped together. Thus, once the first element of each cluster is retrieved, all the others can be obtained via sequential accesses. Since the column is sorted, the first element can be found by using a binary search algorithm.

\textbf{Run Length Encoding}
To further improve the lookup performance, Run-Length Encoding (RLE) is employed by OMC for the sorted columns. RLE compresses runs of the same value to a compact singular representation \cite{journals/ftdb/AbadiBHIM13,abadi2006integrating}. The state-of-the-art RLE is to replace  a run of the same values with a triple : ($t_i$, $s_i$, $l_i$), where $t_i$ is a value, $s_i$ is the first position/row id of $t_i$, and $l_i$ is the number of occurrences of $t_i$.
\begin{gotofull}
For example, if  in $D_1$ column of $R$, $d_3$ appears $100,000$ times and the tuple/row id of the first one is $3,000,000$, then the $100,000$ terms are represented as one triple ($d_3$, $3,000,000$, $100,000$).
\end{gotofull}
OMC uses binary search to find corresponding RLE ($t_i$, $s_i$, $l_i$) for value $t_i$
% , the row ids are continuous, i.e., [$s_i$,$s_i+l_i$]. Then OMC obtains the corresponding values in column $D_2$ by using one random access to locate position $s_i$ and sequential accesses to find the remaining values in position  ($s_i$,$s_i+l_i$]. Note that PMC needs to use many random accesses to find all the values.
% Also note that we apply RLE for all possible columns.

\textbf{Two copies}
One limitation of sorting and clustering is that columns within one relation can only be sorted  and clustered  according to one column, since the other columns should maintain the same order to guarantee the correctness of row ids \smallfootnote{We assume that each column is store separately without maintaining a additional column of row ids. The row ids are omitted and the tuple positions are instead used}. Therefore, the existing sorting and clustering can only speedup lookup performance for one column. Motivated by the idea of storing different copies to support fault-tolerance in \cite{journals/ftdb/AbadiBHIM13,avizienis1984fault}, OMC stores two copies for each relationship table $R$\smallfootnote{Recall that, here we only focus on 2 foreign keys relationship table.} In the experiments section, we will show that the two copies optimization will significantly improve performance, though the penalty is more space usage.

\noindent\textbf{Modifications from PMC to OMC}
Though PMC and OMC have the same operator-at-a-time execution model, they have different generated codes for each operator. The main difference between OMC and PMC are summarized as follows:
\begin{figure}
\center
\includegraphics[width=0.4\textwidth]{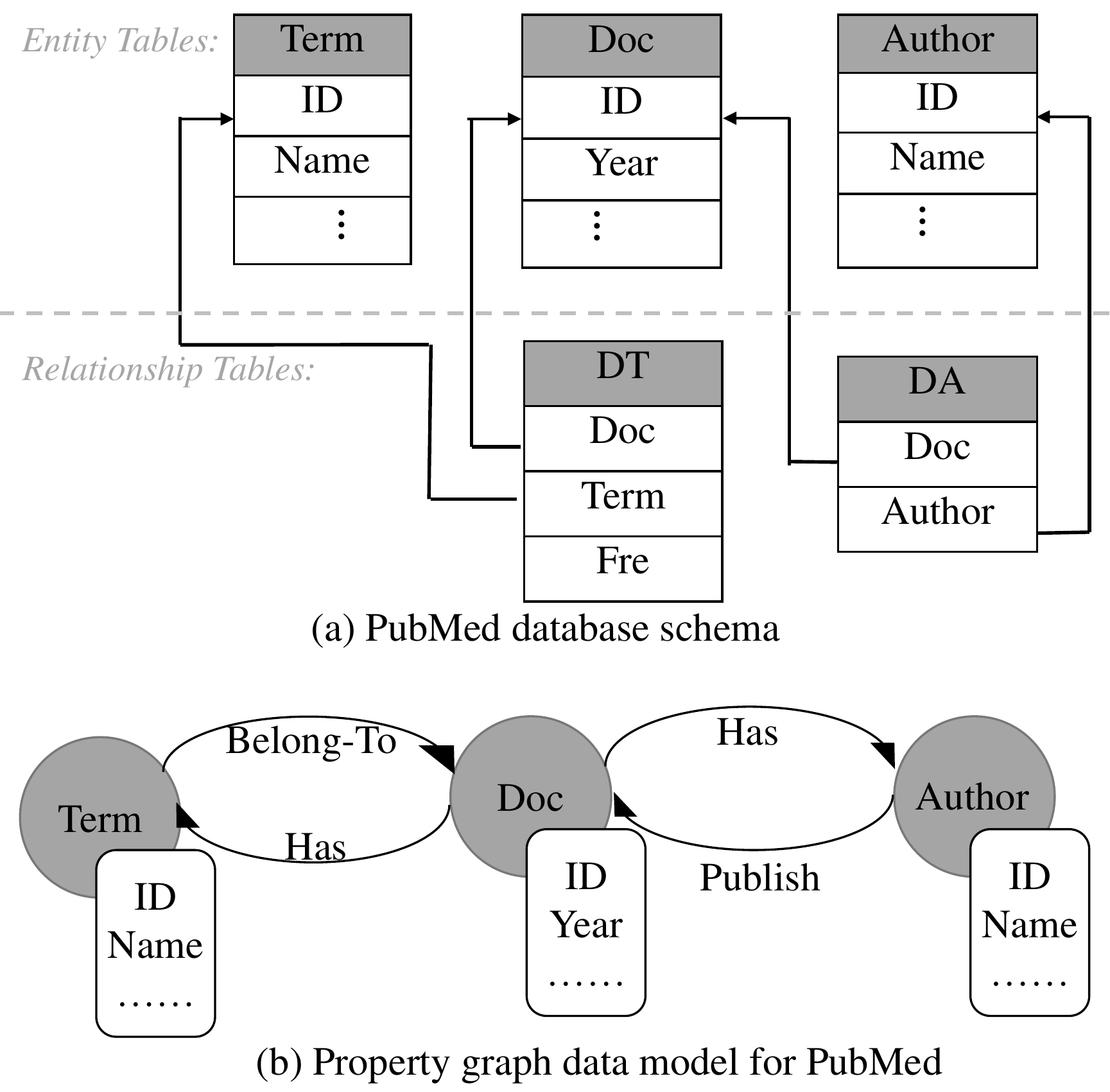}
\vspace{-3mm}
\caption{\textbf{PubMed database schema and property graph data model for PubMed}. Relationship tables \gt{DT} and \gt{DA} have reference keys pointing to entity tables \gt{Doc}, \gt{Term} and \gt{Author}. For its graph data model, there are three types of vertices \gt{Term}, \gt{Doc} and \gt{Author}, which map to the entities in PubMed. In addition, there are four types of edges IN, HAS, INCLUDE, and PUBLISH. Both IN and HAS edges have one attribute \gt{Fre}.}
\label{figure:pubmed-schema-2}
\end{figure}

\begin{compact_enum}
\item{\em Binary search vs. whole column scan.} Columns are sorted and clustered, thus OMC can apply binary search instead of full column scan to find row IDs in Algorithm \ref{alg:PMC_selection} and Algorithm \ref{alg:PMC_join}. But PMC has to scan whole columns.

\item{\em Avoiding row ids vs. maintaining row ids.} OMC encodes the soted columns by RLE, thus each lookup result can be represented as a triple (id, start, end) rather than a list of row IDs in Algorithm \ref{alg:PMC_selection} and Algorithm \ref{alg:PMC_join}. Therefore, OMC directly fetch the lookup results by using ``start'' and ``end'' positions, while PMC needs to store each row id as they are randomly distributed.

\item {\em Merge-based intersection vs. nested-loop intersection. } Values in each column are sorted in OMC, thus OMC can call a merge-based intersection instead of using a nested-loop intersection in Algorithm \ref{alg:PMC_intersection}. However, PMC has to use nested-loop intersection as the values are not sorted.
\end{compact_enum}

\subsection{Graph database}
\label{sec:appendix_graph}
Figure~\ref{figure:pubmed-schema-2} shows the PubMed database schema and its corresponding graph data model. Note that, each tuple in a relationship table is translated into an edge. More precisely, a tuple $(s,t, m_1,...,m_n)$ in a relationship table now is an edge between vertices $s$ and $t$ with attributes $m_1,...,m_n$ on the edge.

Since Neo4j provides its own declarative graph query language, i.e., Cypher. In order to run our queries on Neo4j, we translate all the queries into Cypher as follows:

\begin{figure*}[h!]
\centering
\includegraphics[width=0.9\textwidth]{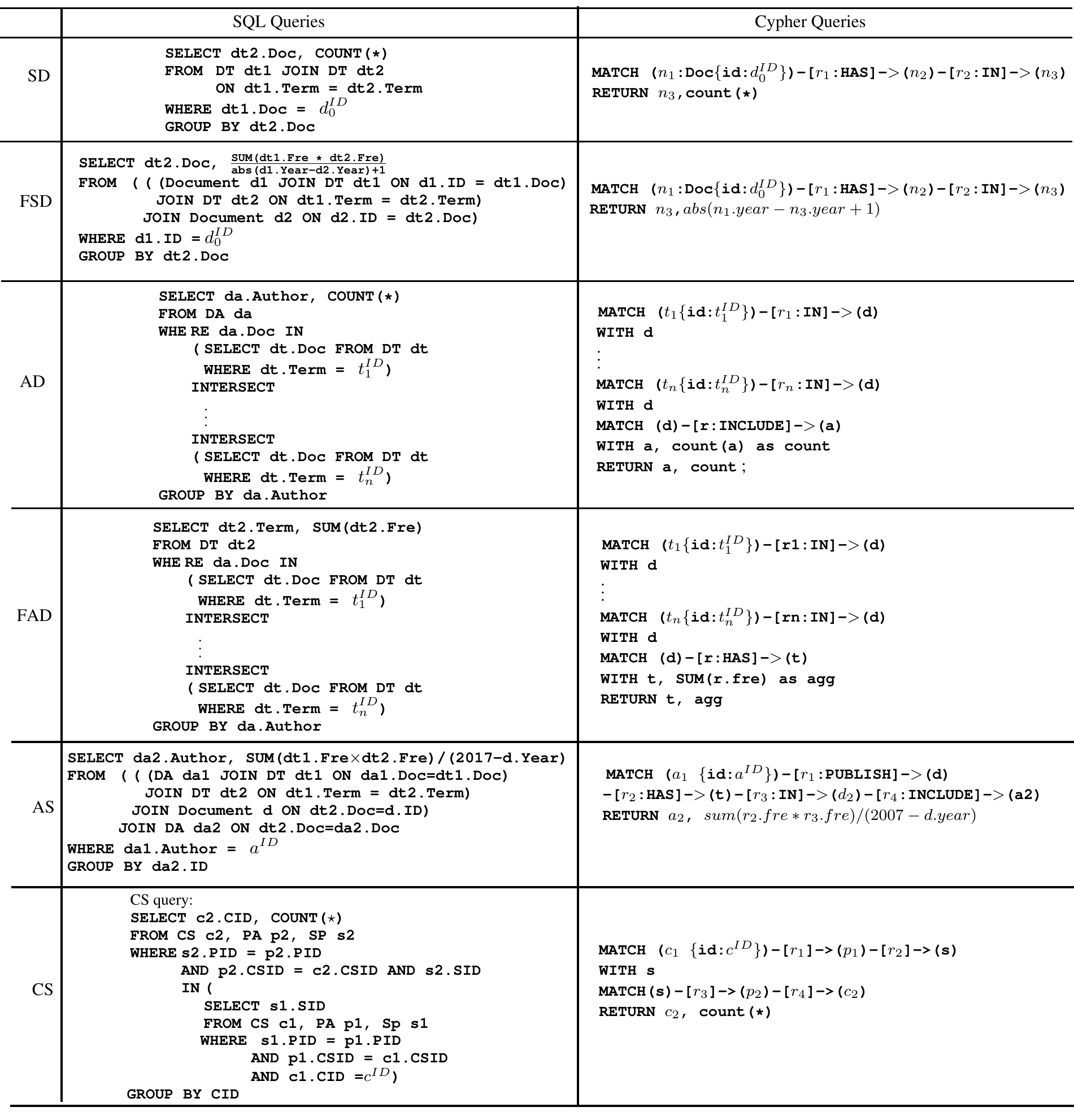}
\caption{\textbf{The equivalent Cypher expressions of the relationship queries employed in this paper.}}\label{exp:fig_avoid_intermediate}
\end{figure*}

\subsection{Additional Experiments}
We conducted experiments to evaluate the performance of different strategies on multi-threading environment, i.e., (1) using spinlock to  protect the correctness of sharing the arrays, and (2) maintaining local arrays for each thread, then aggregating them together at last. Currently, \fastr adopts the first design choice. We modified \fastr to use the second choice as \fastrUAS, which is \fastr that only uses uncompressed array encoding and instead of using spinlocks, it maintains local arrays for each thread.
\begin{table}[htbp]
\small
\centering
\renewcommand{\tabcolsep}{2mm}
\begin{tabular}{l|rrrr} \hline\hline
    & Ave \# results & \fastrUAM & \fastrUA & $\theta$\\\hline
SD  &  27,443,100 & 439.63  & \textbf{177.08} & 73.37\%  \\ \hline
FSD &  27,307,529 & 671.94  & \textbf{435.60} & 35.17\%  \\ \hline
AD  &  200,679    & 183.40    & \textbf{30.33} & 83.46\%  \\ \hline
FAD &  56,518     & \textbf{21.65}    & 25.95 & -19.86\%  \\ \hline
AS  &  20,019,297 & \textbf{3577.60}   & 4510.11 & -26.07\%  \\ \hline
CS  & 5,057       & 25.38    & \textbf{8.62} & 66.04\%  \\ \hline\hline
\end{tabular}
 \caption{\textbf{\fastrUA vs. \fastrUAS (in milliseconds)}. The last column shows the improvements where $\theta$=1-$\frac{\mathrm{\fastrUA}}{\mathrm{\fastrUAS}}$.}
 \label{table:spinlock}
\end{table}

As shown in Table \ref{table:spinlock}, none of them always wins the other one. However, \fastrUA generally achieves higher speedup than \fastrUAS (see the last column). Therefore, \fastr selects the first design choice.

\end{gotofull}

{\small
\bibliographystyle{abbrv}
\bibliography{vldb2015}
}

\end{document}